\documentclass{forarxiv}
\usepackage{color}
\usepackage{epstopdf}
\def\lsim{\mathrel{\mathstrut\smash{\ooalign{\raise2.5pt\hbox{$<$}\cr\lower2.5pt\hbox{$\sim$}}}}}
\def\gsim{\mathrel{\mathstrut\smash{\ooalign{\raise2.5pt\hbox{$>$}\cr\lower2.5pt\hbox{$\sim$}}}}}
\newcommand{\Thydro}{(T^{\rm hydro})}
\newcommand{\Alpha}{\alpha}

\newcommand{\Kn}{\hbox{Kn}}
\newcommand{\difxi}{\zeta}

\begin{document}

\markboth{S.~Gubser, S.~Pufu, F.~Rocha, and A.~Yarom}{Energy loss and the gauge-string duality}

\catchline{}{}{}{}{}

\title{ENERGY LOSS IN A STRONGLY COUPLED THERMAL MEDIUM AND THE GAUGE-STRING DUALITY}

\author{\footnotesize STEVEN S. GUBSER, SILVIU S. PUFU, FABIO D. ROCHA, AND AMOS YAROM}

\address{Joseph Henry Laboratories, Princeton University \\
Princeton, NJ  08540,
United States of America\\
ssgubser, spufu, frocha, ayarom@Princeton.EDU}

\maketitle

\begin{history}
\end{history}

\begin{abstract}
We review methods developed in the gauge-string duality to treat energy loss by energetic probes of a strongly coupled 
thermal medium.  After introducing the black hole description of the thermal medium, we discuss the trailing string behind a 
heavy quark and the drag force that it implies.  We then explain how to solve 
the linearized Einstein equations in the presence of the trailing string and extract from the solutions the energy density and 
the Poynting vector of the dual gauge theory.  We summarize some efforts to compare these calculations to heavy ion 
phenomenology.\vfill

\noindent
This manuscript was typeset using a class file inherited from ws-ijmpe.cls.  Ws-imjpe.cls is copyrighted to World Scientific Publishing Company.
\end{abstract}

\clearpage

\tableofcontents
\markboth{S.~Gubser, S.~Pufu, F.~Rocha, and A.~Yarom}{Energy loss and the gauge-string duality}

\section{Introduction}
\label{S:Introduction}

A long-standing hope, as yet unrealized, is 
that quantum chromodynamics (QCD) will be reformulated as a string theory.  The gauge-string duality\cite{Maldacena:1997re,Gubser:1998bc,Witten:1998qj} provides the closest approach to 
that goal so far attained.  It provides useful computational methods for 
studying strongly coupled gauge theories.  The theory that is most accessible via these methods is ${\cal N}=4$ super-Yang-Mills theory (SYM) in the limit of a large number of colors and large 't~Hooft coupling.\footnote{
${\cal N}=4$ super-Yang-Mills theory is a gauge theory whose matter content consists of gluons, four Majorana fermions in the adjoint representation of the gauge group, and six real scalars, also in the adjoint representation.  All the fields are related to one another by the ${\cal N}=4$ supersymmetry, which completely fixes the Lagrangian once the gauge group and gauge coupling are chosen.  Our interest is in the gauge group $SU(N)$.}
Aspects of the progress in using the gauge-string duality to understand QCD have recently been 
reviewed at a pedestrian level.\cite{Gubser:2009md}  The aim of the current article is a more focused review of efforts to 
understand energy loss by energetic probes of a thermal medium in a strongly coupled gauge theory, such as SYM, which has a string theory dual; and to review how energy loss in SYM can be compared to energy loss in QCD.

At least in simple cases like SYM, the response of an infinite, static, strongly coupled thermal 
medium to an energetic probe can be presumed to be hydrodynamical far from the probe, because hydrodynamic 
perturbations are the only long-wavelength modes available.  In an infinite, interacting thermal medium, there is no radiation, because 
there are no asymptotic states.  In a strongly coupled medium, it is not clear that there is a gauge-invariant distinction between collisional 
energy loss and radiative energy loss.  So the main questions are:
 \begin{enumerate}
  \item {\it What is the rate of energy loss from an energetic probe?}\medskip
  \item {\it What is the hydrodynamical response far from the energetic probe?}\medskip
  \item {\it What gauge-invariant information can be extracted using the gauge-string duality about the non-hydrodynamic region 
near the probe?}\medskip
  \item {\it Do the rate and pattern of energy loss have some meaningful connection to heavy ion phenomenology?}
 \end{enumerate}

In section~\ref{S:Medium} we briefly review the dual description of the thermal state of SYM as an AdS${}_5$-Schwarzschild black hole. The reader interested in a more extensive discussion of the AdS/CFT duality is referred to various reviews in the literature.\cite{Aharony:1999ti,Klebanov:2000me,DHoker:2002aw}  In section~\ref{S:Trailing} we explain how to describe heavy quarks in $\mathcal{N}=4$ SYM using strings in AdS${}_5$, and in section~\ref{S:Magnitude} we extract the drag force acting on the quark via an ``obvious'' and ``alternative'' identification of parameters between SYM and QCD. Also in section 4, we consider how the string theory estimates of drag force relate to the measured nuclear modification factor for heavy quarks.

The response of the stress-tensor to the motion of the quark is dual to the metric perturbations around an AdS black hole. These are studied in section~\ref{S:Einstein}, where we also discuss how the metric perturbations map into the stress tensor of the plasma. In section~\ref{S:Asymptotics} we provide analytic approximations to the stress tensor both near to the moving quark and far from it. The full numerical solution is described in section~\ref{S:Stress}, and its application to heavy-ion phenomenology can be found in section~\ref{S:Hadronization}.  In \ref{S:Glossary} we provide a glossary of mathematical notations used in the main text.

\section{The thermal medium as a black hole}
\label{S:Medium}

${\cal N}=4$ $SU(N)$ super-Yang-Mills theory at finite temperature can be described in terms of $N$ D3-branes near 
extremality.\cite{Gubser:1996de} 
D3-branes are $3+1$ dimensional objects on which strings may end.\cite{Dai:1989ua,Polchinski:1995mt}  Each string can end on one of the $N$ D3-branes, which eventually gives rise to the $SU(N)$ gauge symmetry of SYM.\cite{Witten:1995im}  Having a non-vanishing tension, D-branes themselves warp the spacetime around them.
The near horizon geometry produced by the D3-branes is AdS${}_5$-Schwarzschild times a 
five-sphere, threaded by $N$ units of flux of the self-dual Ramond-Ramond five-form of type~IIB supergravity.\cite{Schwarz:1983qr}  For most calculations of interest to us in this review, one can ignore the five-sphere and work just with the five non-compact dimensions.  

Our starting point is the Einstein equations of type~IIB supergravity in the non-compact directions, which can be recovered from 
the action
 \begin{equation}
  S_{\rm bulk} = {1 \over 2\kappa^2} \int d^5 x \, \sqrt{-G}
    \left[ R + {12 \over L^2} \right] \,. \label{E:EinAction}
 \end{equation}
Here $\kappa$ is the five-dimensional gravitational coupling, $G = \det G_{\mu\nu}$, and $L$ is the radius of the $S^5$.  
Standard relations based ultimately on the quantized charge of the D3-brane lead to
 \begin{equation}
  {L^3 \over \kappa^2} = \left( {N \over 2\pi} \right)^2 \,.
    \label{E:LandKappa}
 \end{equation}
AdS${}_5$-Schwarzschild is a solution of the Einstein equations following from (\ref{E:EinAction}).  Its line element is
 \begin{equation}
  ds^2 = G_{\mu\nu} dx^\mu dx^\nu = 
   {L^2 \over z^2} \left( -h(z) dt^2 + d\vec{x}^2 + 
    {dz^2 \over h(z)} \right) \,, \label{E:LineElement}
 \end{equation}
where the ``blackening function'' $h(z)$ is given by
 \begin{equation}
  h(z) = 1 - {z^4 \over z_H^4} \,.  \label{E:Blackening}
 \end{equation}
In this coordinate system, the conformal boundary of AdS${}_5$ is located at $z=0$. 

The AdS${}_5$-Schwarzschild solution has a horizon at $z=z_H$, whose temperature is
 \begin{equation}
  T = {1 \over \pi z_H} \,.  \label{E:Temp}
 \end{equation}
According to the gauge-string duality, the temperature of the horizon (\ref{E:Temp}) is also the temperature of the thermal 
medium in the dual gauge theory.\cite{Witten:1998qj}  This medium is infinite and static.  Its energy density $\epsilon$ equals the mass per unit 
coordinate volume, $d^3 x$, of the black hole, and its pressure $p$ can also be straightforwardly computed as minus the free 
energy of the black hole.  To leading order in the number of colors $N$ and the 't Hooft coupling $\lambda$, one finds
 \begin{equation}
  {\epsilon \over 3} = p = {\pi^2 \over 8} N^2 T^4 \,.
    \label{E:EOS}
 \end{equation}
These relations reflect the conformal invariance of ${\cal N}=4$ super-Yang-Mills theory, which is exact even at finite $N$ 
and $\lambda$.  If we send $z_H \to \infty$, or equivalently $T \to 0$, we end up with a pure AdS${}_5$ geometry and no black hole.

Perturbations of the AdS${}_5$-Schwarzschild black hole with wavelengths much longer than $1/T$ are described by 
relativistic fluid dynamics,\cite{Policastro:2002se,Bhattacharyya:2008jc} and the viscosity is known to be remarkably small:\cite{Policastro:2001yc}
 \begin{equation}
  {\eta \over s} = {1 \over 4\pi} \,,  \label{E:EtaOverS}
 \end{equation}
again to leading order in the limit of large $N$ and $\lambda$.

In summary: A thermal medium of $SU(N)$ ${\cal N}=4$ super-Yang-Mills theory can be represented as a black hole in 
AdS${}_5$ in the limit of large $N$ and large 't Hooft coupling.  Relations between the radius of AdS${}_5$ and the five-dimensional gravitational coupling, between the temperature and the position of the horizon, between the energy density and 
the temperature, and between the viscosity and the entropy density, can all be derived starting from ten-dimensional type~IIB 
string theory.

\section{The trailing string}
\label{S:Trailing}

${\cal N}=4$ super-Yang-Mills theory has no fundamentally charged quarks.  Instead, its field content is the gluon and its 
superpartners under ${\cal N}=4$ supersymmetry: namely, four Majorana fermions and six real scalars, all in the adjoint 
representation of $SU(N)$.  The AdS${}_5$ description of SYM makes no direct reference to these colored dynamical fields.  The magic of AdS/CFT is to replace the strong coupling dynamics of open strings, whose low-energy quanta participate in the gauge theory, with the gravitational dynamics of closed strings (gravitons, for example) in a weakly curved background.  For example, as we saw in section~\ref{S:Medium}, a finite temperature bath in the gauge theory is replaced by a black hole in AdS${}_5$.  The absence of colored degrees of freedom in the gravitational theory makes it less than obvious how to discuss energetic colored probes of the medium.

A clue comes from the treatment of Wilson loops in the gauge-string duality.\cite{Rey:1998ik,Maldacena:1997re}  A static, infinitely massive, fundamentally charged quark can be represented as a string hanging straight down from the boundary of AdS${}_5$, at $z=0$, into the horizon.  There is no contradiction with the previous statement that ${\cal N}=4$ super-Yang-Mills has no fundamentally charged quarks, because the quark is an external probe of the theory, not part of the theory.  An anti-fundamentally charged quark is represented in the same way as a quark, except that the string runs the other way.  (In type~IIB string theory, strings are oriented, so this statement makes sense.)  If a quark and an anti-quark are both present, it is possible for the strings running down into AdS${}_5$ to lower their total energy by connecting into a shape like a catenary.  From the point of view of the boundary theory, the string configuration gives rise to an attractive potential with a $1/r$ dependence, as conformal invariance says it must.

Let's go back to a single isolated quark---which makes sense in ${\cal N}=4$ even at zero temperature because there is no confinement. 
The string dual to the quark can be described by its embedding in AdS${}_5$.  The two-dimensional spacetime manifold swept out by the string is called a worldsheet and can be parameterized by two coordinates. If we use $t$ and $z$ to parameterize the string worldsheet, then the string in AdS${}_5$, or in AdS${}_5$-Schwarzschild, dual to a static quark at $\vec{x}=0$, is described by the equation
 \begin{equation}
  \hbox{static quark:}\qquad \vec{x}(t,z) = 0 \,.  \label{E:Static}
 \end{equation}
This is a solution to the string equations of motion with boundary conditions such that the string endpoint on the boundary is stationary. By symmetry, \eqref{E:Static} is the only possible solution with these boundary conditions.  In the absence of a medium, it must be that the description of a moving quark can be obtained by performing a Lorentz boost on the description of a static quark.  If the boost is in the $x_1$ direction, and we continue to use $t$ and $z$ to parameterize the worldsheet, then the equations describing the string dual to a quark moving in the absence of a medium are
\begin{equation}
  \hbox{quark moving in vacuum:}\qquad
   x_1(t,z) = vt \,,\quad \vec{x}_\perp(t,z) = 0 \,,  \label{E:Boosted}
 \end{equation}
where $v$ is the velocity of the quark and $\vec{x}_\perp = (x_2,x_3)$.\footnote{It would be more proper to use $(x^1,x^2,x^3)$ in place of $(x_1,x_2,x_3)$, and reserve the notation $x_\mu$ for $G_{\mu\nu} x^\nu$.  However, it simplifies notation to set $x_i = x^i$ for $i=1,2,3$, and we will do this consistently.}  Given that (\ref{E:Static}) solves the equations of motion of classical string theory, it's guaranteed that (\ref{E:Boosted}) does too, because the equations are invariant under the boost in the $x_1$ direction, and the line element (\ref{E:LineElement}) with $h=1$ is invariant under any boost acting on the $(t,\vec{x})$ coordinates.

When $h \neq 1$, signaling the presence of a medium, a moving quark is not the same as a static one.  Naively, one might 
nevertheless try to represent the moving quark in terms of the string shape described in (\ref{E:Boosted}).  Let's see why this 
is problematic.  String dynamics is defined in terms of the metric on the worldsheet.  We chose to parameterize the 
worldsheet by $t$ and $z$, but any coordinates $\sigma^\alpha = (\sigma^1,\sigma^2)$ could have been chosen.  For a general 
embedding $x^\mu = x^\mu(\sigma^1,\sigma^2)$ of a string worldsheet into spacetime, the worldsheet metric is
 \begin{equation}
  g_{\alpha\beta} = {\partial x^\mu \over \partial\sigma^\alpha}
   {\partial x^\nu \over \partial\sigma^\beta} G_{\mu\nu} \,.
   \label{E:WorldsheetMet}
 \end{equation}
What (\ref{E:WorldsheetMet}) says is that times and distances along the string are measured the same way as in the ambient 
spacetime.  Using $\sigma^\alpha = (t,z)$ and the ansatz (\ref{E:Boosted}), one immediately finds
 \begin{equation}
  g_{\alpha\beta} = {L^2 \over z^2} \begin{pmatrix} -h+v^2 & 0 \\
    0 & {1 \over h} \end{pmatrix} \,.  \label{E:BoostedMetric}
 \end{equation}
This shows that the $z$ direction on the worldsheet is always spacelike and that the $t$ direction is timelike only when 
$h>v^2$.  Using (\ref{E:Blackening}), this condition is equivalent to
 \begin{equation}
  z < z_* \equiv z_H \sqrt[4]{1-v^2} \,.  \label{E:WShorizon}
 \end{equation}
The part of the string worldsheet with $z>z_*$ is purely spacelike, which means that the string is locally moving faster than the speed of light.  This does not make sense if we are aiming to describe the classical dynamics of a string moving in real Minkowski time.

There is a simpler way to arrive at (\ref{E:WShorizon}): any trajectory of a point particle in the bulk of AdS${}_5$-Schwarzschild with $x_1 = vt$ must be spacelike---that is, the speed of the particle will exceed the local speed of light---if the inequality $z < z_*$ is violated.  This conclusion holds even when there are other components of the velocity, but the inequality is sharp only in the case where there aren't.

Physically, what we learn from (\ref{E:WShorizon}) is that it is harder and harder to move forward as one approaches the 
horizon.  This evokes the idea that there must be some drag force from the medium.  But how does one get at that drag 
force?  An answer was provided by two groups,\cite{Herzog:2006gh,Gubser:2006bz} and closely related work on fluctuations appeared at the same time.\cite{Casalderrey-Solana:2006rq}  The string does not hang straight 
down from the quark: rather, it trails out behind it.  If the shape is assumed not to change as the quark moves forward, and if it 
respects the $SO(2)$ symmetry rotating the $\vec{x}_\perp$ coordinates, then it must be specified by a small variant of 
(\ref{E:Boosted}):
 \begin{equation}
  x_1(t,z) = vt + \xi(z)  \label{E:TrailingAnsatz}
 \end{equation}
for some function $\xi(z)$.  If we insist that the quark's location on the boundary is $x_1 = vt$, then we must have $\xi \to 0$ 
as $z \to 0$.

To determine $\xi(z)$, one must resort to the classical equations of motion for the string.  These follow from the action
 \begin{equation}
  S_{\rm string} = -{1 \over 2\pi\alpha'}
    \int d^2\sigma \, \sqrt{-g} \,,  \label{E:NambuGoto}
 \end{equation}
where $g = \det g_{\alpha\beta}$.  The parameter $\alpha'$ is related to $L$ and the 't Hooft coupling by
 \begin{equation}
  L^4 = \lambda \alpha'^2 \,.  \label{E:tHooft}
 \end{equation}
Here we define $\lambda = g_{\rm YM}^2 N$, so that a quantity analogous to the coupling $\alpha_s$ in QCD is $\alpha_{\rm YM} = 
\lambda / 4\pi N$.  The classical equations of motion following from the action (\ref{E:TrailingAnsatz}) take the form
 \begin{equation}
  \nabla_\alpha p^\alpha{}_\mu = 0 \,,  \label{E:WSconservation}
 \end{equation}
where 
 \begin{equation}
  p^\alpha{}_\mu \equiv -{1 \over 2\pi\alpha'} g^{\alpha\beta}
    G_{\mu\nu} \partial_\beta x^\nu  \label{E:ConjugateCurrent}
 \end{equation}
is the momentum current on the worldsheet conjugate to the position $x^\mu$.  Plugging (\ref{E:TrailingAnsatz}) into 
(\ref{E:ConjugateCurrent}), one straightforwardly finds that
 \begin{equation}
  {d\xi \over dz} = {\pi_\xi \over h}
    \sqrt{h-v^2 \over {L^4 \over z^4} h - \pi_\xi^2} \,,
    \label{E:XiShape}
 \end{equation}
where $\pi_\xi$ is a constant of integration.  In order for $\xi(z)$ to be real, the right hand side of (\ref{E:XiShape}) must be 
real.  There are three ways this can happen
 in a manner consistent with the assumption of steady-state behavior:\footnote{Technically, there is a fourth way, but its significance is obscure to us.  
A string can lead down to $z=z_*$ in the shape of the trailing string, (\ref{E:TrailingString}), and then turn around and retrace 
its path back up again.  The energy localized at the kink must grow linearly with time, which means that this configuration is 
not quite a steady-state solution.}
 \begin{itemize}
  \item One can choose $\pi_\xi = 0$.  This leads to $\xi=0$, which shows that (\ref{E:Boosted}) is formally a solution of the 
equations of motion.  But it is not a physical solution---at least, not in the context of classical motions of a string---because of 
the problem with the signature of the worldsheet metric.  The action for this solution is complex.
  \item One can arrange for the worldsheet never to go below $z=z_*$, and choose $\pi_\xi$ small enough that the 
denominator inside the square root in (\ref{E:XiShape}) is always positive.  There is indeed a one-parameter family of such 
solutions, and they describe a heavy quark and anti-quark in a color singlet state propagating without drag (at the level of the 
current treatment), one behind the other.  Similar states were studied by other groups,\cite{Peeters:2006iu,Liu:2006nn,Chernicoff:2006hi} but 
they do not capture the dynamics of a single quark propagating through the plasma, so we do not consider them further here.
  \item One can choose
 \begin{equation}
  \pi_\xi = \pm {L^2 \over z_*^2} \sqrt{h(z_*)}
    = \pm {v \over \sqrt{1-v^2}} {L^2 \over z_H^2} \,,
    \label{E:PiChoice}
 \end{equation}
so that the denominator inside the square root in (\ref{E:XiShape}) changes sign at the same value of $z$ as the numerator, 
namely $z=z_*$, rendering the ratio inside the square root everywhere positive and finite.  Choosing the sign that makes $
\pi_\xi$ positive means that the string trails out in front of the quark instead of behind it.  Although this is technically a solution, 
it does not describe energy loss and should be discarded.  Choosing $\pi_\xi$ negative leads to the trailing string solution that 
we are interested in.  Eq.~(\ref{E:XiShape}) then straightforwardly leads to
 \begin{equation}
  \xi = -{z_H v \over 4i}
    \left( \log {1 - iy \over 1 + iy} + 
      i \log {1 + y \over 1 - y} \right) \,,
        \label{E:TrailingString}
 \end{equation}
where we have introduced a rescaled depth variable,
 \begin{equation}
  y = {z \over z_H} \,.  \label{E:YDef}
 \end{equation}
 \end{itemize}
There is another way to justify the choice of the minus sign, not the plus sign, in (\ref{E:PiChoice}):\cite{CasalderreySolana:2007qw} The solution (\ref{E:TrailingString}) is non-singular at the future event horizon, whereas the solution one would get 
with the opposite sign choice is singular.  To understand this point, it is convenient to pass to Kruskal coordinates, defined 
implicitly by the equations
 \begin{equation}
  UV = -{1 - y \over 1 + y} e^{-2 \tan^{-1} y} \qquad
  {V \over U} = -e^{4t/z_H} \,.  \label{E:KruskalCoords}
 \end{equation}
In the region outside the horizon, $U<0$ and $V>0$, while in the region inside the future horizon, $U>0$ and $V>0$.  The trailing string solution (\ref{E:TrailingAnsatz}), with $\xi$ given by (\ref{E:TrailingString}), can be extended to a non-singular 
solution over the union of these two regions:
 \begin{equation}
  x_1 = {v \over 2} \log V + v \tan^{-1} y \,.
    \label{E:UnionSolution}
 \end{equation}
Thus, the logarithmic singularity in $\xi$ at $y=1$ (meaning $z=z_H$) is a singularity not at the future horizon, which is at 
$U=0$, but at the past horizon, which is at $V=0$.  Reversing the sign choice in (\ref{E:PiChoice}) would lead to a solution 
that is singular at the future horizon but not the past horizon.  Causal dynamics in the presence of a black hole horizon can 
generally be described in terms of functions which are smooth at the future horizon.

At the level of our presentation, it has been assumed rather than demonstrated that the trailing string is a stable, steady-state 
configuration representing the late time behavior of a string attached to a moving quark on the boundary of AdS${}_5$-
Schwarzschild.  In fact, this has been fairly well checked.\cite{Herzog:2006gh,Gubser:2006nz}

The description we have given of the trailing string is not limited to Schwarzschild black holes in AdS${}_5$. One may extend this analysis to various other black hole geometries which asymptote to AdS${}_5$ near their boundary.\cite{Herzog:2006se,Talavera:2006tj,Caceres:2006dj,MichalogiorgakisPHD,Herzog:2007kh,Gubser:2007ni} 
These geometries describe theories which are deformations of SYM\@.  The literature also includes a discussion of the distribution of energy along the string\cite{Chernicoff:2008sa} and an interpretation of the shape of the string in terms of a rapid cascade of strongly coupled partons.\cite{Hatta:2008tx}

\section{The magnitude of the drag force}
\label{S:Magnitude}

As we explained in the previous section, an infinitely massive, fundamentally charged quark moving at speed $v$ in the $x_1$ direction through an infinite, static, thermal medium of ${\cal N}=4$ super-Yang-Mills theory can be described at strong coupling in terms of the trailing string solution (\ref{E:TrailingString}).  The quark cannot slow down because it has infinite mass.  However, it does lose energy and momentum at a finite, calculable rate.  In five-dimensional terms, this energy can be thought of as flowing down the string toward the black hole horizon.  In four-dimensional terms, energy and momentum emanates from the quark and eventually thermalizes.  To calculate the four-momentum $\Delta p_m$ delivered from the quark to the bath over a time $\Delta t$, one can integrate the conserved worldsheet current $p^\alpha{}_m$ of spacetime energy-momentum over an appropriate line-segment ${\cal I}$ on the worldsheet.  ${\cal I}$ should cover a time interval $\Delta t$, and it can be chosen to lie at a definite depth $z_0$ in AdS${}_5$.  The four-momentum $\Delta p_m$ is\footnote{There is an explicit minus sign in \eqref{E:DeltaP} which doesn't appear in the analogous equation of one of the original works.\cite{Gubser:2006bz}  This is due to use of the $z$ variable, which increases as one goes deeper into AdS${}_5$, instead of the $r=L/z$ variable, which increases as one goes out toward the boundary.}
 \begin{equation}
  \Delta p_m = -\int_{\cal I} dt \sqrt{-g} \, p^z{}_m \,.  \label{E:DeltaP}
 \end{equation}
Because the trailing string is a steady-state configuration, four-momentum is lost at a constant rate:
 \begin{equation}
  {dp_m \over dt} = -\sqrt{-g} \, p^z{}_m \,.
 \end{equation}
In particular, the drag force can be defined as
 \begin{equation}
  F_{\rm drag} = {dp_1 \over dt} = -\sqrt{-g} p^z{}_1 = 
    -{L^2 \over 2\pi z_H^2 \alpha'} {v \over \sqrt{1-v^2}} \,.  \label{E:DFone}
 \end{equation}
Using (\ref{E:Temp}) and (\ref{E:tHooft}), one obtains
 \begin{equation}
  F_{\rm drag} = -{\pi\sqrt\lambda \over 2} T^2 {v \over \sqrt{1-v^2}} \,.
    \label{DragForce}
 \end{equation}
If, instead of an infinitely massive quark, we consider a quark with finite but large mass $m$, then using the standard relativistic expression
 \begin{equation}
  p = {mv \over \sqrt{1-v^2}}  \label{StandardP}
 \end{equation}
leads to
 \begin{equation}
  F_{\rm drag} = -{\pi\sqrt\lambda \over 2} T^2 {p \over m} \,.  \label{FiniteMass}
 \end{equation}
It has been explained\cite{Herzog:2006gh} that (\ref{StandardP}) receives corrections when a finite mass quark is described as a string ending at a definite depth $z=z_*$ on a D7-brane.  While these corrections are interesting, it would take us too far from the main purpose of this review to give a proper explanation of how the D7-branes modify the physics.  Our discussion is formal because we derived the result (\ref{DragForce}) in the strict $m=\infty$ limit and then applied it to finite mass quarks.

From (\ref{FiniteMass}) it is clear that the drag force causes the momentum of a quark to fall off exponentially:
 \begin{equation}
  p(t) = p(0) e^{-t/t_{\rm quark}} \qquad\hbox{where}\qquad 
   t_{\rm quark} = {2 \over \pi\sqrt\lambda} {m \over T^2} \,.
     \label{MomentumFalloff}
 \end{equation}
In order to make a physical prediction for QCD, we must plug in sensible values for $m$, $\lambda$, and $T$.  The effective quark mass in the thermal medium is already non-trivial to specify precisely, but $m_c = 1.5\,{\rm GeV}$ for charm and $m_b = 4.8\,{\rm GeV}$ for bottom are reasonably representative values which were used in a recent phenomenological study.\cite{Akamatsu:2008ge}  We will review here two approaches\cite{Gubser:2006qh} to specifying $\lambda$ and $T$.  The first approach, used earlier in a calculation of the jet-quenching parameter $\hat{q}$ from a lightlike Wilson loop in ${\cal N}=4$ super-Yang-Mills theory,\cite{Liu:2006ug} is to identify the temperature $T_{\rm SYM}$ with the temperature $T_{\rm QCD}$, and then identifying the gauge coupling $g_{\rm YM}$ of super-Yang-Mills with the gauge coupling $g_s$ of QCD evaluated at temperatures typical of RHIC.  Because of the proximity of the confinement transition, $g_s$ has substantial uncertainty.  A standard choice is $\alpha_s = 0.5$, corresponding to $g_s \approx 2.5$.  With the number of colors $N$ set equal to $3$, one finds $\lambda \approx 6\pi$.  We will refer to this as the ``obvious scheme.''

The second approach, called the ``alternative scheme,'' is based on two ideas.  The first idea is that it may make more sense to compare SYM to QCD at fixed energy density than fixed temperature.  SYM has $\epsilon \propto T^4$, and so does QCD, approximately: this approximation is surprisingly good for $T \geq 1.2 T_c$, according to lattice data.\cite{Karsch:2001cy}  But the constant of proportionality is about $2.7$ times bigger for SYM than for QCD.\footnote{This mismatch has previously been stated\cite{Gubser:2006qh} as a factor of $3$ rather than $2.7$.  Some uncertainty exists on both the SYM and the QCD sides, because of finite coupling effects and time discretization, respectively; but $2.7$ is probably closer to the true figure.}  That is, SYM has about $2.7$ times as many degrees of freedom as QCD above the confinement transition.  So $\epsilon_{\rm SYM} = \epsilon_{\rm QCD}$ implies $T_{\rm SYM} \approx T_{\rm QCD}/(2.7)^{1/4}$.  This identification leads to a suppression of $F_{\rm drag}$ by a factor of $\sqrt{2.7}$ relative to the obvious scheme.  The second idea behind the alternative scheme is that the 't~Hooft parameter $\lambda$ in string theory can be determined by comparing the static force between a quark and an anti-quark, as calculated in string theory, to the same force calculated in lattice gauge theory.  The string theory calculation is based on a U-shaped string connecting the quark and the anti-quark.  This string pulls on the static quarks in a fashion that is similar to how the trailing string pulls on a moving quark.  The lattice calculation is based on computing the excess free energy due to the presence of an external quark and anti-quark in a thermal bath.  There is a significant difficulty: in the simplest string theory calculation, based only on the U-shaped string, the force between the quark and anti-quark vanishes for separations larger than some limiting distance $r_*$, and this distance is quite small: $r_* \approx 0.24\,{\rm fm}$ when $T_{\rm SYM} \approx 195\,{\rm MeV}$ (corresponding to $T_{\rm QCD} = 250\,{\rm MeV}$).  It has been pointed out\cite{Bak:2007fk} that exchange of closed strings between two long strings describing the quark and anti-quark at separations $r > r_*$ contribute to the quark-anti-quark force at the same order in $N$ as the U-shaped string.  Unfortunately, it is hard to compute the contribution of closed string exchange.  The approach\cite{Gubser:2006qh} is therefore to match the U-shaped string computation to lattice data\cite{Kaczmarek:2005ui} near the limiting distance $r_*$.  The result of this matching is $\lambda \approx 5.5$.
 \begin{table}[pt]
  \tbl{\label{T:schemes} The obvious and alternative schemes for comparing SYM and QCD.\smallskip} 
  {
   \begin{tabular}{@{}cccccccccc@{}} \toprule
    Scheme & $T_{\rm QCD}$ & $T_{\rm SYM}$ & 
      $\epsilon_{\rm QCD}$ & $\epsilon_{\rm SYM}$ & $\lambda$ &
      $m_c$ & $m_b$ & $t_c$ & $t_b$  \\
     & MeV & MeV & ${\rm GeV}/{\rm fm}^3$ & ${\rm GeV}/{\rm fm}^3$ & 
      GeV & GeV & fm & fm  \\ \colrule
    obvious & $250^a$ & 250 & $5.6^b$ & 15 & $6\pi$ & 1.5 & 4.8 & 
      0.69 & 2.2 \\
    alternative & $250^a$ & 195 & $5.6^b$ & 5.6 & 5.5 & 1.5 & 4.8 & 
      2.1 & 6.8 \\ \botrule
   \end{tabular}
  }
  \begin{tabfootnote}
   \ \\[5pt]
   \tabmark{a} We set $T_{\rm QCD} = 250\,{\rm MeV}$ because this is a typical temperature scale for heavy ion collisions at $\sqrt{s_{NN}} = 200\,{\rm GeV}$.\\[5pt]
   \tabmark{b} We use $\epsilon/T^4 \approx 11$ for QCD.
  \end{tabfootnote}
 \end{table}

As we show in table \ref{T:schemes}, heavy quark relaxation times $t_c$ and $t_b$ are remarkably short when one uses the obvious scheme, and somewhat larger in the alternative scheme.  The uncertainties in $t_c$ and $t_b$ are substantial: even if one accepts the ideas behind the alternative scheme, one should probably regard the resulting relaxation times as uncertain by a factor of $1.5$.

An experimental study\cite{Adare:2006nq} favors a model\cite{vanHees:2005wb} in which $t_c$ is roughly $4.5\,{\rm fm}$ at $T_{\rm QCD} = 250\,{\rm MeV}$, as estimated from plots from a detailed exposition of that model.\cite{vanHees:2004gq}  
This seems to indicate that the string theory estimates of $t_c$ and $t_b$, even in the alternative scheme, are too short.  
However, the results of a recent phenomenological study\cite{Akamatsu:2008ge} favor a range of parameters that is consistent with the string theory predictions translated to QCD using the alternative scheme.

Let us briefly review the recent study.\cite{Akamatsu:2008ge}  The starting point is the Langevin equation, which in the It\^o discretization scheme takes the form
 \begin{equation}
  \Delta \vec{x}(t) = {\vec{p} \over E} \Delta t \qquad
  \Delta \vec{p}(t) = -\Gamma \vec{p} \Delta t + \vec\xi(t) \,.
    \label{ItoLangevin}
 \end{equation}
Here $\vec\xi(t)$ is a stochastic force, assumed to be Gaussian and uncorrelated from one time-step to the next.  The strength of the stochastic force is related to the drag coefficient $\Gamma$ by demanding that the relativistic Maxwell-Boltzmann distribution is preserved by the time evolution (\ref{ItoLangevin}).  Ordinary relativistic kinematics, $E = \sqrt{\vec{p}^2 + m^2}$, are assumed.  It is also assumed that
 \begin{equation}
  \Gamma = \gamma {T^2 \over m} \,,  \label{HiranoGamma}
 \end{equation}
where $\gamma$ is a dimensionless quantity with no $p$ dependence.  Evidently, $t_{\rm quark} = 1/\Gamma$.  The temperature in (\ref{HiranoGamma}) is $T_{\rm QCD}$, whereas the temperature in (\ref{MomentumFalloff}) is $T_{\rm SYM}$.  Comparing these two equations, one finds that the string theory prediction is
 \begin{equation}
  \gamma = {\pi\sqrt\lambda \over 2} 
    \left( {T_{\rm SYM} \over T_{\rm QCD}} \right)^2 = 
   \left\{ \vcenter{\openup1\jot
     \halign{\strut \span\hbox{#} & \span\hbox{#} \hfil \cr
      6.8 & obvious scheme \cr
      2.2 & alternative scheme. \cr
     }}
   \right.  \label{gammaValues}
 \end{equation}
The alternative scheme value in (\ref{gammaValues}) is fractionally larger than the one quoted in the study under discussion,\cite{Akamatsu:2008ge} due to the use here of the factor $2.7$ for the ratio of degrees of freedom between SYM and QCD, as compared to $3$ in previous work.\cite{Gubser:2006qh}  (It is a numerical coincidence that the dimensionless factor $\gamma$ is the same, in the alternative scheme, as $t_c$ in femtometers when $T_{\rm QCD}=250\,{\rm MeV}$.)

\begin{figure}[th]
  \centerline{\includegraphics[width=4in]{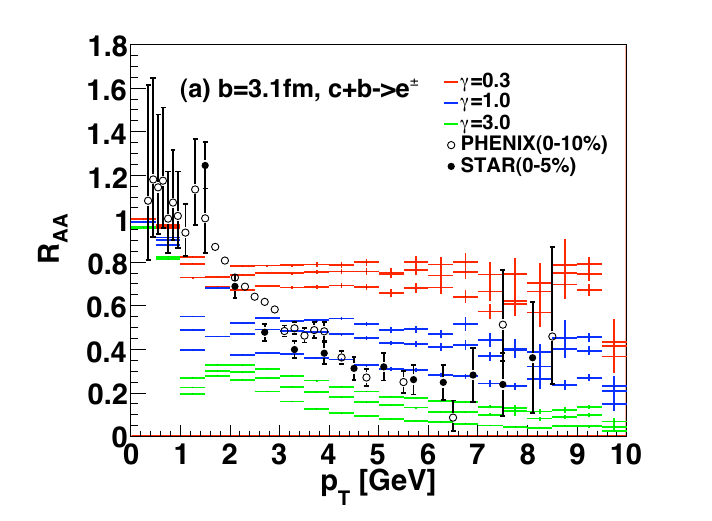}}
  \caption[Short]{(Color online.)  Comparison\cite{Akamatsu:2008ge} between experimental data for gold-gold collisions with impact parameter $b = 3.1\, {\rm fm}$ and theoretical predictions based on Langevin dynamics of heavy quarks in a hydrodynamically expanding plasma.  The open circles correspond to central collisions in a PHENIX experiment, \cite{Adare:2006nq} while the solid circles come from a STAR experiment.\cite{Abelev:2006db}  Theoretical predictions are plotted for $\gamma = 0.3$ (red), $1$ (blue), and $3$ (green), with $\gamma$ is as defined in \eqref{HiranoGamma}.  Reprinted with permission from the authors from Y.~Akamatsu, T.~Hatsuda, and T.~Hirano, ``Heavy Quark Diffusion with Relativistic Langevin Dynamics in the Quark-Gluon Fluid,'' {\tt 0809.1499}.}\label{F:RAAfigure}
 \end{figure}

The next step of the study\cite{Akamatsu:2008ge} is to compare Langevin dynamics of heavy quarks in a hydrodynamically expanding plasma to PHENIX\cite{Adare:2006nq} and STAR\cite{Abelev:2006db} data on the nuclear modification factor $R_{AA}$ for non-photonic electrons---meaning electrons and positrons coming from decays of heavy-quark mesons.  Because of the treatment of hadronization, the theoretical results are deemed trustworthy only when the transverse momentum $p_T$ of the non-photonic electron is at least $3\,{\rm GeV}$\@.\footnote{The electron carries only a fraction of the $p_T$ of the charm quark that led to its production.  This fraction varies, but a reasonable rule-of-thumb value is $1/2$.}  For fairly central collisions (impact parameter $b = 3.1\,{\rm fm}$), agreement between theory (using the alternative scheme) and experiment is best for $\gamma$ between $1$ and $3$: see figure \ref{F:RAAfigure}.
Thus the prediction (\ref{gammaValues}) of string theory in the alternative scheme can reasonably be said to agree with data to within the uncertainties of the calculations.  These uncertainties stem in large part from the difficulty of comparing SYM to QCD; however, it is also clear that the treatment of hadronization is a significant hurdle.\cite{Akamatsu:2008ge}

Fluctuations of the trailing string\cite{Casalderrey-Solana:2006rq,Gubser:2006nz,CasalderreySolana:2007qw,deBoer:2008gu} provide direct access to the stochastic forces in (\ref{ItoLangevin}).  In the non-relativistic limit, the size of these forces, relative to the drag force, is exactly what is needed to equilibrate to a thermal distribution.  Indeed, the original calculations of drag force\cite{Herzog:2006gh,Gubser:2006bz} and stochastic forces\cite{Casalderrey-Solana:2006rq} were done independently.  For relativistic quarks, the stochastic forces are enhanced by powers of $1/(1-v^2)$, including enhancement of longitudinal stochastic forces by the startlingly large factor $1/(1-v^2)^{5/4}$.  This is larger by $1/(1-v^2)^{3/4}$ than what is needed to equilibrate to a thermal distribution.  Another issue is that the correlation time $t_{\rm cor}$ for these stochastic forces grows with velocity: based on results for the relevant Green's function\cite{Gubser:2006nz} one may estimate
 \begin{equation}
  t_{\rm cor} \approx {1 \over \pi T (1-v^2)^{1/4}} \,.  \label{EstimateTCor}
 \end{equation}
If we use $t_{\rm cor} < t_{\rm quark}$ as a criterion of validity for the Langevin dynamics, then we get a limit on the Lorentz boost factor:
 \begin{equation}
  {1 \over \sqrt{1-v^2}} < {4 \over \lambda} {m^2 \over T^2} \,.
    \label{BoostLimit}
 \end{equation}
Approximately the same inequality arises from demanding that when the string ends on a D7-brane, its endpoint should not move superluminally.  The same inequality, but with $m$ replaced by some fixed scale $\mu$, also arises from demanding that the worldsheet horizon\cite{Gubser:2006nz,CasalderreySolana:2007qw} should be at a depth in AdS${}_5$ corresponding to a scale $\mu$ where the dynamics of QCD is far from weakly coupled.  Plugging in numbers in the alternative scheme, with $\mu = 1.2\,{\rm GeV}$ and $T_{\rm QCD} = 250\,{\rm GeV}$, one finds that $1/\sqrt{1-v^2} \lsim 30$.  This inequality should be understood as quite a rough estimate, because of the quadratic dependence on the quantity $\mu$ which is only qualitatively defined.  For a charm quark, the corresponding limit on the transverse momentum of the non-photonic electron is $p_T \lsim 20\,{\rm GeV}$.  This is well in excess of the highest momentum for which there are statistically significant data; moreover, for $p_T$ more than a few ${\rm GeV}$, bottom quarks dominate, and for them the bound $1/\sqrt{1-v^2} \lsim 30$ translates to $p_T \lsim 70\,{\rm GeV}$ for non-photonic electrons.  The upshot is that heavy quarks at RHIC do not obviously fall outside the regime of validity of a self-consistent Langevin treatment based on the trailing string---except for the troublesome scaling of longitudinal stochastic forces as $1/(1-v^2)^{5/4}$, whose consequences, we feel, are ill-understood.

In another study,\cite{Horowitz:2007su} it is argued that energy loss in the perturbative and strongly coupled regimes have experimentally distinguishable signatures for large enough transverse momenta, $p_T \gg m_b$, which will be attained at the LHC\@.
A convenient observable that distinguishes between predictions of string theory and perturbative QCD is the ratio of the nuclear modification factors for $b$ and $c$ quarks,
 \begin{equation}
  \label{E:RbcDef}
  R^{cb} = {R^{c}_{AA}(p_T) \over R^{b}_{AA}(p_T)} \,.
 \end{equation}
In particular, the drag force formula \eqref{FiniteMass} implies that at large enough $p_T$, one has
 \begin{equation}
  \label{E:RcbAdS}
  R_{\rm AdS}^{cb} \approx {m_c \over m_b} \approx 0.3\,.
 \end{equation}
where we used the bottom and charm masses quoted right after equation \eqref{MomentumFalloff}.  In contrast, perturbative QCD predicts that
 \begin{equation}
  \label{E:RcbpQCD}
  R^{cb}_{\rm pQCD} \approx 1 - {p_{cb} \over p_T}
 \end{equation}
at large $p_T$, where $p_{cb}$ is a relevant momentum scale.  So according to perturbative QCD, $R^{cb}$ should approach unity at large $p_T$.  At sufficiently large $p_T$, the trailing string treatment presumably fails, and perturbative QCD presumably is correct.  But as discussed following (\ref{BoostLimit}), it is difficult to give a good estimate of the characteristic value of $p_T$ where the trailing string fails.  Absent a reliable estimate of the characteristic $p_T$, the upshot is that if the measured $R^{cb}$ is significantly below the perturbative prediction, the trailing string should be considered as a candidate explanation.

\section{The perturbed Einstein equations}
\label{S:Einstein}

Given that an external quark dual to the trailing string described in section~\ref{S:Trailing} experiences drag, one might ask what happens to the energy that the quark deposits in the medium.  At scales much larger than the inverse temperature, one expects the excitations present in the medium due to interactions with the moving quark to be well-described by linearized hydrodynamics.  Earlier investigations of linearized hydrodynamics revealed the presence, for generic sources, of both a sonic boom and a diffusion wake.\cite{CasalderreySolana:2006sq} The sonic boom is a directional structure, which, in an ideal fluid, is concentrated on the Mach cone, but in a real fluid there is broadening because of viscous effects.  It appears only when the probe is moving faster than the speed of sound in the medium, and it comes from constructive interference among spherical waves sourced by the quark along its trajectory.  The diffusion wake is a flow of the medium behind the quark in the direction of the quark's motion.  It too is broadened by viscous effects.  We will discuss the hydrodynamic limit in more detail in section~\ref{SSS:hydro}.

\begin{figure}
  \centerline{\includegraphics[width=4in]{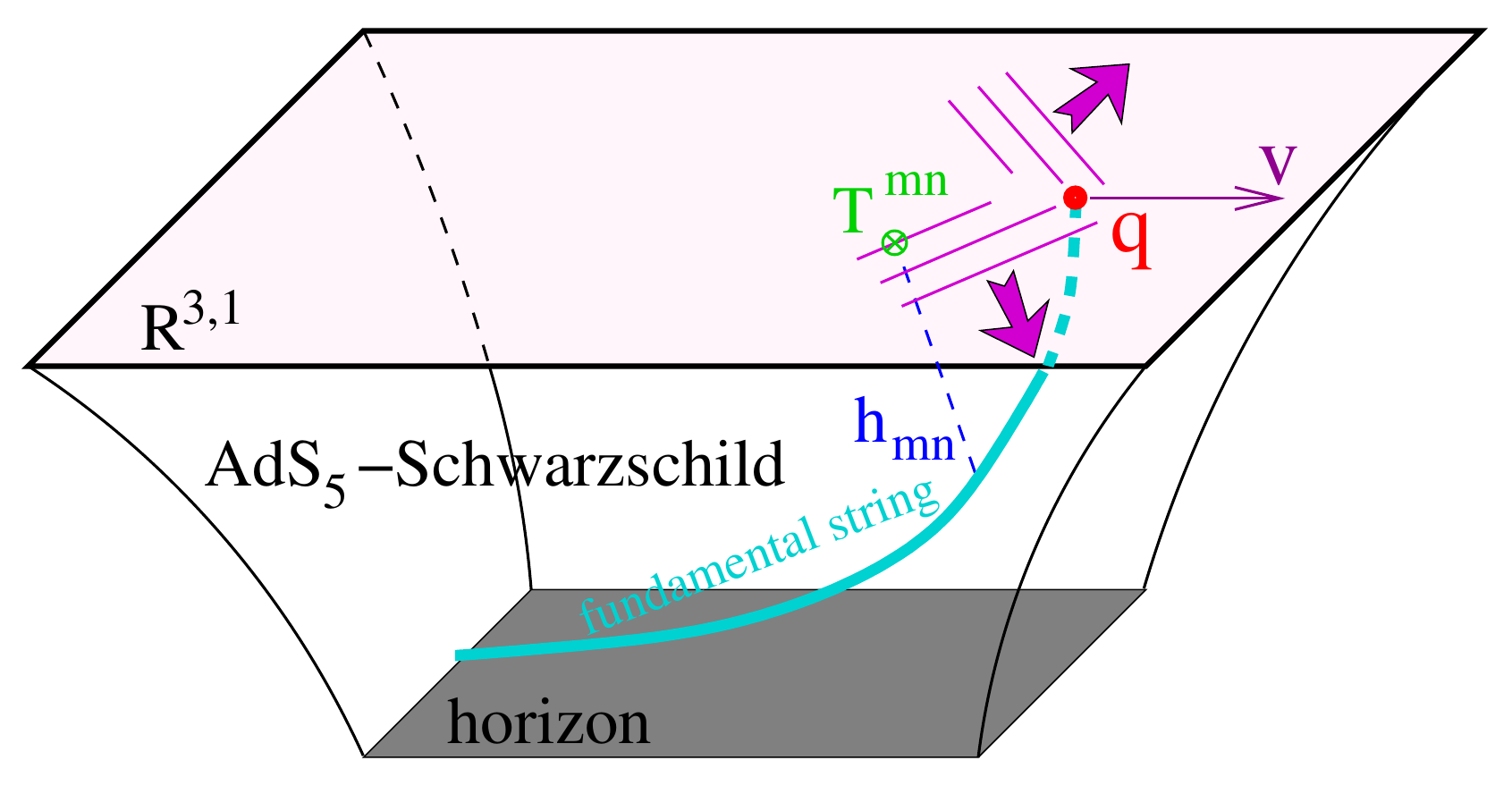}}
  \caption[Short]{(Color online.)  A visual summary\cite{Friess:2006fk} of the calculation of the response $\langle T_{mn} \rangle$ of a strongly coupled thermal medium to a heavy quark moving with a speed $v$ via AdS/CFT.  $h_{mn}$ is a perturbation of the metric caused by the trailing string.}\label{F:Wake}
 \end{figure}

Apart from a qualitative understanding of energy loss at large distances, little is known {\em a priori} about what happens, for example, at small distances close to the quark.  An all-scales description can be achieved using the gauge-string duality, where one computes the disturbances in the stress-energy tensor due to the presence of the quark.  This was done in a series of papers.\cite{Friess:2006fk,Gubser:2007nd,Gubser:2007zr,Yarom:2007ni,Gubser:2007xz,Chesler:2007an,Gubser:2007ga,Chesler:2007sv}  The linearized response of the lagrangian density in the dual field theory was also computed in Fourier space.\cite{Friess:2006aw,Gao:2006se}  The purpose of this section and the next two is to present a reasonably self-contained summary of how the gauge theory stress-energy tensor is computed starting from the trailing string.  A visual summary of the main elements of the computation is shown in figure~\ref{F:Wake}.

In the context of the AdS/CFT duality, the stress-energy tensor $\langle T_{mn} \rangle$ of the boundary theory is dual to fluctuations of the metric in the bulk.  So in order to compute the expectation value of $\langle T_{mn} \rangle$, one first needs to compute to linear order the backreaction of the string describing the quark on the metric. Non-linear corrections to the Einstein equations will be suppressed by $\sqrt{\lambda}/N$.  The total action describing both the string and the metric is $S = S_{\rm bulk} + S_{\rm string}$, which one can write as
 \begin{equation}
  \label{E:TotalAction}
  S = \int d^5 x \left[ {\sqrt{-G} (R + 12/ L^2) \over 2 \kappa^2}
    -{1\over 2 \pi \alpha'} \int d^2 \sigma\, \sqrt{-g} \,
    \delta^5 (x^\mu - x_*^\mu(\sigma)) \right] \,,
 \end{equation}
where $x_*^\mu(\sigma)$ is the embedding function of the string in AdS${}_5$-Schwarzschild.  In a gauge where we parameterize the string worldsheet by $\sigma^\alpha = (t, z)$, $x_*^\mu(\sigma)$ is given by
 \begin{equation}
  \label{E:StringEmbedding}
  x_*^\mu = \begin{pmatrix}
    t & v t + \xi(z) & 0 & 0 & z
    \end{pmatrix} \,,
 \end{equation}
with $\xi(z)$ as given in \eqref{E:TrailingString}.  The equations of motion following from \eqref{E:TotalAction} are just Einstein's equations:
 \begin{equation}
  \label{E:EinsteinEquations}
  R^{\mu\nu} - {1\over 2} G^{\mu\nu}R - {6 \over L^2} G^{\mu\nu} = \tau^{\mu\nu} \,,
 \end{equation}
where
 \begin{equation}
  \label{E:Gottaumunu}
  \tau^{\mu\nu} = - {\kappa^2 \over 2\pi \alpha'} 
    {z_H^3\over L^3} y^3 \sqrt{1 - v^2} \, \delta(x^1 - vt - \xi(z)) 
    \, \delta(x^2) \delta(x^3) \partial_\alpha x_*^\mu 
      \partial^\alpha x_*^\nu 
 \end{equation}
is the bulk stress-energy tensor of the trailing string.

To compute the backreaction of the string on the metric, we write
 \begin{equation}
  \label{E:DeltaGmunu}
  G_{\mu\nu} = G_{\mu\nu}^{(0)} + h_{\mu\nu}
 \end{equation}
where $G_{\mu\nu}^{(0)}$ is the unperturbed AdS${}_5$-Schwarzschild metric given in \eqref{E:LineElement}, and plug this into \eqref{E:EinsteinEquations} to obtain the linearized equations of motion for the metric perturbations $h_{\mu\nu}$.  The resulting equations take the form
 \begin{equation}
  \label{E:Lichnerowicz}
  {\cal L}  h^{\mu\nu} = \tau^{\mu\nu} \,,
 \end{equation}
where $h^{\mu\nu} = G^{\mu \rho}_{(0)} G^{\nu \sigma}_{(0)} h_{\rho \sigma}$ and ${\cal L}$ is the 
differential operator given by\cite{martinezmorales}
 \begin{equation}
  \label{E:LichDef}
  \begin{split}
   {\cal L} h^{\mu\nu} &= - \square h_{\mu\nu}
   - 2 R_{\mu\rho\nu\sigma} h^{\rho\sigma}
  + 2 R_{(\mu}^{\rho} h_{\nu)\rho} 
  - \nabla^\mu\nabla^\nu h 
  +2 \nabla^{(\mu}\nabla_\rho h^{\nu)\rho} \\ 
  &\qquad{}+ G^{\mu\nu}_{(0)} \left( -\nabla_\rho\nabla_\sigma h^{\rho\sigma} +\square h + R_{\rho\sigma}h^{\rho\sigma} \right)
  - \left({6\over L^2} -  R\right) h^{\mu\nu}  \,.
  \end{split}
 \end{equation}
The covariant derivatives, the Riemann  and Ricci tensors, and the Ricci scalar appearing in \eqref{E:LichDef} are computed using the background metric $G_{\mu\nu}^{(0)}$, and we have denoted $h=G^{(0)}_{\mu\nu}h^{\mu\nu}$.

For a steady-state solution of \eqref{E:Lichnerowicz}, $h_{\mu\nu}$ depends on $x^1$ and $t$ only through the combination 
$x^1 - v t$.  We therefore pass to co-moving Fourier space variables by writing
 \begin{equation}
  \label{E:FourierSpace}
  \tau^{\mu\nu}(t, x^1, x^2, x^3, z) = \int {d^3 K \over (2\pi)^3} 
    \tau^{\mu\nu}_K(z) e^{i \left[K_1 (x^1 - vt) + K_2 x^2 + K_3 x^3 \right]/z_H} \,,
 \end{equation}
where we have defined the dimensionless wavevector $\vec{K} \equiv  \vec{k} z_H = \vec{k}/\pi T$.  We make a similar expansion for $h_{\mu\nu}$.  In Fourier space, \eqref{E:Lichnerowicz} can then be written as 
 \begin{equation}
  \label{E:EomsFourier}
  {\cal L}^K h_K^{\mu\nu} = \tau_K^{\mu\nu} \,,
 \end{equation}
with $\tau^{\mu\nu}_K$ being given by
 \begin{equation}
  \label{E:GottaumunuK}
  \tau^{\mu\nu}_K = {\kappa^2 \over 2 \pi \alpha'}
    {e^{-i K_1 \xi(z)/z_H} \over \sqrt{1 - v^2}}
    {z_H^2 y^5 \over L^5}
    \begin{pmatrix}
      \displaystyle{h + v^2 y^4 \over h^2} & \displaystyle{v \over h} & 0 & 0 & \displaystyle{v^2 y^2 \over h} \\[10pt]
      \displaystyle{v \over h} & v^2 & 0 & 0 & v y^2 \\[5pt]
      0 & 0 & 0 & 0 & 0 \\
      0 & 0 & 0 & 0 & 0 \\[2pt]
      \displaystyle{v^2 y^2 \over h} & v y^2 & 0 & 0 & v^2 - h
    \end{pmatrix} \,,
 \end{equation}
where $y$ is the rescaled depth coordinate defined in \eqref{E:YDef}, but the tensor components are given in the $(t,x^1,x^2,x^3,z)$ coordinate system. The explicit form of ${\cal L}^K$ is too complicated to be reproduced here. 

We will decouple equations \eqref{E:EomsFourier} by passing to a gauge where $h_{\mu z} = 0$, which we will refer to as ``axial gauge.''  Note that $h_{\mu z}=0$ leaves some residual gauge freedom. We will discuss this shortly.

The rest of this section is organized as follows.  In section~\ref{S:Axial} we explain how to decouple equations \eqref{E:EomsFourier}.  In section~\ref{S:BoundaryConditions} we explain the boundary conditions needed to solve these equations.  Lastly, in section~\ref{SS:Boundarystress} we explain how the one-point function of the  SYM stress-energy tensor is related to the asymptotic behavior of the metric perturbations near $y = 0$.

\subsection{Metric perturbations in axial gauge}
\label{S:Axial}

As mentioned above, we choose a gauge where $h_{\mu z} = 0$, and let
 \begin{equation}
  \label{E:hmunuFourier}
  h_{\mu\nu}^K = {\kappa^2 \over 2 \pi \alpha'} {1\over \sqrt{1 - v^2}} {L \over z^2}
    \begin{pmatrix}
    H_{00} & H_{01}  & H_{02} & H_{03} & 0 \\
    H_{10} & H_{11}  & H_{12} & H_{13} & 0 \\
    H_{20} & H_{21}  & H_{22} & H_{23} & 0 \\
    H_{30} & H_{31}  & H_{32} & H_{33} & 0 \\
    0 & 0 & 0 & 0 &0 
    \end{pmatrix} \,.
 \end{equation}
Rotational symmetry around the direction of motion of the quark allows us to set $\vec{K} = \begin{pmatrix} K_1 & K_\perp & 
0 \end{pmatrix}$ with $K_\perp > 0$.  Defining 
 \begin{equation}
  \label{E:KThetaDef}
  K = \sqrt{K_1^2 + K_\perp^2} \qquad \vartheta = \tan^{-1} {K_\perp \over K_1} \,,
 \end{equation}
one can form the following linear combinations of metric perturbations:
 \begin{subequations}
  \label{E:HmnDef}
  \begin{align}
    \label{E:HmnDefA}
    A &= {-H_{11} + 2 \cot \vartheta H_{12} - \cot^2 \vartheta H_{22} + \csc^2 \vartheta H_{33} \over 2 v^2} \\
    \label{E:HmnDefB}
    B_1 &= {H_{03} \over v} \qquad
    B_2 = -{H_{13} + \tan \vartheta H_{23} \over v^2} \\
    \label{E:HmnDefC}
    C &= {-\sin \vartheta H_{13} + \cos \vartheta H_{23}} \\
    \label{E:HmnDefD}
    D_1 &= {H_{01} - \cot \vartheta H_{02} \over 2 v^2} \qquad
    D_2 = {-H_{11} + 2 \cot 2 \vartheta H_{12} + H_{22} \over 2 v^2} \\
    \label{E:HmnDefE12}
    E_1 &= {1 \over 2} \left(-{3 \over h} H_{00} + H_{11} + H_{22} + H_{33}  \right) \qquad
    E_2 = {H_{01} + \tan \vartheta H_{02} \over 2 v} \\
    \label{E:HmnDefE3}
    E_3 &= {H_{11} + H_{22} + H_{33} \over 2} \\
    \label{E:HmnDefE4}
    E_4 &= {-H_{11} - H_{22} + 3 \cos 2 \vartheta (-H_{11} 
      + H_{22}) + 2 H_{33} - 6 \sin 2 \vartheta H_{12} \over 4} \,.
  \end{align}
 \end{subequations}
Using these new variables, the Einstein equations \eqref{E:EomsFourier} decouple into five sets:\cite{Friess:2006fk}
\begin{subequations}
\label{E:OriginalABCDEset}
  \begin{equation}
    \label{E:ASet}
    \left[ \partial_y^2 + \left(-\frac{3}{y}+\frac{h'}{h}\right) \partial_y
      +\frac{K^2}{h^2}\left(v^2 \cos^{2}\vartheta - h\right) \right] A 
      = \frac{y}{h}e^{-i K_1 \xi / z_H}
  \end{equation}
  \begin{equation}
    \label{E:BSet}
    \left[ \partial_y^2 + \begin{pmatrix}
    -{3 \over y} & 0 \\
    0 & -{3 \over y} + {h'\over h}
    \end{pmatrix}
    + {K^2 \over h^2} \begin{pmatrix}
    -h & v^2 h \cos^2 \vartheta \\
    -1 & v^2 \cos^2 \vartheta
    \end{pmatrix}
    \right] \begin{pmatrix}
    B_1\\ B_2
    \end{pmatrix}
    = \begin{pmatrix} 0 \\ 0 \end{pmatrix}
  \end{equation}
  \begin{equation}
    \label{E:BConstraint}
    B_1' - h B_2' = 0
  \end{equation}
  \begin{equation}
    \label{E:CSet}
    \left[ \partial_y^2 + \left(-\frac{3}{y}+\frac{h'}{h}\right) \partial_y
      +\frac{K^2}{h^2}\left(v^2 \cos^{2}\vartheta - h\right) \right] C 
      = 0
  \end{equation}
  \begin{equation}
     \label{E:DSet}
    \left[ \partial_y^2 + \begin{pmatrix}
    -{3 \over y} & 0 \\
    0 & -{3 \over y} + {h'\over h}
    \end{pmatrix}
    + {K^2 \over h^2} \begin{pmatrix}
    -h & v^2 h \cos^2 \vartheta \\
    -1 & v^2 \cos^2 \vartheta
    \end{pmatrix}
    \right] \begin{pmatrix}
    D_1\\ D_2
    \end{pmatrix}
    = {y \over h} e^{- i K_1 \xi /z_H} 
    \begin{pmatrix} 1 \\ 1 \end{pmatrix} 
  \end{equation}
  \begin{equation}
    \label{E:DConstraint}
    D_1' - h D_2' = 0
  \end{equation}
  \begin{multline}
    \label{E:ESet}
    \left[ \partial_y^2 + \begin{pmatrix}
    -{3 \over y} + {3 h' \over 2 h} & 0 & 0 & 0 \\
    0 & -{3 \over y} & 0 & 0 \\
    0 & 0 & -{3 \over y} + {h' \over 2 h} & 0 \\
    0 & 0 & 0 & {-3 \over y} + {h' \over h}
    \end{pmatrix} \partial_y \right.\\
    \left. + {K^2 \over 3 h^2} \begin{pmatrix}
    - 2h & 12 v^2 \cos^2 \vartheta & 6 v^2 \cos^2 \vartheta + 2 h & 0\\
    0 & 0 & 2 h & h \\
    0 & 0 & - 2h & -h \\
    2 h & -12 v^2 \cos^2 \vartheta & 0 & 3v^2 \cos^2 \vartheta + h
    \end{pmatrix} \right] \begin{pmatrix}
    E_1 \\ E_2 \\ E_3 \\ E_4
    \end{pmatrix} \\
    = {y \over h} e^{-i K_1 \xi / z_H} \begin{pmatrix}
    1 + {v^2 \over h} \\
    1\\
    -1 + v^2 - {v^2 \over h}\\
    {1 + 3 \cos 2 \vartheta \over 2} v^2
    \end{pmatrix}
  \end{multline}
  \begin{multline}
    \label{E:EConstraint}
    \left[ \begin{pmatrix}
    0 & 1 & 1 & 0 \\
    -h & 0 & -3v^2 \cos^2 \vartheta - h & -h \\
    h & 0 & 2 & 0 
    \end{pmatrix} \partial_y \right.\\
     + \left. {1 \over 6 h} \begin{pmatrix}
    0 & -6 h' & - 3 h' & 0 \\
    -3 h h' & 18 v^2 \cos^2 \vartheta h' & 3 ( 3v^2 \cos^2 \vartheta + h) h' & 0 \\
    2 K^2 y h & - 2 K^2 v^2 y \cos^2 \vartheta &
    - 2 K^2 y (3 v^2 \cos^2 \vartheta - h) & 2 K^2 y h
    \end{pmatrix} \right] \begin{pmatrix}
    E_1 \\ E_2 \\ E_3 \\ E_4
    \end{pmatrix} \\
    = {h' \over 4 K y h} e^{-i K_1 \xi /z_H} \begin{pmatrix}
    -i v y \sec \vartheta \\
    3 i v y \cos \vartheta (v^2 + h) \\
    K (v^2 - h)
    \end{pmatrix} \,.
  \end{multline}
 \end{subequations}
In \eqref{E:OriginalABCDEset}, $A$, $B_i$, $C$, $D_i$, and $E_i$ are all functions of the rescaled AdS depth $y \equiv z / z_H$, and primes denote derivatives with respect to $y$.

A few comments are in order.   Let's start by counting the equations.  Einstein's equations \eqref{E:EomsFourier} consist of fifteen linearly independent equations that split between the $A$, $B$, $C$, $D$, and $E$ sets as follows:  the $A$ and $C$ sets each consist of one second order equation for one unknown function;  the $B$ and $D$ sets each consist of two second order equations and one first order constraint for two unknown functions;  lastly, the $E$ set consists of four second order equations and three first order constraints for four unknown functions.  At first glance, the $B$, $D$, and $E$ systems of equations might seem overdetermined.  A more careful analysis shows that the constraints are consistent with the second order equations in the sense that if they hold at a particular value of $y$, they continue to hold at all $y$.  We can therefore think of the constraint equations as reducing by five the number of integration constants in the second order equations.  There are fifteen remaining integration constants that are fixed by the boundary 
conditions which we discuss in section~\ref{S:BoundaryConditions}.

Because the $B_i$ and $C$ equations, \eqref{E:BSet}--\eqref{E:CSet}, have no source terms, one can consistently set $B_1=B_2=C=0$.  This identification is in fact enforced if we insist that the response of the medium should respect the same axial symmetry around the direction of motion of the quark that the trailing string does.  In order to keep the discussion of perturbations, boundary conditions, and integration constants general, let us not discard the $B_i$ and $C$ fields just yet.  A more general source would force them to be non-zero.

Equations \eqref{E:OriginalABCDEset} can be reduced to just five equations using the residual gauge symmetry.\cite{Kovtun:2005ev,Chesler:2007sv}  The action \eqref{E:TotalAction} is reparameterization invariant, and infinitesimal gauge transformations act by sending 
 \begin{equation}
  \label{E:GaugeTransf}
  h_{\mu\nu} \to h_{\mu\nu} + \nabla_\mu \difxi_\nu + \nabla _\nu \difxi_\mu
 \end{equation}
for any one-form $\difxi_\mu dx^\mu$.  This allowed us to pass to axial gauge in the first place:  if we had started with some arbitrary metric perturbations $h_{\mu\nu}$ we could solve, at least locally,
 \begin{equation}
  \label{E:GaugeFix}
  \nabla_z \difxi_\mu + \nabla_\mu \difxi_z = -h_{\mu z} \,,
 \end{equation}
so applying the infinitesimal gauge transformation \eqref{E:GaugeTransf} gives $h_{\mu z} = 0$.  Residual gauge symmetry arises from the fact that \eqref{E:GaugeFix} specifies $\difxi_\mu$ only up to five integration constants, which in general are functions of $t$ and $\vec{x}$.  Put differently, there are five linearly independent gauge transformations that preserve the axial gauge condition $h_{\mu z} = 0$, and the corresponding $\difxi_\mu$ are given by the linearly independent solutions to \eqref{E:GaugeFix} with $h_{\mu z} = 0$.  

For the steady-state solution of \eqref{E:EomsFourier} the allowed gauge transformations are the ones where the $t$ and $\vec{x}$ dependence of $\xi_\mu$ is of the form $e^{i K_m x^m /z_H}$, where we have defined
 \begin{equation}
  \label{E:KmDef}
  K_m = \begin{pmatrix}
    -K_1 v & K_1 & K_2 & K_3
    \end{pmatrix} \,.
 \end{equation}
As in previous sections, our convention is that lower case Roman indices $m,n,\ldots$ take values in $(t,x^1,x^2,x^3)$. The first four of these gauge transformations are parameterized by
 \begin{equation}
  \label{E:PureGaugeXi4}
  \difxi^\mu_{(a)} = {\kappa^2 \over {2 \pi \alpha'}} {z_H e^{i K_m x^m /z_H} \over L \sqrt{1- v^2}}   \delta_a^\mu \,,
 \end{equation}
where $a = 0$, $1$, $2$, or $3$, while the fifth is given by
 \begin{equation}
  \label{E:PureGaugeXi5}
  \difxi^\mu_{(5)} = {\kappa^2 \over {2 \pi \alpha'}} {z_H e^{i K_m x^m /z_H} \over L \sqrt{1- v^2}}
    \left[2 y \sqrt{h} \delta_5^\mu  - i K^0 {y^2 \over h} \delta_0^{\mu} - i K^j \arcsin y^2 \delta_j^\mu \right] \,.
 \end{equation}
The corresponding pure gauge solutions are given by
 \begin{align}
  \label{E:PureGaugeHa}
  H_{mn(a)} &= - 2 i K_{(m} G_{n)a}^{(0)} {z^2 \over L^2} 
    \qquad a = 0, 1, 2, 3 \\
  \label{E:PureGaugeH5}  
  H_{mn(5)} &= 4 \sqrt{h}\, \eta_{mn} - 2 K_m K_n  \arcsin y^2 
    - 4 y^4 \sqrt{h}\,  \delta_m^0 \delta_n^0 \notag \\
    {}&+ 2 K^0 K_{(m} \eta_{n)0} \left[\arcsin y^2 - y^2 \sqrt{h} \right]
     \,,
 \end{align}
where $\eta_{mn}$ is the Minkowski metric.  One can straightforwardly check that \eqref{E:PureGaugeHa}--\eqref{E:PureGaugeH5} satisfy equations \eqref{E:EomsFourier}.

Using the definitions \eqref{E:HmnDef}, we can work out the pure gauge solutions in the $ABCDE$ variables.  We find that $A$ and $C$ are invariant, which was to be expected since their equations of motion are already fully decoupled. The other variables, however, do transform non-trivially under \eqref{E:GaugeTransf} with \eqref{E:PureGaugeXi4} and \eqref{E:PureGaugeXi5}: for example the $B_i$ and $D_i$ variables vary by
\begin{align}
\label{E:BDGaugeTrans}
 \delta B_1 &=   i \lambda_3 K_1 \quad &
 \delta D_1 &=  {i K_1(\lambda_1 -\lambda_2 \cot\vartheta)\over 2 v}  \nonumber \\
 \delta B_2 &=  { i \lambda_3 K_1 \over v^2 \cos^2\vartheta} \quad & 
 \delta D_2 &=  { i K_1 (\lambda_1 - \lambda_2\cot\vartheta)\over 2v^2 \cos^2\vartheta \sin^2\vartheta} \,,
 \end{align}
where $\lambda_a$ are arbitrary constants multiplying the pure gauge solutions parameterized by $\difxi^\mu_{(a)}$.
 
From these transformation laws it is easy to see that there is a linear combination of $B_1$ and $B_2$ that is gauge-invariant (and the same is true for $D_1$ and $D_2$). A choice of gauge invariants is given by
\begin{align}
\label{E:BDInvDef}
B  &= B_1 -\cos^2 \vartheta \, v^2 B_2 \nonumber\\
D &= D_1 - \cos^2 \vartheta \, v^2 D_2 \,.
\end{align} 
The transformation law for the $E_i$ variables is simple to derive but its exact form is not very enlightening. The corresponding gauge-invariant can be taken to be
\begin{equation}
\label{E:EInvDef}
E = 4 E_3 + E_4 + (12 E_2-3 E_4) v^2 \cos^2\vartheta -(2E_1+E_4)h(z) \,.
\end{equation}

From \eqref{E:OriginalABCDEset} we obtain the equations of motion for the $B$, $D$, and $E$ invariants:
\begin{equation}
\label{E:BInvEom}
\left[\partial_y^2 + \left(
{ 3 h^2 + (h-4) v^2 \cos^2\vartheta \over y h (v^2\cos^2\vartheta-h)}
\right)\partial_y + 
{K^2 (v^2\cos^2\vartheta-h)\over h^2} \right] B =0
\end{equation}
\begin{equation}
\label{E:DInvEom}
\begin{split}
\left[\partial_y^2 + \left(
{ 3 h^2 + (h-4) v^2 \cos^2\vartheta \over y h (v^2\cos^2\vartheta-h)}
\right)\partial_y + 
{K^2 (h -v^2\cos^2\vartheta)\over h^2} \right] D \\
= e^{-i K_1 \xi / z_H}{y \over h} \left(
1- v^2\cos^2\vartheta + {4 i v y^5 \over K (v^2\cos^2\vartheta - h)} 
\right)
\end{split}\end{equation}
\begin{equation}
\begin{split}
\label{E:EInvEom}
\left[
 \partial_y^2
   + \left(
{1\over y} - {4\over y h} + {16 y^4 \over 4 -6 v^2 \sin^2\vartheta +2 h}
\right)\partial_y  
\right.\qquad\qquad\qquad\qquad\qquad  \\ \left.
{}- {K_1^2 v^2\over h^2}-  {32 y^8\over y^2 h (4 - 6 v^2 \sin^2\vartheta + 2 h)}
+ {K_1\over h \cos^2\vartheta}
\right]E \qquad \\
= e^{-i K_1 \xi \over z_H} {y  \over  h} { 3 v^2 \cos^2\vartheta - 2- v^2\over
2+h - 3 v^2 \cos^2\vartheta}\qquad\qquad\qquad\qquad\qquad\qquad \\
\times \left[
9v^4\cos^4\vartheta 
- 18v^2 \cos^2\vartheta\left( 1 + {8 i y^5\over 3 v K_1}\right)
- y^8+9
\right]. 
\end{split}
\end{equation}

The gauge invariants we just described are unique up to an overall $z$-dependent factor, provided we only consider linear combinations of the $B_i$, $D_i$, and $E_i$ variables. If we also allow derivatives of the latter there are many other choices of gauge invariants that might prove useful.  For instance,
\begin{align}
\label{E:CheslerInvDef}
Z_0 &= {1\over 3} K^2 E \nonumber \\
\overrightarrow{Z_1} &=   \begin{pmatrix}
 2 v^2 \sin\vartheta z_H D_1'  \\ v z_H B_1' 
 \end{pmatrix} \nonumber \\
 \overleftrightarrow{Z_2} &=
\begin{pmatrix}
- A \sin^2\vartheta v^2  & -C \\
- C & A \sin^2\vartheta v^2 
 \end{pmatrix}.
\end{align}
is another set of gauge invariants which has been used in the literature.\cite{Chesler:2007sv} 
Yet another set, consisting of ``master fields,''\cite{Gubser:2007nd} is: 
\begin{equation}
  \label{E:masterfields}
  \begin{split}
  \psi_T^{\rm odd} &=-\frac{z_H^3 \alpha_v}{L^6} C \\
  \psi_T^{\rm even} &=-\frac{z_H^3 v^2 \alpha_v}{L^6} A \sin^2\vartheta\\
  \psi_V^{\rm odd} &=   \frac{z_H^2 v\alpha_v}{2 L^4} h B_2'\\
  \psi_V^{\rm even} &= \frac{z_H^2 v \alpha_v}{L^4} h D_2' \sin\vartheta \\
  \psi_S &= - \frac{K^2 z_H \alpha_v}{6 L^2 h y (K^2 + 6 y^2)} \left[
    3 h^2 y E_4''+ 6 h^2 E_2'  - 3 h (3 +y^4) E_4' \right. \\
    & +\left.
    2 h y K^2 \left(E_1-E_3\right) + 3 y \left(2 h y^2 - K_1^2 v^2 \right) \left(4E_2- E_4\right) \right],
  \end{split}
 \end{equation}
where $\alpha_v=1/2\pi\alpha'\sqrt{1-v^2}$.   The gauge invariants \eqref{E:masterfields} will prove useful for the asymptotic analysis of section \ref{SS:Smallrasymptotics}. Below, we list their equations of motion:
\begin{subequations}
\label{E:EOMmasterfields}
 \begin{eqnarray}
 \label{E:EOMmasterT}
   \psi_T'' + \left( -{3\over y} + {h' \over h} \right) \psi_T' 
     + {K^2 (v^2 \cos^2 \vartheta - h)\over h^2} \psi_T &= - J_T \\
 \label{E:EOMmasterV}
   \psi_V'' + \left( -{3 \over y} + {h' \over h} \right) \psi_V'  + 
     {K^2 v^2 y^2 \cos^2 \vartheta + h (3 - K^2 y^2 + 9 y^4)
     \over y^2 h^2} \psi_V &= - J_V \\
 \label{E:EOMmasterS}
    \psi_S'' + \left({h' \over h} - {3\over y}\right) \psi_S' 
     \qquad\qquad\qquad\qquad\qquad\qquad\qquad\qquad  \nonumber \\
     + \left[{K_1^2 v^2 \over h^2}
     + {K^2 (4 - K^2 y^2) + 12 y^2 (6 - y^4) \over y^2 h (K^2 + 6 y^2)^2} \right]
     \psi_S &= -J_S
 \end{eqnarray}
 \end{subequations}
where 
 \begin{align}
  J_T^{\rm odd} &= J_V^{\rm odd} = 0 \label{E:oddSource} \\
  J_T^{\rm even} &=  {z_H^3 \alpha_v  \over L^6}
    {v^2 y \sin^2 \vartheta \over h} e^{-i K_1 \xi /z_H} \\
  J_V^{\rm even} &= {z_H^2 \alpha_v  \over L^4}
    {e^{-i K_1 \xi /z_H}\over h} \left[v \sin\vartheta (5 y^4 - 1 - i K v y^3 \cos \vartheta)
    + i K y^3 \tan \vartheta \right]  \label{E:evenVsource} \\
    J_S &= {z_H \alpha_v \over 6 L^4} { K^2 y e^{-i K_1 \xi /z_H} \over (K^2+6 y^2)^2} \Big[ 
      (2+v^2)\left( 2 y^2 (K^4-45) -3 K^2\right) \notag \\
      {}&+ 3 K^2 (2-5 v^2) y^4
      + 18 (2+3 v^2) y^6  - 3 i K v^3 y^3 (K^2 - 12 y^2) \cos \vartheta \notag\\
      {}&+ 3 v^2 \left[90 y^2 - K^4 (4 + v^2) y^2 - 
      54 y^6 + 3 K^2 ( 1 + y^4 - 2 v^2 y^4) \right]  \cos^2 \vartheta \notag\\
      {}&+ 9 i K v^3 y^3 (K^2 - 18 y^2) \cos^3 \vartheta
      + 9 K^2 v^4 y^2 (K^2 + 6 y^2) \cos^4 \vartheta \notag \\
      {}&-6 i v y^3 K (K^2 - 2 y^2) \sec \vartheta \Big] \,.
 \end{align}
By omitting the ``even'' and ``odd'' superscripts in (\ref{E:EOMmasterT}) and (\ref{E:EOMmasterV}) we mean that these equations take the same form (up to the different source terms as indicated in (\ref{E:oddSource})--(\ref{E:evenVsource})).

\subsection{Boundary conditions}
\label{S:BoundaryConditions}

In axial gauge, the system of equations \eqref{E:EomsFourier} consists of ten second order differential equations in $y$ and five first order constraints, so we need to specify fifteen integration constants.  The purpose of this section is to show that five of these are fixed by imposing boundary conditions at the horizon ($y=1$) and ten of them are fixed from the boundary conditions at the conformal boundary ($y = 0$).

Consider first the tensor field $A$, defined in \eqref{E:HmnDefA}.  The near horizon solution to \eqref{E:ASet} takes the form
\begin{equation}
\label{E:Anearhorizon}
	A = \frac{e^{-\frac{i v K_1}{8}\left(\pi - \ln 4\right)}}{4 \left(1-\frac{i v K_1}{2}\right)}
		\left(1-y\right)^{1-i v K_1/4}+ U_A (1-y)^{-i v K_1/4} + V_A(1-y)^{i v K_1/4} + \ldots \,,
\end{equation}
where $\ldots$ indicates terms which are subleading to one of the ones shown.  The first term in \eqref{E:Anearhorizon} is a particular solution to \eqref{E:ASet} characterizing the response of $A$ to the trailing string source.  The second and third terms are solutions to the homogeneous equation, and $U_A$ and $V_A$ are constants of integration.  The correct boundary condition at the horizon is $V_A=0$.  This corresponds to requiring that there are no outgoing modes at the horizon, and it can be justified by the fact that classical horizons don't radiate.  The result of choosing purely infalling conditions is that in the dual gauge theory, we describe a causal response of the medium to the probe. 

To see that the $U_A$ term in \eqref{E:Anearhorizon} is infalling, let's define a new coordinate $y_*= \log (1-y)$ that ranges from $-\infty$ when $y = 1$ (the horizon) to zero when $y = 0$ (the conformal boundary of AdS${}_5$).  Recalling that the time dependence of metric perturbations was assumed to be $e^{-i K_1 v t /z_H}$, one sees immediately that the $(1-y)^{-i v K_1/4}$ term in \eqref{E:Anearhorizon} corresponds to certain metric components behaving as
 \begin{equation}\label{E:InfallingForm}
  h_{mn} \sim e^{ - i v K_1 (y_* + 4 t/z_H)/4}
 \end{equation}
at large negative $y_*$.  This behavior describes a wave traveling towards negative $y_*$, i.e.,~falling into the black hole horizon.  Similarly, the $V_A$ term in \eqref{E:Anearhorizon} corresponds to an outgoing mode.  The same story goes through for the $C$ combination of metric components defined in \eqref{E:HmnDefC}, except that since there is no source term, there will be no analog to the first term in~\eqref{E:InfallingForm}.

A subtlety arises in the horizon boundary conditions for the $B$, $D$, and $E$ sets: for each set, in addition to a single infalling solution and a single outgoing solution, there are the pure gauge solutions discussed around~\eqref{E:PureGaugeHa} and~\eqref{E:PureGaugeH5}.  The pure gauge solutions are neither infalling nor outgoing at the horizon.  The correct boundary conditions are to exclude the outgoing solution and to permit both the infalling solution and the pure gauge solutions.  If one passes to a description only in terms of gauge-invariants, then this subtlety is avoided: each gauge invariant field has only an infalling and outgoing solution, and the latter is excluded by the horizon boundary conditions. 

Having fixed five integration constants (one for each of the $ABCDE$ sets) using infalling boundary conditions at the horizon, we now discuss the boundary conditions at the boundary of AdS\@.  Close to $y=0$, Einstein's equations can be solved in a series expansion in $y$.  The two homogeneous solutions are
\begin{equation}
	H_{mn}^{(1)}(y) = R_{mn}\left(1+\mathcal{O}(y^2)\right)
	\qquad
	H_{mn}^{(2)}(y) = Q_{mn}\left(y^4 + \mathcal{O}(y^6)\right)
\end{equation}
where $R_{mn}$ and $Q_{mn}$ are arbitrary constants.
The full solution then takes the form
 \begin{equation}\label{E:HmnBoundary}
  H_{mn} =  H_{mn}^{(1)}(y) +H_{mn}^{(2)}(y)  + \frac{P_{mn}}{3} y^3+ \mathcal{O}(y^5)\,.
\end{equation}
The components of $P_{mn}$ are given by
\begin{equation}
  \label{E:PXval}
  P_{mn} = 
  \begin{pmatrix}
    {2 \over 3} (2 + v^2) & -2 v & 0 & 0 \\
    -2 v & {2\over 3} (1 + 2 v^2) & 0 & 0 \\
    0 & 0 & {2 \over 3} (1 - v^2) & 0  \\
    0 & 0 & 0 & {2 \over 3} ( 1- v^2)
  \end{pmatrix} \,.
\end{equation}
The boundary conditions we impose are that 
\begin{equation}
\label{E:boundaryBC}
	R_{mn}=0,
\end{equation}
thus fixing the remaining ten integration constants. Allowing non-zero $R_{mn}$ would correspond to deformations of the gauge theory lagrangian. With this condition imposed, the $Q_{mn}$ are related to the expectation value of the stress tensor in the gauge theory, as we will see in section~\ref{SS:Boundarystress}.

Equation \eqref{E:boundaryBC} can be translated into boundary conditions for the $ABCDE$ variables.  Once the horizon boundary conditions are also taken into account, it follows that each set of equations in \eqref{E:OriginalABCDEset} is supplemented by just enough boundary conditions to uniquely fix a solution.  For example, \eqref{E:boundaryBC} implies boundary conditions on $A$ which fix one constant of integration.  Another integration constant is fixed by the horizon boundary conditions. Since the underlying equation \eqref{E:HmnDefA} is second order and linear, the solution is unique.  A more complicated example is the $E$ set, where the boundary conditions from \eqref{E:boundaryBC} fix four integration constants and the horizon boundary conditions fix one more.  Since the four second order equations of motion \eqref{E:ESet} are subject to three constraints \eqref{E:EConstraint}, the number of integration constants available is five.  Thus one again finds a unique solution.  If we pass to a description in terms of gauge invariants, then \eqref{E:boundaryBC} fixes a single constant of integration for each gauge-invariant field.  Because the horizon fixes another constant and the underlying equation for the gauge-invariant is always second order, we again recover a unique solution.

\subsection{The boundary stress-energy tensor}
\label{SS:Boundarystress}

The AdS/CFT duality offers a prescription\cite{Gubser:1998bc,Witten:1998qj}  for computing the stress tensor of the boundary theory from the bulk action:\cite{Balasubramanian:1999re,Balasubramanian:1999re,Balasubramanian:1999re,deHaro:2000xn}
\begin{equation}
\label{E:Tmnprescription}
	\langle T_{mn} \rangle = \lim_{\epsilon \to 0} \frac{2}{\sqrt{-g}} \frac{\delta S_{\rm total}}{\delta g^{mn}} \,.
\end{equation}
We now explain what $S_{\rm total}$, $g_{mn}$ and $\epsilon$ are. $g_{mn}$ is the metric on the conformal boundary of AdS${}_5$.  After taking the variational derivative, $g_{mn}$ is set equal to the metric of the boundary theory, i.e., the Minkowski metric $\eta_{mn}$ in our case.  $S_{\rm total}$ is given by
\begin{equation}
	S_{\rm total} = S_{\rm bulk}+S_{\rm GH}+S_{\rm counter} \,.
\end{equation}
Here $S_{\rm bulk}$ is the bulk action \eqref{E:EinAction}; $S_{\rm GH}$ is the Gibbons-Hawking boundary term
\begin{equation}
	S_{\rm GH} = \frac{1}{\kappa^2} \int d^4 x \sqrt{G^{\Sigma}} K^{\Sigma} \,,
\end{equation}
where $\Sigma$ is a co-dimension one surface close to the AdS boundary with outward normal $n_{\mu}$, induced metric
\begin{equation}
\label{E:GSigma}
  G_{\mu\nu}^{\Sigma} \equiv G_{\mu\nu} - n_\mu n_\nu \,,
\end{equation}
and extrinsic curvature tensor
\begin{equation}
  \label{E:KSigma}
  K_{\mu\nu}^\Sigma \equiv - G_{\mu\rho}^\Sigma \nabla^\rho n_\nu \,;
\end{equation}
and $S_{\rm counter}$ is an additional boundary term which renders the on-shell action finite for geometries which do not induce a trace anomaly in the boundary theory.  An explicit expression for this term is
\begin{equation}
\label{E:counteraction}
	S_{\rm counter} = \frac{1}{\kappa^2} 
	\int d^4 x \sqrt{-G^{\Sigma} } \left(\frac{2}{L}
	-\frac{L^2} {4}R^{\Sigma}\right) \,,
\end{equation}
with $R^{\Sigma}$ the Ricci scalar constructed from $G^{\Sigma}_{mn}$. Usually, there are additional terms in \eqref{E:counteraction} coming from the matter action. Since we are working in the probe limit, we do not need to worry about these extra terms.\cite{Gubser:2008vz} The coordinate $\epsilon$ in \eqref{E:Tmnprescription} specifies a hypersurface $\Sigma(\epsilon)$ which coincides with the boundary of AdS${}_5$ as we take the $\epsilon \to 0$ limit.

Since we are working with a flat boundary metric, the variation of $S_{\rm counter}$ will not contribute to $\langle T_{mn} \rangle$, and we can choose $\Sigma$ to be a surface of constant $z$. The outward normal form is
 \begin{equation}
  \label{E:NormalForm}
  n_\mu d x^\mu = -{L \over z \sqrt{h}} dz \,,
 \end{equation}
and then
 \begin{equation}
  \label{E:KSimp}
  K_{mn}^\Sigma = {z \sqrt{h} \over 2L} {\partial G_{mn} \over \partial z} \,.
 \end{equation}
Equation \eqref{E:Tmnprescription} now reads
\begin{equation}
\label{E:BdyStress}
	\langle T_{mn} \rangle = \lim_{z \to 0} \frac{L^2}{z^2 \kappa^2} \left(K_{mn}^{\Sigma} - K^{\Sigma} G_{mn}^{\Sigma} \right).
\end{equation}

Using \eqref{E:BdyStress}, the expectation value of the stress-energy tensor in the absence of the quark is that of a thermal bath
\begin{equation}
 \label{E:TmnBath}
   \langle T_{mn} \rangle_{\rm bath} = {\pi^2 \over 8} N^2 T^4 
    {\rm diag} \{3, 1, 1, 1 \} \,.
\end{equation}
The presence of the quark generates two additional contributions:  writing
 \begin{equation}
  \label{E:TmnKDef}
  \langle T_{mn} \rangle = \langle T_{mn} \rangle_{\rm bath} + 
   \int {d^3 K \over (2 \pi)^3} \left[ \langle T_{mn}^K \rangle_{\rm div} 
   + \langle T_{mn}^K \rangle \right]
   e^{i [K_1 (x^1 - vt) + K_2 x^2 + K_3 x^3]/z_H} \,,
 \end{equation}
and using the definitions \eqref{E:hmunuFourier} and \eqref{E:HmnBoundary}, as well as the boundary condition $R_{mn} = 0$ and the AdS/CFT identities \eqref{E:LandKappa} and \eqref{E:tHooft}, one obtains
 \begin{align}
  \label{E:TmnFromPmn}
  \langle T_{mn}^K \rangle_{\rm div} &={1 \over 4 \epsilon}  {\pi^2 T^3 \sqrt{\lambda}  \over \sqrt{1 - v^2}} 
      \left(P_{mn} - \eta_{mn} P_l^{\phantom{l}l} \right) \\
  \label{E:TmnFromQmn}    
  \langle T_{mn}^K \rangle &=  {\pi^3 T^4\sqrt{\lambda} \over \sqrt{1 - v^2}} 
   \left(Q_{mn} - \eta_{mn} Q_l^{\phantom{l}l}  \right) \,.
 \end{align}

Let's first understand the divergent contribution.  Plugging \eqref{E:PXval} into \eqref{E:TmnFromPmn}, it is not hard to see that in position space $\langle T_{mn} \rangle_{\rm div}$ takes the form of a contact term
 \begin{equation}
  \label{E:TmnFromPmnAgain}
  \langle T_{mn} \rangle_{\rm div} = {\sqrt{\lambda} \over 2 \pi \epsilon} 
    u_m u_n \sqrt{1 - v^2} \delta(x^1 - vt) \delta(x^2) \delta(x^3)
 \end{equation}
where $u^m = {1\over \sqrt{1 - v^2}} (1, \vec{v})$ is the four-velocity of the quark.  A divergence of this form was to be expected, and it is associated to having an infinitely massive quark.  In the dual gravity language, the mass of the quark can be identified with the energy of the trailing string.  But this energy is both IR and UV divergent.  The UV divergence corresponds to the ``bare mass'' of the quark and is exactly given by
 \begin{equation}
  \label{E:MDefinition}
  M = {\sqrt{\lambda} \over 2 \pi \epsilon} \,,
 \end{equation}
where $\epsilon$ is the UV cutoff. Equation \eqref{E:TmnFromPmnAgain} then takes the form of the stress-energy tensor of a particle with mass $M$ moving with velocity $v$ along the $x^1$-direction.  It can be shown that any string configuration whose endpoint lies on the boundary of AdS, whether it is stationary, moving with constant velocity relative to the plasma, or accelerating, will generate a divergent contribution of the form \eqref{E:TmnFromPmnAgain}.\cite{Gubser:2008vz}

The finite contribution to the stress-energy tensor given in \eqref{E:TmnFromQmn} can be further simplified by using the $55$ Einstein equation.  Using the series expansion \eqref{E:HmnBoundary} with $R_{mn} = 0$, the $55$ Einstein equation imposes a tracelessness relation on the $Q_{mn}$:
 \begin{equation}
  \label{E:Tracelessness}
  -Q_{00} + Q_{11} + Q_{22} + Q_{33} = 0 \,.
 \end{equation}
This equation implies further that
 \begin{equation}
  \label{E:TmnFromQmnAgain}    
  \langle T_{mn}^K \rangle =  {\pi^3 T^4\sqrt{\lambda} 
    \over \sqrt{1 - v^2}} Q_{mn}  \,.  
 \end{equation}
Note that even from \eqref{E:TmnFromQmn} one can see that $\langle T_{mn}^K \rangle$ is traceless, so \eqref{E:Tracelessness} does not imply tracelessness of the boundary stress-energy tensor.  What \eqref{E:Tracelessness} does is that it allows us to write the one-point function of the stress tensor in the simplified form \eqref{E:TmnFromQmnAgain}.

The $5m$ Einstein equations imply four more relations among the $Q_{mn}$:
 \begin{equation}
  \label{E:NonConservationQmn}
  K^m Q_{mn} = {i v \over 2} \begin{pmatrix}
    v & -1 & 0 & 0 \end{pmatrix} \,.
 \end{equation}
Using \eqref{E:TmnFromQmnAgain}, \eqref{E:NonConservationQmn} shows that the boundary stress-energy tensor fails to be conserved:
 \begin{equation}
  \label{E:NonConservationTmn}
  K^m \langle T_{mn} \rangle =  {i v \over 2} {\pi^3 T^4 \sqrt{\lambda} \over \sqrt{1-v^2}}
    \begin{pmatrix} v & -1 & 0 & 0 \end{pmatrix} \,.
 \end{equation}
This non-conservation comes from the fact that the quark is prescribed to move at constant velocity.  The drag force can be interpreted as minus the force exerted by the quark on the medium.  With this interpretation in mind, we can check explicitly that the drag force \eqref{DragForce} can be recovered from \eqref{E:NonConservationTmn}, as follows.  Given the stress-energy tensor, the external force acting on a region $V$ can be computed from 
 \begin{equation}
  \label{E:GotFn}
  F^n = {d \over dt} \int_V d^3 x\, \langle T^{0n} \rangle
    + \oint_{\partial V} d^2 a \, n_i \langle T^{in} \rangle
    = \int_V d^3 x\, \partial_m \langle T^{mn} \rangle \,.
 \end{equation}
The region $V$ is assumed not to depend on time in the asymptotic rest frame of the plasma.  The first term in \eqref{E:GotFn} gives the rate of change of energy-momentum in this region, while the second term corresponds to the energy-momentum flux through the boundary of $V$.  To obtain the last equality we used the divergence theorem.  Using \eqref{E:TmnKDef} and taking the limit where the volume of $V$ goes to infinity, as appropriate for computing the total force on the system, one can see that
 \begin{equation}
  \label{E:FFourier}
  F^n = i z_H^2  \lim_{\vec{K} \to 0} K_m \langle T^{mn}_K \rangle 
   = {v \over 2} {\pi T^2 \sqrt{\lambda} \over \sqrt{1-v^2}}
   \begin{pmatrix} v & 1 & 0 & 0 \end{pmatrix} \,.
 \end{equation}
$F^1$ is then indeed minus the drag force, as can be easily checked by comparing \eqref{E:FFourier} to \eqref{DragForce}. 

The prescription \eqref{E:TmnFromQmnAgain} gives the stress tensor in terms of the coefficients $Q_{mn}$ that appear in the near boundary asymptotics of the components of $H_{mn}$. We can give similar prescriptions in terms of the near boundary asymptotics of the other linear combinations of metric perturbations that we discussed in section~\ref{S:Axial}---all that we need to do is to relate the $Q_{mn}$ to a near boundary expansion of these variables. For the $ABCDE$ variables defined in \eqref{E:HmnDef}, the near boundary behavior is
  \begin{equation}
    \label{E:XBoundary}
    X = R_X+\frac{P_X}{3} y^3 + Q_X y^4   + \mathcal{O}(y^5)\quad {\rm where} \quad X=A,B_i,C,D_i,E_i\,,
  \end{equation}
Here, as before, the $Q_X$ and $R_X$ are arbitrary constants while the $P_X$ are set by \eqref{E:OriginalABCDEset}. The boundary condition $R_{mn}=0$ is equivalent to $R_X=0$.

It is easy to invert the definition \eqref{E:HmnDef} and use the near boundary expansions \eqref{E:XBoundary} and \eqref{E:HmnBoundary} to write the $Q_{mn}$ components  in terms of the $Q_X$.  However, it is useful to notice that we need not specify all of the $Q_X$. The reason for this is that the constraint equations \eqref{E:BConstraint}, \eqref{E:DConstraint}, and \eqref{E:EConstraint} imply relations among the $Q_X$, namely
  \begin{equation}
    \label{E:QXConstraints}
    \begin{split}
       &Q_{B_1} - Q_{B_2} =0 \qquad Q_{D_1} -Q_{D_2}= -\frac{i}{4 v K_1} \qquad Q_{E_1} + 2 Q_{E_3} = 0 \\ 
       &Q_{E_1} - 2Q_{E_2} = -\frac{i v}{2 K_1} \qquad (1-3v^2\cos^2\vartheta) Q_{E_1} +2 Q_{E_4} = \frac{3 i v (1+v^2) \cos\vartheta}{2 K} \,.
     \end{split}
  \end{equation}
Using \eqref{E:QXConstraints}, we can write $Q_{mn}$ in terms of only five of the $Q_X$, one from each set. This will be nothing more than a parameterization of the most general $Q_{mn}$ obeying constraints \eqref{E:Tracelessness} and \eqref{E:NonConservationQmn}. 

Recalling that the response to the trailing string has $B_1 = B_2 = C = 0$ everywhere because of symmetry, we conclude that we can write $Q_{mn}$ in terms of $Q_A$, $Q_D\equiv Q_{D_1}$ and $Q_E\equiv Q_{E_1}$:\footnote{In later sections, we will continue to use $Q_D$ to mean $Q_{D_1}$, and likewise $Q_E = Q_{E_1}$.  If the gauge-invariants $D$ and $E$ defined in \eqref{E:BDInvDef} and~\eqref{E:EInvDef} were expanded in powers of $y$, like in \eqref{E:XBoundary}, the coefficients of $y^4$ would be related to $Q_{D_1}$ and $Q_{E_1}$, but not identically equal.}
  \begin{equation}
    \label{E:GotQmnFromQX}
    Q_{mn} = a_{mn} Q_A + d_{mn} Q_D + e_{mn} Q_E + p_{mn} \,,
  \end{equation}
where
  \begin{equation}
    \label{E:amnDef}
    a_{mn} = \frac{v^2 \sin^2\vartheta}{2}
    \begin{pmatrix}
      0 & 0 & 0 & 0 \\
      0 & -2 \sin^2\vartheta & \sin 2\vartheta & 0 \\
      0 & \sin 2\vartheta & -2 \cos^2\vartheta & 0 \\
      0 & 0 & 0 & 2
    \end{pmatrix} 
  \end{equation}
  \begin{equation}
    \label{E:dmnDef}
    d_{mn} = \frac{v^2}{2}
    \begin{pmatrix}
      0 				& 	4 \sin^2\vartheta 		& 	- 2\sin 2\vartheta 	 & 0 \\      
      4\sin^2\vartheta 	& 	-2 \sin^2 2\vartheta	& 	\sin 4\vartheta 	 & 0 \\
      -2\sin 2\vartheta	&	\sin 4\vartheta		&	2\sin^2 2 \vartheta  & 0 \\
      0				&	0				&	0		           & 0 
    \end{pmatrix} 
  \end{equation}
  \begin{equation}
    \label{E:emnDef}
    \begin{split}
      e_{mn} &= \frac{1}{4}
      \begin{pmatrix}
        -4			& 	4 v\cos^2\vartheta 							& 	2 v\sin 2\vartheta 									& 0 \\      
        4v\cos^2\vartheta 	& 	4 e_{11}								& 	4 e_{12} 				& 0 \\
        2v\sin 2\vartheta	&	4 e_{12}		&	4 e_{22}										& 0 \\
        0			&	0									&	0											& 2 v^2 \cos^2\vartheta -2 
      \end{pmatrix}  \\
   & e_{11} = \frac{1}{2} \left( -1 + (1+v^2) \cos^2\vartheta -3 v^2 \cos^4\vartheta \right) \\ 
   & e_{12} = \frac{1}{4} \sin 2\vartheta \left(1-3v^2\cos^2\vartheta\right) \\
   & e_{22} =\frac{1}{2} \cos^2\vartheta \left(-1 -2 v^2 + 3 v^2\cos^2\vartheta)\right) \
    \end{split} 
  \end{equation}
  \begin{equation}
   \label{E:pmnDef}
   \begin{split}
      p_{mn} &= \frac{i v \cos\vartheta}{8 K}
      \begin{pmatrix}
        0			& 	4 v							& 	4 v\tan\vartheta					& 0 \\      
        4v		 	& 	\wp_{11}	& 	\wp_{12}	& 0 \\
        4v \tan\vartheta	&	\wp_{12}	&	\wp_{22}	& 0 \\
        0			&	0							&	0							& 2+2v^2 
      \end{pmatrix}  \\
     &\wp_{11} = (1-3v^2)\cos 2\vartheta - 5 - v^2  \\
     &\wp_{12} = (1-3v^2)\sin 2\vartheta - 4\tan\vartheta  \\
     &\wp_{22} = (3v^2-1)\cos 2\vartheta  + 3 -v^2 \,.
   \end{split}
  \end{equation}
For the master fields \eqref{E:masterfields}, the near boundary behavior is 
  \begin{subequations}
  \label{E:masterboundary}
    \begin{align}
    \label{E:psiTbdy}
    \psi_T^{\rm even} &= \frac{P_T}{L^3} y^3 + \frac{Q_T}{L^8} y^4 + \mathcal{O}(y^5)  \\
    \label{E:psiVbdy}
    \psi_V^{\rm even} &=  \frac{P_V}{L^4} y^2 + \frac{Q_V}{L^6} y^3 + \mathcal{O}(y^4)  \\
    \label{E:psiSbdy}
    \psi_S &=  \frac{P_S}{L^2} y + \frac{Q_S}{L^4} y^2 + \mathcal{O}(y^3) \,. 
    \end{align}
  \end{subequations}
 Here we write asymptotics after having imposed the condition $R_{mn}=0$. Similarly to before, the $Q$ coefficients are arbitrary while the $P$ are set by \eqref{E:EOMmasterfields}. The odd master fields $\psi_T^{\rm odd}$ and $\psi_V^{\rm odd}$ have similar expansions, but since their equations of motion are homogeneous, we can set them to be identically zero and worry about them no longer. Using the definitions \eqref{E:masterfields} and \eqref{E:QXConstraints} it is easy to relate the $Q$ coefficients for the master fields to the $Q_X$. The result is
   \begin{subequations}
   \label{E:GotQXFromPsi}
      \begin{align}
      \label{E:GotQAFromPsi}
      Q_A &= -\frac{Q_T}{\alpha_v v^2 L^2 z_H^3 \sin^2\vartheta}  \\
      \label{E:GotQDFromPsi}
      Q_D &= -\frac{i}{4 v K_1} + \frac{Q_V}{4\alpha_v  v L^2 z_H^2 \sin\vartheta}  \\ 
      \label{E:GotQEFromPsi}
      Q_E &= -\frac{i v}{2 K_1} - \frac{Q_S}{2\alpha_v L^2 z_H} \,.
      \end{align}
   \end{subequations}
Similar formulas relating the $Q_X$ to the asymptotics of the other gauge-invariants discussed can be easily derived, but we will not need them.

\section{Asymptotics}
\label{S:Asymptotics}

Equations \eqref{E:OriginalABCDEset} or \eqref{E:EOMmasterfields} are difficult to solve exactly and we eventually resort to numerics to obtain a full solution. However, there are various approximations which can be used in order to get a handle on the large and small momentum asymptotics of the metric fluctuations. These may be Fourier transformed to real space, giving us approximations to the near- and far-field behavior of the boundary theory stress-energy tensor. In section~\ref{SS:Largerasymptotics} we focus on the small momentum asymptotics of the solution:  in section \ref{SSS:smallKasymptotics} we construct a small momentum series expansion of the metric perturbations, while in subsection \ref{SSS:hydro}  we explain how this maps into  hydrodynamic behavior on the boundary theory. In section~\ref{SS:Smallrasymptotics} we focus on the near-field of the stress-energy tensor.

\subsection{Long distance asymptotics}
\label{SS:Largerasymptotics}

\subsubsection{Momentum space analysis}
\label{SSS:smallKasymptotics}
The small momentum asymptotics of the solution can be obtained by formally expanding the appropriate fields in power series in the momentum $K$. In what follows we will go over such an expansion in the $A$, $B$, $C$, $D$, $E$ variables of \eqref{E:HmnDef}.\cite{Friess:2006fk}  A similar construction can be carried out for the other parameterizations of the Einstein equations\cite{Gubser:2007nd,Chesler:2007an} given in \eqref{E:CheslerInvDef} and \eqref{E:masterfields}. We start by formally expanding the $A$ variable in \eqref{E:HmnDefA} such that
\begin{equation}
\label{E:AexpansionLowK}
	A = \sum_{n=0}^{\infty} \alpha_n K^n \,,
\end{equation}
where the $\alpha_n$ are functions of $\vec{K}$ which are invariant under rescalings $\vec{K} \to \lambda \vec{K}$.  That is, the $\alpha_n$ only depend on the direction of $\vec{K}$, not its magnitude $K = |\vec{K}|$.  Plugging the expansion \eqref{E:AexpansionLowK} into \eqref{E:ASet} and collecting terms with identical powers of $K$, we obtain a set of equations for $\alpha_n$ of the form
\begin{equation}
\label{E:alphaequations}
	\frac{y^3}{h}\partial_y \frac{h}{y^3} \partial_y \alpha_n = S_n \,,
\end{equation}
where $S_n$ can depend on $\alpha_m$ with $m<n$. For example, 
\begin{equation}
\label{E:alphaSources}
	S_0 = \frac{y}{h} \quad 
	S_1 = -i\frac{y}{h}\frac{\cos\vartheta \xi}{h z_H} \quad
	S_2 = -\frac{y}{h} \left(\frac{\cos\vartheta}{z_H} \right)^2 - \frac{v^2 \cos^2\vartheta - h}{h^2}\alpha_0.
\end{equation}
The most general solution to \eqref{E:alphaequations} is 
\begin{equation}
\label{E:alphasolution}
	\alpha_n  = \int_{y_0}^y d\tilde{y} \frac{\tilde{y}^3}{h(\tilde{y})} \int^{\tilde{y}}_{y_1} d\tilde{\tilde{y}} \frac{h(\tilde{\tilde{y}})}
{\tilde{\tilde{y}}^3} S_n(\tilde{\tilde{y}}) \,.
\end{equation}
In order to satisfy the boundary conditions \eqref{E:boundaryBC} at the asymptotically AdS boundary, we require that $y_0 = 0$. The other integration constant in \eqref{E:alphasolution} is obtained by matching the near horizon behavior of \eqref{E:alphasolution} to the series approximation  \eqref{E:Anearhorizon} with $V_A = 0$, expanded at small $K$. For example, at order $n=0$, we find that
 \begin{equation}
  \label{E:alpha0sol}
	\alpha_0 = \frac{1}{4}\left(2 \tan^{-1} y+
	  \log {1-y \over 1+ y}  +y_1 \log {1+y^2 \over 1 - y^2}\right) \,.
 \end{equation}
Expanding \eqref{E:alpha0sol} near the horizon and matching it to a small $K$ expansion of \eqref{E:Anearhorizon},
\begin{equation}
	A = (1-y) + U_A + \mathcal{O}(K) \,,
\end{equation}
we find that we need to set $y_1 = 1$.

According to \eqref{E:TmnFromQmnAgain} and \eqref{E:GotQmnFromQX}, to obtain the stress tensor we will eventually need $Q_A$. Expanding \eqref{E:alpha0sol} for small $y$ and recalling \eqref{E:XBoundary}, we can read off that $Q_A=1/4 + \mathcal{O}(K)$.  The next order corrections to $Q_A$ can be obtained in the same manner. We find that to order $\mathcal{O}(K)$,
\begin{equation}
\label{E:QAsolution}
	Q_A = \frac{1}{4} - \frac{i \ln 2}{8} v K_1 +\mathcal{O}(K^2)
	  \,.
\end{equation}
A similar treatment gives us the asymptotic values of $D_1$ and $E_1$. The result is 
\begin{align}
\label{E:QDsolution}
	Q_D &= -\frac{ i \sec \vartheta}{4 v K} - \frac{\sec^2 \vartheta - 4 v^2}{16 v^2} + \mathcal{O}(K^2) \\
\label{E:QEsolution}
	Q_E &=  \frac{3 i v (1+v^2)\cos\vartheta}{2 K (1- 3 v^2 \cos^2\vartheta)} -
		\frac{3 v^2 \cos^2\vartheta (2 + v^2 (1-3 \cos^2 \vartheta))}{2 (1- 3 v^2 \cos^2\vartheta)^2} \,.
\end{align}
Note that $Q_D$ exhibits a pole structure at $K=0$, while $Q_E$ exhibits a pole structure at $K^2 = 3 K_1^2 v^2$. We will see shortly that these correspond to the diffusion pole and sound pole expected of the hydrodynamic behavior of the plasma far from the moving quark.

\subsubsection{Relating the large momentum asymptotics to hydrodynamics}
\label{SSS:hydro}
At scales much larger than the mean free path, we expect to be able to describe a thermal gauge theory by effective,  hydrodynamic, slowly varying degrees of freedom. Let's first see what such a description entails, and then check how well our asymptotics match it. Consider a conformal theory
for which the hydrodynamic energy-momentum tensor $\Thydro_{mn}$ is traceless:
\begin{equation}
\label{E:traceless}
      \Thydro^m_m  = 0 \,.
\end{equation}
A static configuration of the fluid will be given by
\begin{equation}
	\Thydro_{mn} = {\epsilon_0 \over 3} {\rm diag} 
	\left\{3, 1, 1, 1 \right\} \,.
\end{equation}
If we now perturb this fluid slightly, and choose our hydrodynamic variables to be $\epsilon = \Thydro^{00}$ and $S_i = \Thydro^{0i}$ with $i=1$, $2$, or $3$, then we should be able to write the remaining space-space components of the energy-momentum tensor, $\Thydro_{ij}$, in terms of gradients of $\epsilon$  and $S_i$. The only possible combination consistent with the tracelessness condition \eqref{E:traceless} is
\begin{equation}
\label{E:constitutive}
	\Thydro_{ij} = \frac{1}{3}\epsilon \delta_{ij} -\frac{3}{2}\Gamma ik_{(i} S_{j)} + \mathcal{O}(k^2) \,,
\end{equation}
where the free parameter $\Gamma$ is the sound attenuation length, related to the shear viscosity, temperature, and entropy density through $\Gamma = 4 \eta/3 s T$. Using \eqref{E:EtaOverS} we find that $\Gamma = 1/3\pi T$ in the theories we are considering. Brackets denote the symmetric traceless combination, i.e.,
\begin{equation}
M_{(mn)} = \frac{1}{2} \left( M_{mn} + M_{nm}\right) - \frac{1}{4}\eta_{mn} \eta^{pq} M_{pq} \,.
\end{equation}
The energy density $\epsilon$ and the energy flux $S_i$ can now be computed from the conservation equations,
\begin{equation}
\label{E:nonconservation}
	i k_n \Thydro^{n}_{m} = f^{\rm hydro}_m \,,
\end{equation}
where $f^{\rm hydro}_m$ is the source which perturbs the hydrodynamic stress-energy tensor. Separating the spatial part of the source, $\vec{f}^{\rm hydro}$, into longitudinal and transverse components, $\vec{f}^{\rm hydro}_L = (\vec{f}^{\rm hydro} \cdot \hat{k}) \hat{k}$ and $\vec{f}^{\rm hydro}_{T} = \vec{f}^{\rm hydro}-\vec{f}^{\rm hydro}_L$, with $(\hat{k})_i = k_i/k$, we find
from \eqref{E:constitutive} and \eqref{E:nonconservation} that
\begin{subequations}
\label{E:hydrosol}
\begin{align}
\label{E:epsilonhydro}
	\epsilon &= \frac{-3 i \left(\vec{f}^{\rm hydro} \cdot \vec{k}-f^{\rm hydro}_0 \omega\right)-3 f_0 k^2 
	\Gamma} {k^2-3\omega^2-3 i  \Gamma  k^2\omega} + \mathcal{O}(f k)\\
\label{E:SLhydro}
	\vec{S}\cdot \vec{k} &= \frac{-f^{\rm hydro}_0 k^2 + 3 \vec{f}^{\rm hydro}\cdot\vec{k}\omega}{k^2-3\omega^2-3 i  
\Gamma  k^2\omega} + \mathcal{O}(f k)\\
\label{E:SThydro}
	\vec{S}-(\vec{S}\cdot\hat{k}) \hat{k} &= \frac{-i\vec{f}^{\rm hydro}_T}{\omega+\frac{3}{4}\Gamma i k^2}+ \mathcal{O}(f 
k) \,.
\end{align}
\end{subequations}
In real space, the pole at approximately $\omega\sim 0$ corresponds to a diffusive wake behind the source, and the pole at roughly $k^2 \sim 3 \omega^2$ corresponds to the shock wave which appears along the Mach cone behind the moving probe. The displacement of the two poles from the real axis due to the shear viscosity of the fluid is responsible for viscous broadening of the wake and Mach cone. 

We can now compare the small momentum results of section~\ref{SSS:smallKasymptotics} to the hydrodynamic behavior \eqref{E:hydrosol}. To do so, we use \eqref{E:QAsolution}--\eqref{E:QEsolution} in \eqref{E:GotQmnFromQX} and \eqref{E:TmnFromQmnAgain} to obtain the small momentum stress-energy tensor. We find
\begin{equation}
\label{E:TmnSmallK}
	\langle T^K_{mn} \rangle =
		\frac{(\pi T)^4 \sqrt{\lambda}}{\sqrt{1-v^2}}\left[\begin{pmatrix}
			\epsilon & -S_1 & -S_{\bot} & 0 \\
			-S_1 & & & \\
			-S_{\bot} & & T_{ij} & \\
			0 & & & \\
		\end{pmatrix}
		+
		\mathcal{T}_{mn}
		+\mathcal{O}(K)
		\right]
\end{equation}
with
\begin{subequations}
\label{E:Holographicdata}
\begin{align}
\label{E:EnergysmallK}
	\epsilon  =& -\frac{3 i K_1 v(1+v^2)}{2\pi\left(K^2-3 K_1^2 v^2\right)}	 
	+\frac{3 K_1^2 v^2\left(-3 K_1^2 v^2+K^2(2+v^2)\right)}{2\pi\left(K^2-3 K_1^2 v^2\right)^2}\\
	\notag
		S_1 = &-i\frac{i K_1 \left(v^2+1\right)}{2 \pi  \left(K^2-3 K_1^2 v^2\right)}
		+\frac{-6 K_1^4 v^5-K^4 v+K_1^2 K^2 \left(6 v^2+1\right) v}{2 \pi 
   		\left(K^2-3 K_1^2 v^2\right)^2}\\
   	\label{E:S1smallK}
  		 &+\frac{i}{2 K_1 \pi}+\frac{K_{\bot}^2}{32 K_1^2\pi v^2}\\
	\notag
   		S_{\bot}  = & -\frac{i K_2(1+v^2)}{2\pi\left(K^2-3 K_1^2 v^2\right)}+\frac{K_1 K_2 v\left( K^2+3 	
		K_1^2 v^4\right)}{2\pi (K^2 - 3 K_1^2 v^2)^2}\\
	\label{E:SpsmallK}
		&-\frac{K_2}{8\pi K_1 v}
\end{align}
\end{subequations}
and
\begin{align}
\label{E:Tijasymptotics}
	T_{ij} & = \frac{1}{3} \epsilon \delta_{ij} - \frac{1}{6} i K_{(i}S_{j)}\\
\label{E:nonhydroasymptotics}
	\mathcal{T}_{mn} &= \frac{v^2}{6\pi} {\rm diag} \left\{ 0,  -2, 1, 1\right\} \,.
\end{align}
Note that we may always carry out a resummation
\begin{equation}
\label{E:soundpole}
	\frac{\alpha_i K_i}{K^2 - 3 K_1^2 v^2} + \frac{\mathcal{O}(K^4)}{(K^2 - 3 K_1^2 v^2)^2}
	=
	\frac{\alpha_i K_i + \mathcal{O}(K^2)}{K^2 - 3 K_1^2 v^2- i K_1 K^2 v}
\end{equation}
(where $\alpha_i$ are constants) or 
\begin{equation}
\label{E:diffusionpole}
	\frac{1}{ v K_1} +\frac{\mathcal{O}(K^2)}{v^2 K_1^2} = \frac{1+\mathcal{O}(K)}{v K_1 + \frac{1}{4}i K^2} \,.
\end{equation}
After such a resummation, the expressions for $\epsilon$ and $\vec{S}$ in \eqref{E:Holographicdata} take the form 
\eqref{E:hydrosol}
with
\begin{equation}
	f^{\rm hydro}_n =
		\frac{1}{2\pi}\begin{pmatrix}-v^2 & v & 0 & 0\end{pmatrix}_n
		- i K^m \mathcal{T}_{mn}
\end{equation}
and $\Gamma = 1/3\pi T$. 
Thus, if we identify the first term in the parenthesis on the right hand side of \eqref{E:TmnSmallK} with the hydrodynamic contribution to the stress tensor of the SYM theory, $(T^{\rm hydro, SYM})_{mn}$,
then we find that $(T^{\rm hydro, SYM})_{mn}$ satisfies
\begin{equation}
	i K^m (T^{\rm hydro, SYM})_{mn}= f^{\rm hydro, SYM}_n \,.
\end{equation}
This should be compared to the full conservation law \eqref{E:FFourier},
\begin{equation}
	i K^m \langle T^K_{mn} \rangle = f_n \,.
\end{equation}
The extra term $\mathcal{T}_{mn}$ in \eqref{E:TmnSmallK} holds information on the deviation of the stress-energy tensor from its hydrodynamic form. Alternately, $J_n - i k^m \mathcal{T}_{mn}$ gives us an effective hydrodynamic four-force which sources the hydrodynamic stress-energy tensor. It is no coincidence that the large distance asymptotics of the stress-energy tensor agree with a hydrodynamic expansion. In fact, it can be shown that generic probe-sources excite the metric in such a way that the resulting large distance asymptotics of the boundary theory stress tensor will have  hydrodynamic behavior.\cite{Gubser:2008vz} There is also mounting evidence that such a connection between hydrodynamics and gravity goes beyond the linearized approximation.\cite{Bhattacharyya:2008jc}

To see that the pole structures in \eqref{E:soundpole} and in \eqref{E:diffusionpole} really correspond to a laminar wake and a shock wave, we Fourier transform them to real space. Consider the last two terms in \eqref{E:S1smallK}, resummed as in \eqref{E:diffusionpole}.
Using
\begin{equation}
\label{E:MainFT}
	\int \frac{d^dK}{(2\pi)^d} \frac{e^{i \vec{K} \cdot \vec{X}}}{(K^2 + \mu^2)^n} = \frac{2}{(4\pi)^{d/2} \Gamma(n)} \left(\frac{X}{2\mu} \right)^{n-d/2} \mathbf{K}_{n-d/2}(\mu X) \,,
\end{equation}
we can Fourier transform the resummed expression to position space:
\begin{multline}
	\int \frac{d^3 K}{(2\pi)^3}\frac{2 v}{\pi (K^2 - 4 i K_1 v)} e^{i \left(K_1 (x^1-v t)+K_2 x^2 + K_3 x^3\right)/z_H}
	\\
	=
	\frac{e^{-\frac{2 v}{z_H} \left( (x^1-v t)+\sqrt{(x^1-v t)^2+x_{\bot}^2}\right)}v z_H}{2 \pi^2 \sqrt{(x^1-v t)^2+x_{\bot}^2}} \,.
\label{E:realwake}
\end{multline}
As expected of a diffusion wake, we find that the configuration \eqref{E:realwake} exhibits a directional energy flow, with a parabolic shape far behind the moving quark.

Fourier transforming the resummed sound pole \eqref{E:soundpole} is difficult due to the cubic terms in the denominator, and we eventually resort to numerics to convert such expressions to real space. To see that the pole at $K^2 \sim 3 K_1^2 v^2$ really corresponds to a shock wave, it is sufficient to Fourier transform only the leading order contributions to these poles. Since we are neglecting the viscous contribution to the pole structure, this corresponds to the inviscid limit. Using contour integration and the identities\cite{Gradshteyn}
\begin{align}
 	\int_0^{\infty}  \mathbf{J}_0(a x) \sin(b x) d x  = \begin{cases}
		0 & 0<b<a \\
		\frac{1}{\sqrt{b^2 - a^2}} & 0<a<b
		\end{cases} \\
 	\int_0^{\infty}  \mathbf{J}_0(a x) e^{-b x} d x  =
		\frac{1}{\sqrt{a^2 + b^2}}\quad {\rm Re}(a \pm i b)>0 \,,
\end{align}
where $\mathbf{J}$ is a Bessel function of the first kind, we find that, for example, the leading contribution to the energy density \eqref{E:EnergysmallK} reads
\begin{multline}
	-\int \frac{d^3 K}{(2\pi)^3}\frac{3 i K_1 v (1+v^2)}{2\pi (K^2 - 3 K_1^2 v^2)} e^{i \left(K_1 (x^1-v t)+K_2 x^2 + K_3 x^3\right)/z_H}
	\\
	= \begin{cases}
		-\frac{3 v \left(v^2+1\right) x^1 z_H^2}{8 \pi ^2 \left((x^1)^2+\left(1-3 v^2\right) x_{\bot}^2\right)^{3/2}} 
		& v^2<1/3 \\
		-\frac{3 v \left(v^2+1\right) x^1 z_H^2}{4 \pi ^2 \left((x^1)^2+\left(1-3 v^2\right) x_{\bot}^2\right)^{3/2}} 
		& v^2>1/3, \quad -x_\perp \sqrt{3v^2-1} < x^1 < 0 \\
		0 & {\rm otherwise} \,.
	\end{cases}\label{E:SingMach}
\end{multline}
In obtaining \eqref{E:EnergysmallK} we assumed that the sound poles are slightly below the real $K_1$ axis for $v^2>1/3$.  This assumption is motivated by the fact that the viscous corrections that appear at the next order do shift the poles to the lower-half complex $K_1$ plane.  As expected, since we have treated viscous contributions as infinitesimal in the leading order result \eqref{E:SingMach}, the real space expression for the energy density is singular along the Mach cone.

\subsection{Short distance asymptotics}
\label{SS:Smallrasymptotics}
As was the case for the large distance asymptotics, the short distance asymptotics of the solution can be obtained by  formally expanding all variables in large momenta.  Starting from the $z$ coordinate system in \eqref{E:LineElement}, we look  at momenta which are much larger than the inverse temperature scale, $K = k z_H \gg 1$. In this case, it is more practical to use the dimensionless radial coordinate $Z = z k$ instead of $y = z/z_H$.  Setting $K \gg 1$ means that we're pushing the black hole horizon off to $Z \to \infty$, implying that we're nearing the zero temperature limit.   To see this explicitly, consider the expansion
\begin{equation}
\label{E:PsiTexpansionLargeK}
	\psi_T^{\rm even} = K^{-3}\sum_{n=0}^{\infty} t_n K^{-n}
\end{equation}
of the tensor modes defined in \eqref{E:masterfields}, similar to \eqref{E:AexpansionLowK}. Plugging the expansion \eqref{E:PsiTexpansionLargeK} into \eqref{E:EOMmasterT} and collecting terms with similar powers of $K$, we find that the $t_n$'s satisfy
\begin{equation}
\label{E:tequations}
	\partial_Z^2 t_n - \frac{3}{Z} \partial_Z t_n - \left(1 - \frac{k_1^2}{k^2} v^2 \right) t_n = -(J_T)_n
\end{equation}
where, as before, we need to compute $(J_T)_n$ order by order in a large $K$ expansion. For $n=0$, $1$, and $2$ we find
\begin{align}
\label{E:tsources}
	-(J_T)_0 &= -\frac{z_H^3 v^2 \alpha_v}{L^6} \sin^2\vartheta Z\\
	 -(J_T)_1 &= 0 \\
	 -(J_T)_2 &= -\frac{z_H^3 v^2 \alpha_v}{L^6} \sin^2\vartheta \frac{i K_1 v Z^4}{3 K} \,.
\end{align}

To solve \eqref{E:EOMmasterT} we use the method of Green's functions. The homogeneous version of \eqref{E:tequations}  can be easily solved. The solutions are 
\begin{equation}
\label{E:hom}
	t^{(1)}(Z) =  Z^2 \mathbf{I}_2\left(\sqrt{1-\frac{k_1^2 v^2}{k^2}}Z\right) 
	\quad 
	t^{(2)}(Z) =  Z^2 \mathbf{K}_2\left(\sqrt{1-\frac{k_1^2 v^2}{k^2}}Z\right)
\end{equation}
where $\mathbf{I}_2$ and $\mathbf{K}_2$ are modified Bessel functions of the first and second kind.  In \eqref{E:hom} we suppressed the index $n$ because the homogeneous parts of \eqref{E:tequations} are the same for all $n$.   To simplify the notation we define $\Alpha = \sqrt{1- k_1^2 v^2/k^2}$.  The solution which vanishes near the boundary, and therefore does not correspond to a deformation of the theory, is $t^{(1)}$.  The other solution, $t^{(2)}$, is the only solution which does not diverge exponentially in the deep interior of AdS\@. Clearly, $t^{(1)}$ is the solution which captures the boundary asymptotics we have in mind, and it seems physically reasonable to disallow a solution which is exponentially  divergent at large $Z$.  A more rigorous approach would be to find a uniform approximation to the two linearly independent solutions of the homogeneous version of \eqref{E:EOMmasterT} and to show that the solution that behaves like $t^{(2)}$ near the boundary is not purely infalling at the horizon. This can be carried out via a WKB approximation.\cite{Yarom:2007ap} Thus, the solution to
\begin{equation}
	\partial_Z^2 G(Z,Z^{\prime}) - \frac{3}{Z} G(Z,Z^{\prime}) - \left(1 - \frac{k_1^2}{k^2} v^2\right) G(Z,Z^{\prime}) = \delta(Z-
Z^{\prime})
\end{equation}
is
\begin{equation}
	G = \begin{cases}
		(Z^{\prime})^{-3}t^{(2)}(Z^{\prime}) t^{(1)}(Z)\,& Z<Z^{\prime} \\
		(Z^{\prime})^{-3}t^{(1)}(Z^{\prime}) t^{(2)}(Z)\,& Z>Z^{\prime}  \,,
	\end{cases}
\end{equation}
and then
\begin{equation}
	t_n(Z) = \int dZ^{\prime} G(Z,Z^{\prime}) (J_T)_n(Z^{\prime}) \,.
\end{equation}
For $n=0$ and $n=1$ these integrals may be carried out exactly.\cite{Gradshteyn}  We find
\begin{align}
\notag
	 t_0(Z) & = -\frac{z_H^3 v^2 \alpha_v}{L^6} \sin^2 \vartheta \left[
	 Z^2 \mathbf{L}_2 \left(\Alpha Z\right)
	 -Z^2 \mathbf{I}_2\left(\Alpha Z\right)
	 +\frac{2 \Alpha}{3\pi}Z^3 \right]\\
\label{E:texpansion}
	 t_1(Z) & = 0\\
\notag
	  t_2(Z) & = -\frac{z_H^3 v^2 \alpha_v}{L^6} \sin^2  \vartheta
	  -\frac{i K_1 v Z^4}{3 K \Alpha^2} \,,
\end{align}
with $\mathbf{L}_2$ a modified Struve function.  Recalling \eqref{E:psiTbdy} we can read off
\begin{equation}
\label{E:QAlong}
	Q_T  = -z_H^3 v^2 \alpha_v \sin^2 \vartheta L^2 \left[ \frac{1}{16}\pi\sqrt{K^2-K_1^2 v^2} - \frac{i v K_1}{3 (K^2-K_1^2 v^2)}\right]
\end{equation}
from \eqref{E:texpansion}. The equation of motion for $t_0$ coincides with the one that would have been obtained starting from a string hanging straight down from the boundary of AdS space, boosted to a velocity $v$ in the $x_1$ direction. The solution $t_0$ then corresponds to the tensor mode metric perturbation in response to this string and, as we will see shortly, it captures the near-field physics of the stress-energy tensor in response to a massive quark.   At scales much smaller than the mean free path, one may effectively ignore the interaction of the quark with the plasma. The function $t_2$ corresponds to the first thermal corrections to the near-field of the quark.

The computation of the large $K$ asymptotics for the vector and scalar modes follows in a similar manner. Let
\begin{equation}
\label{E:SexpansionLargeK}
	\psi_V^{\rm even} =K^{-4} \sum_{n=-\infty}^0 v_n {K}^{-n} \,,
\end{equation}
similar to \eqref{E:PsiTexpansionLargeK}. Expanding \eqref{E:EOMmasterV} at large $K$, we find
that the $v_n$'s satisfy
\begin{equation}
\label{E:VectorEOMLargeK}
	\partial_Z^2 v_n -\frac{3}{Z} \partial_Z v_n+(3-Z^2\Alpha^2)v_n =-(J_V)_{n} \,.
\end{equation}
The first few terms in $-(J_V)_{n}$ are given by
\begin{align}
	-(J_V)_0 &= \frac{\alpha_v v K_{\bot} Z^4}{K^5} \\
	-(J_V)_1 &= 0 \\
	-(J_V)_2 &= -i\frac{\alpha_v  K_{\bot}(3 K^2 - 4 K_1^2 v^2)}{3 K^2 K_1 } \,.
\end{align}
The homogeneous solutions to \eqref{E:VectorEOMLargeK} are
\begin{equation}
	v^{(1)} = Z^2 \mathbf{I}_1(\Alpha Z)  \qquad v^{(2)} = Z^2 \mathbf{K}_1 (\Alpha Z) \,,
\end{equation}
and using the Green's function method we find
\begin{multline}
	\psi_V^{\rm even} = \frac{\alpha_v K_{\bot} \pi v}{2 K \Alpha^2} 
	Z^2 \left( \mathbf{L}_1(\Alpha Z)-\mathbf{I}_1(\Alpha Z)\right)\\ 
	+\frac{i \alpha_v K_{\bot}(3 K^2 - 4 K_1^2 v^2)
	Z^3(8+Z^2\Alpha^2)}{3 K_1 K^2 \Alpha^4 }Z^3 \,,
\end{multline}
which, recalling \eqref{E:psiVbdy}, implies
\begin{equation}
	Q_V = -\frac{\alpha_v v k_{\bot}}{k }
		\left[
			-\frac{\pi \sqrt{k^2 - v^2 k_1^2}L^2}{4}
			+ \frac{3 k^2 - 4 k_1^2 v^2}{3 i v k k_1 (k^2 - k_1^2 v^2)}\frac{L^4}{z_H^2} +\mathcal{O}(z_H^{-4})
		\right] \,.
\end{equation}
The details of the computation of the scalar modes can be found elsewhere.\cite{Gubser:2007nd,Yarom:2007ni,Gubser:2007xz} The final result is
\begin{equation}
\begin{split}
	Q_S &=\alpha_v
		\Bigg[
		-\frac{\pi\left(2+v^2\right)(k^2-k_1^2 v^2)-v^2 (1-v^2)k_1^2}{12 \sqrt{k^2-k_1^2 v^2}}
		\\
		{}& -\frac{iv}{9 z_H^2}  \left( {9 \over k_1} - {k_1 (5- 11v^2) \over k^2- v^2 k_1^2}
		  - {2 v^2(1-v^2)k_1^3 \over (k^2- k_1^2 v^2)^2} \right)
		 +\mathcal{O}(z_H^{-4})
	\Bigg]
\end{split}
\end{equation}
where
\begin{equation}
	\psi_S = P_S L^{-2} y + Q_S L^{-4} y^2 + \mathcal{O}(y^5) \,.
\end{equation}

With $Q_T$, $Q_V$, and $Q_S$ at hand we can use \eqref{E:GotQXFromPsi}, \eqref{E:GotQmnFromQX}, and \eqref{E:TmnFromQmnAgain} to obtain the leading large momentum asymptotics of the stress-energy tensor. The momentum space expressions for the energy density and Poynting vector are
\begin{equation}
\begin{split}
\label{E:EnergylargeK}
		\langle T_{00}^K \rangle  &= 
		-\frac{\pi^3 T ^4 \sqrt{\lambda}}{\sqrt{1-v^2}}
		\Bigg[
		- \frac{(2+v^2)\sqrt{K^2 - K_1^2 v^2}}{24}
		+\frac{K_1^2 v^2 (1-v^2)}{24 \sqrt{K^2 - K_1^2 v^2} }
		\\
		{}&-\frac{i K_1 v( 11 v^2 - 5)}{18 \pi (K^2 -K_1^2 v^2)}
		+\frac{14+7 v^2}{24 \left(K^2 - K_1^2 v^2\right)^{3/2}} 
		+\frac{i K_1^3 v^3 (1-v^2)}{9 \pi (K^2 - K_1^2 v^2)^2} \\
		{}&+\frac{K_1^2 v^2 \left(10 v^2 - 1\right)}{24(K^2 - K_1^2 v^2)^{5/2}}
		- \frac{K_1^4 v^4 (1 - v^2)}{8(K^2 - K_1^2 v^2)^{7/2}}
		\Bigg]
\end{split}
\end{equation}
\begin{equation}
\label{E:S1largeK}
\begin{split}
		\langle T_{01}^K \rangle  &= 
	-\frac{\pi^3 T ^4 \sqrt{\lambda}}{\sqrt{1-v^2}}
	\Bigg[	
	-\frac{v\sqrt{K^2 - K_1^2 v^2}}{8} 
	+ \frac{K_1^2 v (1-v^2)}{24 \sqrt{K^2 - K_1^2 v^2}} \\
	{}&-\frac{i K_1 v^2}{3 \pi (K^2 - K_1^2 v^2)}
	+\frac{9 v}{16 (K^2 - K_1^2 v^2)^{3/2}}
	+ \frac{i K_1^3 v^2 (1-v^2)}{9 (K^2 - K_1^2 v^2)^2 \pi} \\
	{}&+ \frac{K_1^2 v (1+17 v^2)}{48 (K^2 - K_1^2 v^2)^{5/2}}
	- \frac{K_1^4 v^3 (1-v^2)}{8 (K^2 - K_1^2 v^2)^{7/2}}
	\Bigg]
\end{split}
\end{equation}
and
\begin{equation}
\label{E:S2largeK}
\begin{split}
		\langle T_{02}^K \rangle  &= 
	-\frac{\pi^3 T ^4 \sqrt{\lambda}}{\sqrt{1-v^2}}	
	\Bigg[
	\frac{K_1 K_2 v (1-v^2)}{24 \sqrt{K^2 - K_1^2 v^2}}
	-\frac{i K_2 v^2}{2 \pi (K^2 - K_1^2 v^2)}  \\
	{}&+\frac{i K_1^2 K_2 v^2 (1-v^2)}{9 \pi (K^2 - K_1^2 v^2)^2} 
	+\frac{K_1 K_2 v (1+14 v^2)}{48 (K^2 - K_1^2 v^2)^{5/2}} 
	-\frac{K_1^3 K_2 v^3 (1-v^2)}{8 (K^2 - K_1^2 v^2)^{7/2}}
	\Bigg] \,.
\end{split}
\end{equation}
The expression for $\langle T_{03}^K \rangle$ can be obtained from \eqref{E:S2largeK} by exchanging $K_2$ with $K_3$.
In real space, using the notation in \eqref{E:TmnFromQmn}, we find that to leading order in $x T$,  
\begin{equation}
\label{E:Tmnquark}
	\langle T_{mn}^K \rangle = 
		\left(\Lambda^{-1} T^{\rm quark} \Lambda\right)_{mn}
\end{equation}
where $\Lambda_{mn}$ represents a Lorentz transformation with boost parameter $v$ in the $x_1$ direction and
$T^{\rm quark}_{mn}$ is given by
\begin{equation}
\label{E:TmnquarkStatic}
	T^{\rm quark}_{mn} =
		\frac{\sqrt{\lambda}}{12\pi^2}\begin{pmatrix}
			\displaystyle{\frac{1}{x^4}} & 0 & 0 & 0 \\[10pt]
			0 & 	\displaystyle{\frac{x_{\bot}^2-x_1^2}{x^6}} 
			  & -\displaystyle{\frac{2 x_1 x_2}{x^6}}
			  &  -\displaystyle{\frac{2 x_1 x_3}{x^6}}\\[10pt]
			0 &   -\displaystyle{\frac{2 x_1 x_2}{x^6}} 
			  & \displaystyle{\frac{x_{1}^2+x_3^2-x_2^2}{x^6}}
			  & -\displaystyle{\frac{2 x_2 x_3}{x^6}}  \\[10pt]
			0 &   -\displaystyle{\frac{2 x_1 x_3}{x^6}} 
			  &  -\displaystyle{\frac{2 x_2 x_3}{x^6}} 
			  & \displaystyle{\frac{x_{1}^2+x_2^2-x_3^2}{x^6}}
		\end{pmatrix}.
\end{equation}
$T^{\rm quark}_{mn}$ is the stress-energy tensor of a stationary heavy quark. Up to the overall multiplicative factor, it can be determined by the requirement that it is conserved and satisfies conformal symmetry. Thus, the leading short distance behavior of the near field of our quark is a boosted version of the stress-energy tensor of a stationary quark. At distances much shorter than the typical length scale of the fluid, the quark does not see the plasma it is moving through, and behaves as if it were in vacuum.

Of more interest are the subleading corrections to the stress-energy tensor. These are given by
\begin{subequations}
\label{E:nearfield}
\begin{align}
\label{E:energydensity}
	\langle T_{tt} \rangle &= \frac{\sqrt{\lambda} T^2}{\sqrt{1-v^2}}
	   \frac{v(x-v t)\left[x_{\bot}^2(-5+13 v^2-8 v^4)
	   +(-5+11 v^2)(x-v t)^2\right]}
	   {72 \left[x_{\bot}^2(1-v^2)+(x-v t)^2\right]^{5/2}} \\
	\langle T_{t x_1}  \rangle &= -\frac{\sqrt{\lambda} T^2}{\sqrt{1-v^2}}
	  \frac{v^2 (x-v t) \left[(1-v^2)x_{\bot}^2+2(x-vt)^2\right]}
	  {24 [x_{\bot}^2 (1-v^2)+(x-v t)^2]^{5/2}}  \\ 
	\langle T_{t x_{\bot}}  \rangle &= -\frac{\sqrt{\lambda} T^2}{\sqrt{1-v^2}} 
	  \frac{x_{\bot} (1-v^2)v^2 \left[8 x_{\bot}^2(1-v^2)+11 (x-vt)^2\right]}
		{72 \left[x_{\bot}^2 (1-v^2)+(x-vt)^2\right]^{5/2}} \\
	\langle T_{x_1 x_1}  \rangle &= \frac{\sqrt{\lambda} T^2}{\sqrt{1-v^2}}
	  \frac{v(x-vt)\left[x_{\bot}^2 (8- 13 v^2+5 v^4)+(11-5 v^2) (x-vt)^2\right]}
	  {72 \left[x_{\bot}^2(1- v^2)+(x-vt)^2\right]^{5/2}}\\
	\langle T_{x_{1} x_{\bot}}  \rangle &= \frac{\sqrt{\lambda} T^2}{\sqrt{1-v^2}}
	  \frac{v (1-v^2)\left[8x_{\bot}^2 (1-v^2)+11(x-vt)^2\right]}
		{72 \left[x_{\bot}^2(1-v^2)+(x-vt)^2\right]^{5/2}}\\
	\langle T_{x_{\bot} x_{\bot}}  \rangle &= -\frac{\sqrt{\lambda} T^2}{\sqrt{1-v^2}}
	  \frac{v(1-v^2)(x-vt)\left[5 x_{\bot}^2(1-v^2)+8 x_{\bot}^2\right]}
	  {72 \left[x_{\bot}^2(1- v^2)+(x-vt)^2\right]^{5/2}}\\
	\langle T_{\varphi \varphi}  \rangle &= - \frac{\sqrt{\lambda} T^2}{\sqrt{1-v^2}}
	  \frac{v(1-v^2)(x-vt)x_{\bot}^2}
	  {9\left[r^2(1-v^2)+(x-vt)^2\right]^{5/2}}
\end{align}
\end{subequations}
where $(x_\perp,\vartheta)$ are polar coordinates for the $x_2 x_3$ plane. A strange feature of \eqref{E:energydensity} is that it exhibits a transition from a region of energy depletion behind the quark, to a region of energy depletion in front of it as the quarks velocity decreases. 
When $v^2>5/8$ there is a buildup of energy density ahead of it, forming a ``bulldozer effect.'' See figure \ref{F:energy1}. As it slows down extra lobe-like features appear until $v^2 < 5/13$ where the energy buildup is behind the quark, creating an ``inverse-bulldozer'' effect. See figure \ref{F:energy2}. Recall that the speed of sound in a conformal fluid is $v^2 = 1/3$, so that this transition occurs at velocities which are higher than the speed of sound. This indicates that the features we are seeing are not hydrodynamic in nature. A more detailed analysis of the deviation of the energy density from linearized hydrodynamics can be found in the literature.\cite{Noronha:2007xe}   We will see in section~\ref{S:Hadronization} that it is probably the near field of the stress-energy tensor which dominates high-angle emission of hadrons.  It would certainly be interesting to understand the physical mechanism behind this near-field behavior.

\section{Numerical results for the holographic stress tensor}
\label{S:Stress}

\begin{figure}[hbtp]
\begin{center}
	\scalebox{1.2}{\includegraphics{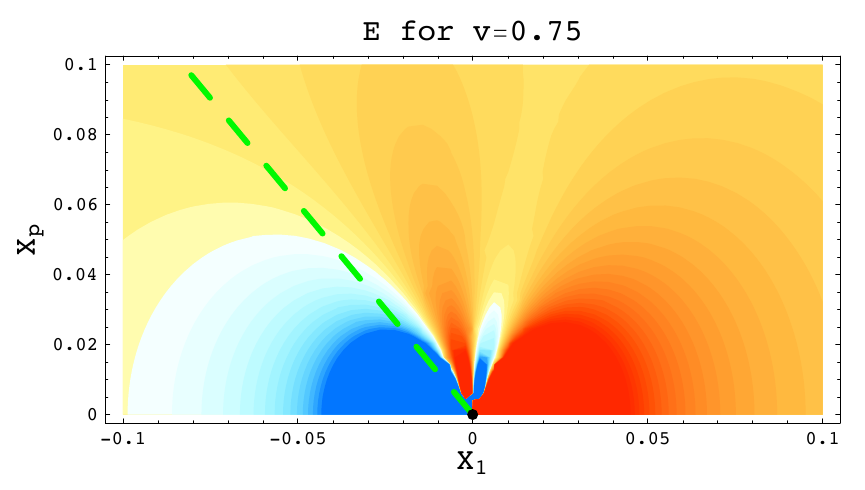}}
	\caption[Short]{\label{F:energy2} (Color online.)  A contour plot of the energy density near the moving quark, with the bath and the Coulombic contributions subtracted.\cite{Gubser:2007xz} Red signifies energies above the background value of the plasma while blue signifies energies below the background value of the plasma. The black dot at $X_1 = X_p = 0$ marks the location of the quark which is moving at a constant velocity of $v=0.75$, well above the speed of sound.  We work in dimensionless units where $X_1 = \pi T x_1$ and $X_p = \pi T x_{\bot}$. The dashed green line shows the presumed location of the Mach cone.}\end{center}
\end{figure}

Expression \eqref{E:nearfield} and the Fourier transform of \eqref{E:TmnSmallK} capture the near-field and far-field asymptotics of the stress-energy tensor. In \ref{SSS:hydro}, we have seen an indication that far from the moving source the energy-momentum tensor exhibits hydrodynamic behavior.  In \ref{SS:Smallrasymptotics}, we have seen that the near-field stress tensor exhibits non-hydrodynamic behavior with interesting features, like the multi-lobe structure in figure~\ref{F:energy2}.  In the intermediate regime, there is a transition region between hydrodynamics and whatever short-distance physics governs the near field. To probe this region, one needs solutions to \eqref{E:Lichnerowicz} for values of $K$ where no analytic asymptotic treatment is available. We have obtained such solutions numerically. First, \eqref{E:ASet}, \eqref{E:BSet}, \eqref{E:CSet}, \eqref{E:DSet}, and \eqref{E:ESet}  were solved, and $Q_A$, $Q_D$, and $Q_E$ were obtained.\cite{Friess:2006fk} Then, the resulting momentum space stress-energy tensor was passed through an FFT to position space using a $128^3$ grid. Such a computation has been carried out for the energy density\cite{Gubser:2007xz,Chesler:2007an} and for the Poynting vector.\cite{Gubser:2007ga,Chesler:2007sv}

\begin{figure}[hbtp]
\begin{center}
	\scalebox{1.2}{\includegraphics{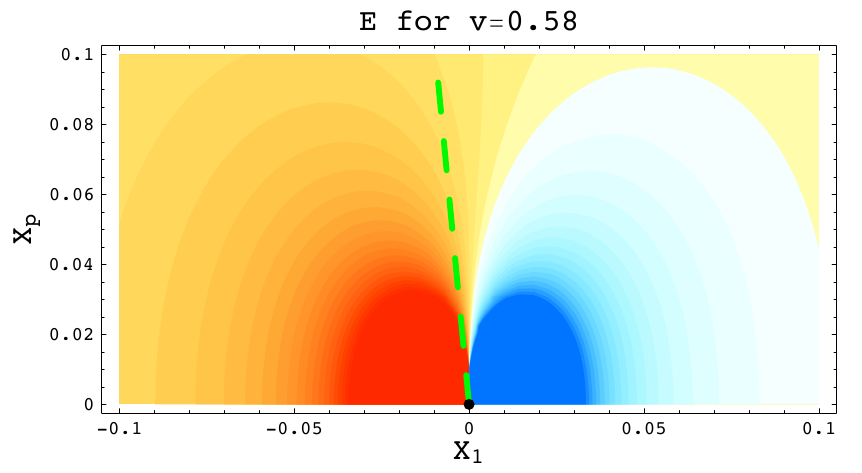}}
	\caption[Short]{\label{F:energy1} (Color online.)  A contour plot of the energy density near the moving quark, with the bath and the Coulombic contributions subtracted.\cite{Gubser:2007xz} Red signifies energies above the background value of the plasma while blue signifies energies below the background value of the plasma. The black dot at $X_1 = X_p = 0$ marks the location of the quark which is moving at a constant velocity of $v=0.58$, just above the speed of sound.  We work in dimensionless units where $X_1 = \pi T x_1$ and $X_p = \pi T x_{\bot}$. The dashed green line shows the presumed location of the Mach cone.}
\end{center}
\end{figure}

\begin{figure}[hbtp]
\begin{center}
	\scalebox{0.8}{\includegraphics{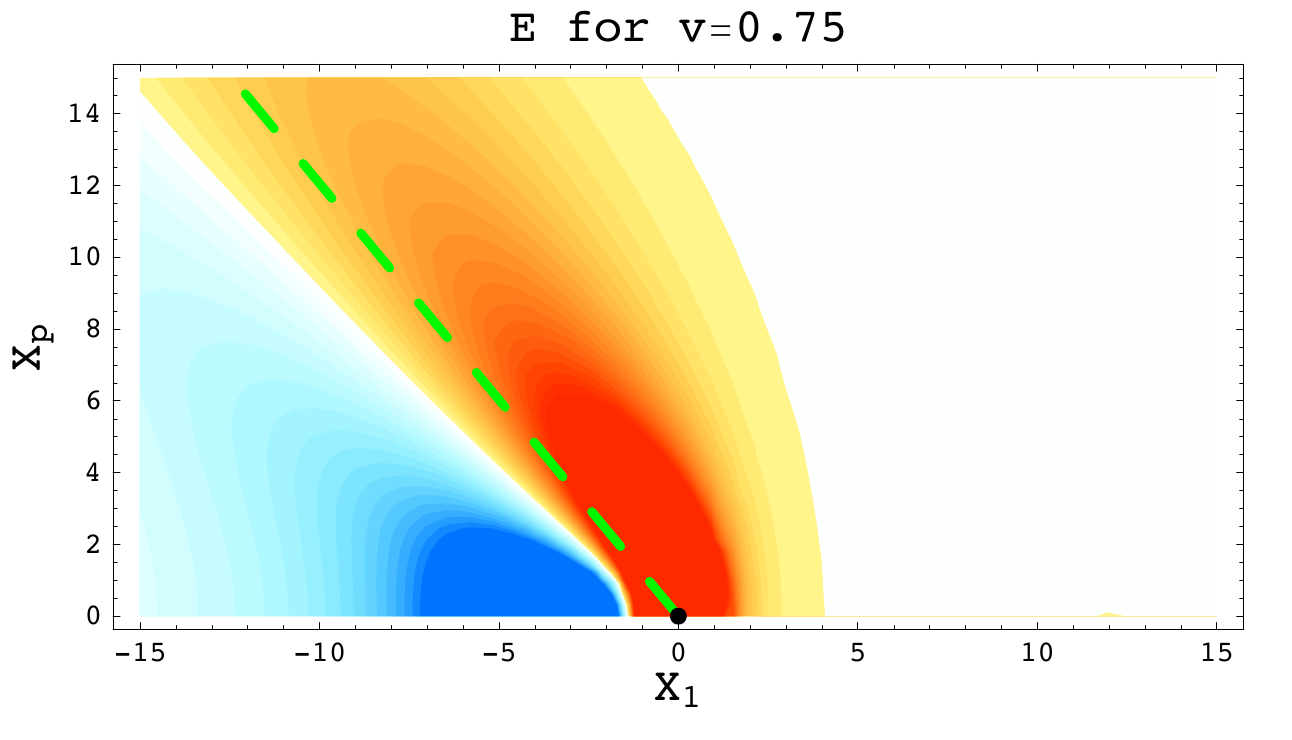}}
	\caption[Short]{\label{F:energy3} (Color online.)  A contour plot of the energy density far from the moving quark, with the bath and the Coulombic contributions subtracted.\cite{Gubser:2007xz} Red signifies energies above the background value of the plasma while blue signifies energies below the background value of the plasma. The black dot at $X_1 = X_p = 0$ marks the location of the quark which is moving at a constant velocity of $v=0.75$, well above the speed of sound.  We work in dimensionless units where $X_1 = \pi T x_1$ and $X_p = \pi T x_{\bot}$. The dashed green line shows the presumed location of the Mach cone.}
\end{center}
\end{figure}

Consider the normalized energy density
\begin{equation}\label{E:calEDef}
	\mathcal{E} = \frac{\sqrt{1-v^2}}{(\pi T)^4 \sqrt{\lambda}}
		 \langle T_{00}^K \rangle \,,
\end{equation}
where $\langle T_{mn}^K \rangle$ is defined in \eqref{E:TmnKDef} as the stress-energy tensor of the system, minus the stress-energy of the thermal bath, minus the divergent delta-function contribution \eqref{E:TmnFromPmnAgain} at the position of the moving quark.  It is convenient to decompose this rescaled energy density into a Coulombic term, a near-field (large momentum) term, a far-field (small momentum) term, and a residual term: 
\begin{equation}\label{E:calESum}
	\mathcal{E} = \mathcal{E}_{\rm Coulomb}+\mathcal{E}_{\rm UV}+\mathcal{E}_{\rm IR}+\mathcal{E}_{\rm res} \,.
\end{equation}
The Coulombic term represents the contribution coming from the near field of the quark:
\begin{equation}
	\mathcal{E}_{\rm Coulomb} = -\left(2 + v^2\right)\frac{\sqrt{K^2 - K_1^2 v^2}}{24 \pi}
		+ (1-v^2) \frac{v^2 K_1^2}{24 \pi} \,,
\end{equation}
which is what we found in \eqref{E:Tmnquark} converted to momentum space. It can be read off of the $\mathcal{O}(K)$ terms in \eqref{E:EnergylargeK}.
The far-field term, $\mathcal{E}_{\rm IR}$, scales like $\mathcal{O}(K^{-5})$ at large momentum and asymptotes to \eqref{E:EnergysmallK} at small momenta.\footnote{Actually, we require that it asymptote to \eqref{E:EnergysmallK} up to a momentum-independent constant.} There are many possible expressions which satisfy the above criteria. Taking note of the resummation \eqref{E:subtractedEIR}, we used
\begin{equation}
\label{E:subtractedEIR}
	\mathcal{E}_{\rm IR} = 
		-\frac{1}{2\pi} \frac{3 i v K_1(1+v^2)-3 v^2 K_1^2}{K^2 - 3 v^2 K_1^2 - i v K^2 K_1}
		+\frac{1}{2\pi} \frac{3 i v K_1(1+v^2)-3 v^2 K_1^2}{K^2 - 3 v^2 K_1^2 - i v K^2 K_1+\mu_{\rm IR}^2}
\end{equation}
where $\mu_{\rm IR}$ is a typical scale where \eqref{E:EnergysmallK} stops being valid. We used $\mu_{\rm IR}=1$. Similarly, $\mathcal{E}_{\rm UV}$ scales like $\mathcal{O}(K^1)$ at small momenta, and asymptotes to \eqref{E:EnergylargeK} at large momenta. To regulate the large momentum expressions in the IR we made the replacement
\begin{equation}
\label{E:UVreplacement}
	\frac{1}{(K^2 - v^2 K_1^2)^{n/2}} = \frac{1}{(K^2 - K_1^2 v^2 + \mu_{\rm UV}^2)^{n/2}}
		\left(1 - \frac{\mu_{\rm UV}^2}{K^2 - K_1^2 v^2 + \mu_{\rm UV}^2}\right)^{-n/2}
\end{equation}
where the last term in \eqref{E:UVreplacement} is expanded to order $\mathcal{O}(K^{-5})$. Similar to \eqref{E:subtractedEIR}, $\mu_{\rm UV}$ is a cutoff scale which we set to $1$. After the replacement \eqref{E:UVreplacement}, the first two terms in a series expansion of the energy density take the form
\begin{equation}
\label{E:UVenergy}
	\mathcal{E}_{\rm UV} = -\frac{(2+v^2) \sqrt{K^2 - K_1^2 v^2 + \mu_{\rm UV}^2}}{24}
		+ \frac{2 K_1^2 v^2 (1-v^2) - (2 + v^2) \mu_{\rm UV}^2}{48 \sqrt{K^2 - K_1^2 v^2 \mu_{\rm UV}^2}}+ \ldots \,.
\end{equation}
Once $\mathcal{E}_{\rm UV}$, $\mathcal{E}_{\rm IR}$, and $\mathcal{E}_{\rm Coulomb}$ are known
we can numerically compute $\mathcal{E}_{\rm res}$, which can be fed through a three-dimensional FFT with controllable errors because it is absolutely integrable. We then add back to the real space 
numerical expression for $\mathcal{E}_{\rm res}$ the real 
space version of $\mathcal{E}_{\rm IR}$, $\mathcal{E}_{\rm UV}$, and $\mathcal{E}_{\rm Coulomb}$ to obtain the energy density in position space. 
The Fourier transform of the large momentum asymptotic expressions can be carried out using \eqref{E:MainFT}.  As explained in section \ref{SSS:hydro}, Fourier transforming the sound pole is difficult due to the cubic term in the denominator---the term associated with the shear viscosity.
To convert the the sound pole structure of \eqref{E:subtractedEIR} to real space we first rewrote it as
\begin{equation}
	\mathcal{E}_{\rm IR} = \frac{\mathcal{A}(K_1)}{K_{\bot}^2 + m (K_1)^2} \,,
\end{equation}
then Fourier transformed in the $K_{\bot}$ direction using \eqref{E:MainFT}, and resorted to numerics to FFT the remaining $K_1$ coordinate. This was carried out on a line with 1944 points with $K_1$ ranging from $-20$ to $20$.
In figures \ref{F:energy1}--\ref{F:energy3} we show the energy density (with the Coulombic field $\mathcal{E}_{\rm Coulomb}$ subtracted) at various spatial scales.

\begin{figure}[hbt]
\begin{center}
\scalebox{1}{\includegraphics{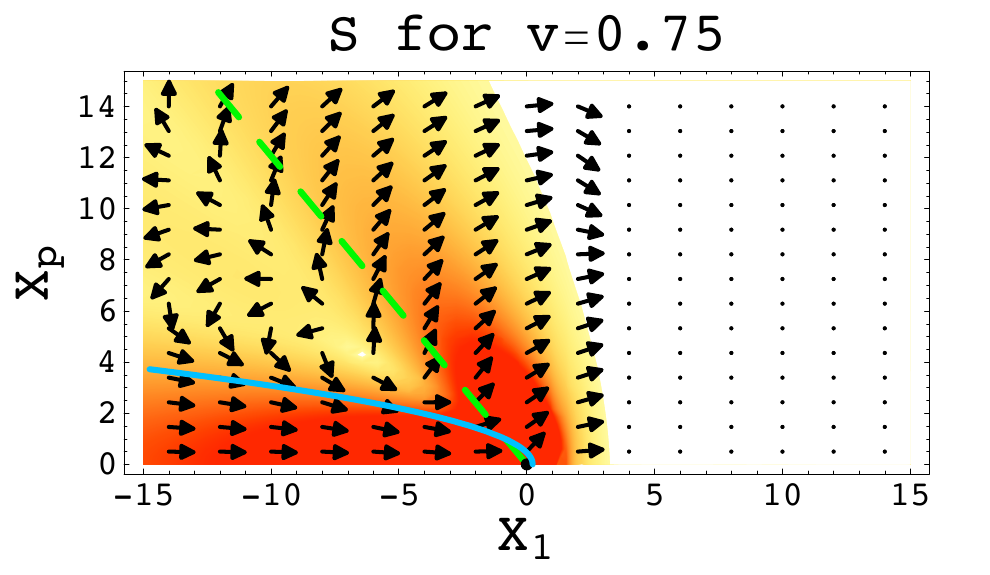}}
\caption[Short]{(Color online.)  Contour plot of the magnitude of the Poynting vector, with the Coulombic contribution subtracted.\cite{Gubser:2007ga} The magnitude of the Poynting vector goes from red (large) to white (zero) while the arrows show its direction. The dashed green line shows the presumed location of the Mach angle, and the blue line shows the location of the laminar wake---as dictated by its large distance asymptotics.
\label{F:SixPlots}}
\end{center}
\end{figure}

The components of the energy flux can be treated in a similar manner: we define 
\begin{equation}\label{E:calSDef}
	{\mathcal{S}_i} = -\frac{\sqrt{1-v^2}}{(\pi T)^4 \sqrt{\lambda}}
		 \langle T_{0i}^K \rangle
\end{equation}
and decompose $\vec{\mathcal{S}}$ into
\begin{equation}
	\vec{\mathcal{S}} = \vec{\mathcal{S}}_{\rm Coulomb}+\vec{\mathcal{S}}_{\rm UV}+\vec{\mathcal{S}}_{\rm IR}+\vec{\mathcal{S}}_{\rm res} \,.
\end{equation}
The Coulombic expression for the Poynting vector is given by the $\mathcal{O}(K)$ terms in \eqref{E:S1largeK} and \eqref{E:S2largeK}.
The small momentum expressions are given by
\begin{align}
\label{E:subtractedS1}
\notag
	\mathcal{S}_{1\,\rm IR} =& -\frac{1}{2\pi} \frac{i(1+v^2)K_1+v K^2 - 2 v^3 K_1^2}{K^2 - 3 v^2 K_1^2-i v K^2 K_1}
	+\frac{1}{2\pi} \frac{i(1+v^2)K_1+v K^2 - 2 v^3 K_1^2}{K^2 - 3 v^2 K_1^2-i v K^2 K_1+\mu_{\rm IR}^2} \\
	&+\frac{2 v }{\pi} \frac{1+i K_1/4 v}{K^2 - 4 i v K_1}
	-\frac{2 v }{\pi} \frac{1+i K_1/4 v}{K^2 - 4 i v K_1+\mu_{\rm IR}^2} \\
\label{E:subtractedSp}
	\mathcal{S}_{\bot\,\rm IR} =& - \frac{1}{2\pi} \frac{i(1+v^2)K_\bot + b^2 K_1 K_{\bot})}
		{K^2-3 v^2 K_1^2-i v K^2 K_1}
		+\frac{1}{2\pi} \frac{i K_{\bot}}{K^2- 4 i v K_1} + \left( \hbox{regulators} \right)
\end{align}
where we have set $\mu_{\rm IR}=1$ and by ``$(\hbox{regulators})$'' we mean terms containing the regulator $\mu_{\rm IR}$, analogous to those in \eqref{E:subtractedEIR} and \eqref{E:subtractedS1}.  The large momentum expressions are given by applying \eqref{E:UVreplacement} to \eqref{E:S1largeK} and \eqref{E:S2largeK}. As was the case for the energy density, we used $\mu_{\rm UV}=1$.
The real space results for the Poynting vector for $v^2 = 3/4$ are shown in figure \ref{F:SixPlots}.
\clearpage

\section{Hadronization, jet-broadening, and jet-splitting}
\label{S:Hadronization}

There is a significant gap between the results of sections~\ref{S:Einstein}--\ref{S:Stress} and experimental data.  Before reviewing recent attempts to bridge this gap, let's briefly summarize some of the relevant data.  There are of course more authoritative summaries in the experimental literature.\cite{Adams:2005ph,Adler:2005ee,Adare:2008cqb,Ulery:2008pj,Abelev:2008nd,Ulery:2008rw}

The data seem to reveal a phenomenon of ``jet-splitting,'' whereby an energetic parton traversing the medium deposits so much of its energy through high-angle emission that---with appropriate momentum cuts and subtractions---the extra particle production due to the parton is at a minimum in the direction of its motion, and reaches a maximum at an angle roughly $1.2\,{\rm radians}$ away.  Jet-splitting is most simply illustrated through two-point histograms of the azimuthal angle $\Delta\phi$ separating a pair of energetic hadrons close to mid-rapidity.  To understand the phenomena better, it is useful to examine two landmark studies of these histograms: one from STAR\cite{Adams:2005ph} and one from PHENIX.\cite{Adler:2005ee}

\begin{figure}[hbtp]
  \begin{center}
   \includegraphics[width=3.7in]{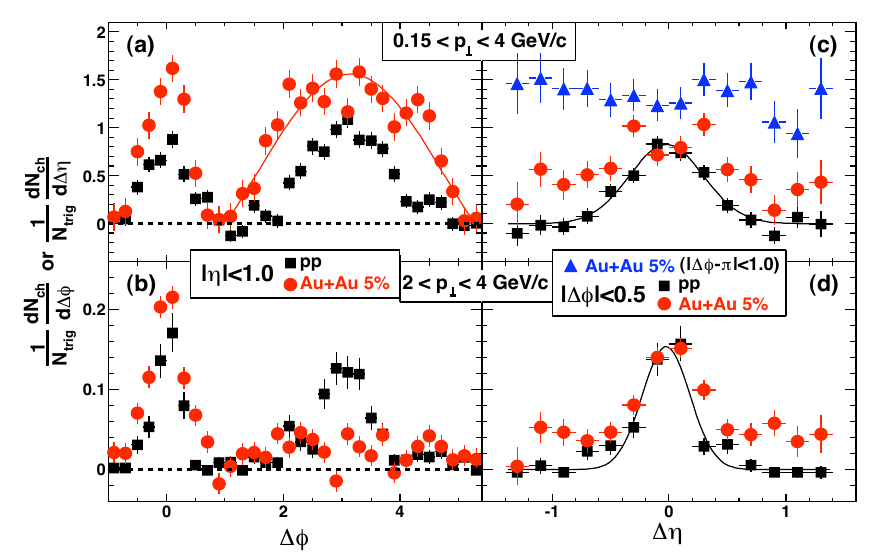}\qquad
   \includegraphics[width=3.7in]{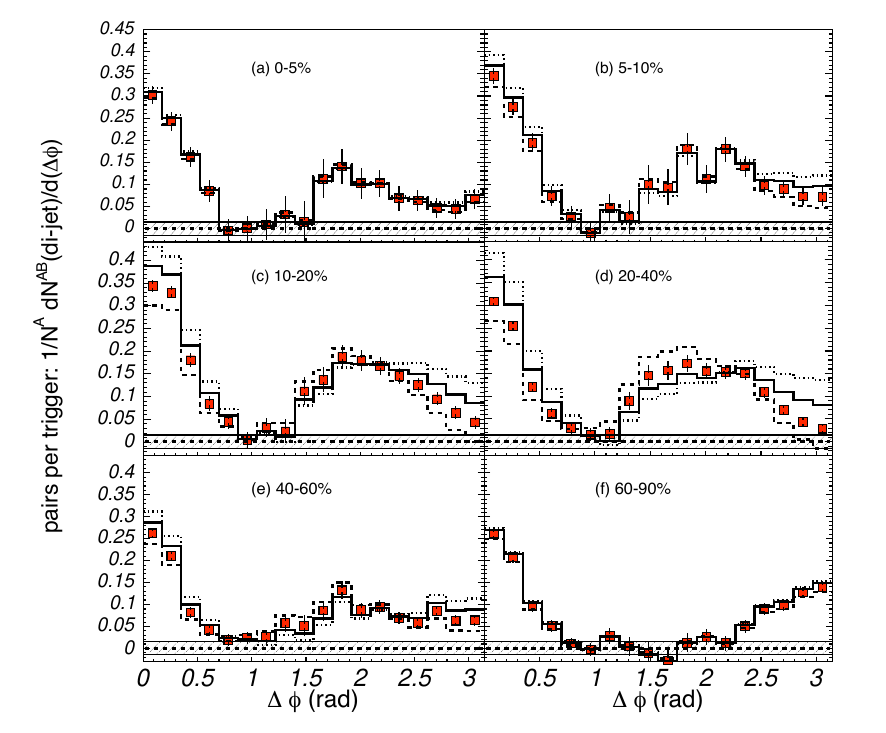}
	\caption[Short]{\label{F:BroadenOrSplit} (Color online.)  Top: The STAR analysis\cite{Adams:2005ph} shows substantial broadening of the away-side jet.  Reprinted figure~1 with permission from J.~Adams {\em et al.},~ Phys.~Rev.~Lett.~95, 152301 (2005), http://link.aps.org/abstract/PRL/v95/p152301.  Copyright 2005 by the American Physical Society.\newline  Bottom: The PHENIX analysis\cite{Adler:2005ee} shows jet-splitting for sufficiently central events. Reprinted figure~2 with permission from S.~S.~Adler {\em et al.},~Phys.~Rev.~Lett.~97, 052301 (2006), http://link.aps.org/abstract/PRL/v97/p052301.  Copyright 2006 by the American Physical Society.\newline
In reference to the images in this figure, we note in accord with the publisher's guidelines that ``Readers may view, browse, and/or download material for temporary copying purposes only, provided these uses are for noncommercial personal purposes. Except as provided by law, this material may not be further reproduced, distributed, transmitted, modified, adapted, performed, displayed, published, or sold in whole or part, without prior written permission from the American Physical Society.''
}
  \end{center}
 \end{figure}

In the STAR analysis, fairly inclusive momentum cuts were considered: under one set of cuts, the less energetic of the two hadrons was required to have transverse momentum greater than $150\,{\rm MeV/c}$.  The resulting data show a peak for nearly collinear hadrons that is approximately the same shape for central gold-gold collisions as for proton-proton collisions: see figure~\ref{F:BroadenOrSplit}.  This ``near-side jet'' feature can reasonably be supposed to arise from vacuum fragmentation effects.  The two-point histogram also shows an ``away-side jet'' feature around $\Delta\phi = \pi$ which is substantially broader for central gold-gold collisions than for proton-proton.  Neither the ``near-side jet'' nor the ``away-side jet'' are reconstructed jets in the usual sense; instead, they are ideas that help explain the main features of histograms assembled from millions of events.  It is usually assumed that the typical event contributing to the histograms involves a hard scattering event where one parton escaped the medium without much interaction, producing the highly energetic ``trigger hadron,'' and the other parton interacted substantially with the medium before generating an ``associated hadron'' in the vicinity of $\Delta\phi = \pi$.  The upshot is that with inclusive momentum cuts, there is substantial broadening of the away-side jet, but not jet-splitting: associated hadron production is still maximized, or statistically indistinguishable from its maximum, at $\Delta\phi = \pi$.  With tighter momentum cuts on the associated hadron, the data used in the particular STAR analysis under discussion show striking jet-broadening, but the scatter in the data is sufficient to prevent firm conclusions from being drawn---from this particular study---about whether there is jet-splitting.  (Subsequent STAR analyses of both two- and three-point hadron correlators provide strong evidence in favor of jet-splitting.\cite{Ulery:2008pj,Abelev:2008nd,Ulery:2008rw})

The PHENIX analysis\cite{Adler:2005ee} is similar to the STAR analysis,\cite{Adams:2005ph} but with more restrictive cuts: in particular, the less energetic hadron was required to have transverse momentum greater than $2.5\,{\rm GeV}/c$.  The resulting histograms, with a zero-yield-at-minimum (ZYAM) subtraction, show a distinct minimum in associated hadron production at $\Delta\phi = \pi$, with a broad maximum in the ballpark of $\Delta\phi = \pi - 1.2$.  This jet-splitting persists down to roughly $50\%$ centrality, meaning that it occurs for events where the impact parameter is less than about $10\,{\rm fm}$.  The ZYAM subtraction is an important part of the analysis, especially for the less central events.  The reason a subtraction is needed is that for a non-central collision, there is an angular modulation of single-particle yields, approximately proportional to $1 + v_2 \cos 2\phi$, where the zero of $\phi$ coincides with the azimuthal direction of the impact parameter, and the elliptic flow coefficient $v_2$ depends on the transverse momentum and species.  Two-point hadron correlators receive a contribution from single-hadron yields.  The ZYAM scheme is to subtract a multiple of the appropriate product of single-particle yields.  The multiple is chosen so that the resulting histogram has one bin with zero net events, while all other bins have a positive net number of events.

A natural hypothesis is that the high-angle emission leading to either jet-splitting or jet-broadening can be described in terms of a sonic boom in the medium.\cite{Stoecker:2004qu,CasalderreySolana:2004qm}  Two related difficulties afflict this idea.  First, it's hard to get a sonic boom with a big enough amplitude to account for the data\cite{Adler:2005ee} with reasonable rates of energy loss;\cite{Chaudhuri:2005vc,CasalderreySolana:2006sq} it should be noted however that not all investigators agree on this point,\cite{CasalderreySolana:2005rf} and that there are some phenomenological models based on sonic booms that fit the data.\cite{Renk:2005si,Renk:2006mv}  Second, one usually finds a diffusion wake with comparable strength to the sonic boom.\cite{CasalderreySolana:2004qm,Gubser:2007ga,Gubser:2007ni,Betz:2008ka}  At least in a static medium, it is hard to get jet-splitting in the presence of a significant diffusion wake.

To compute the relative strength of the diffusion wake and sonic boom for the heavy quark in the SYM theory, we go back to the conservation equation \eqref{E:FFourier},
\begin{equation}
\label{E:Drag}
	F^n  = i z_H^2 \lim_{\vec{K} \to 0} K_m \langle T_K^{mn} \rangle \,.
\end{equation}
In section \ref{SSS:hydro} we saw that at small $K$, the components of the stress-energy tensor may be decomposed into  terms containing sound poles at $K^2 \sim 3 K_1^2 v^2$ and terms associated with a wake which have a pole at $K_1 \sim 0$. If $\langle T_{\rm IR}^{mn} \rangle$ is the leading, small $K$ contribution to the stress-energy tensor, then we may decompose
\begin{equation}
\label{E:TmnIRdecomposition}
	\langle T_{\rm IR}^{mn} \rangle = \langle T_{\rm sound}^{mn} \rangle 
	  + \langle T_{\rm wake}^{mn} \rangle\,,
\end{equation}
where
\begin{equation}
	\langle T_{\rm sound}^{mn} \rangle = 
	 -\frac{(\pi T)^4 \sqrt{\lambda}}{\sqrt{1-v^2}}
	\frac{1+v^2}{2\pi (K^2 - 3 K_1^2 v^2)}
	 \begin{pmatrix}
	 	3 i K_1 v & i K_1 & i K_2 & i K_3 \\
		i K_1 & i K_1 v & 0 & 0 \\
		i K_2 & 0 & i K_1 v & 0 \\
		i K_3 & 0 & 0 & i K_1 v 
	\end{pmatrix}
\end{equation}
and
\begin{equation}
	\langle T_{\rm wake}^{mn} \rangle = 
	 \frac{(\pi T)^4 \sqrt{\lambda}}{\sqrt{1-v^2}}
	\frac{i}{2 \pi K_1}
	 \begin{pmatrix}
	 	0 & 1 & 0 & 0 \\
		1 & 0 & 0 & 0 \\
		0 & 0 & 0 & 0 \\
		0 & 0 & 0 & 0		
	\end{pmatrix}.
\end{equation}
Since only the terms in \eqref{E:TmnIRdecomposition} contribute to the total drag force in \eqref{E:Drag}, this gives a natural division of the total drag force: 
\begin{equation}
	F^n_{\rm sound} = i z_H^2 \lim_{\vec{K} \to 0}  K_m \langle T_{\rm sound}^{mn} \rangle  
	\qquad
	F^n_{\rm wake} = i z_H^2 \lim_{\vec{K} \to 0}  K_m \langle T_{\rm wake}^{mn} \rangle\,,
\end{equation}
From \eqref{E:TmnSmallK}, \eqref{E:Holographicdata}, \eqref{E:Tijasymptotics}, and \eqref{E:nonhydroasymptotics}, we find that
\begin{equation}
	F^0_{\rm sound} = -\frac{1}{v^2} F^0
	\qquad
	F^0_{\rm wake} = \left(1+\frac{1}{v^2}\right) F^0 \,.
\end{equation}
In our conventions, the zero component of $F^n$ gives us the total rate of change in the energy density plus any energy flux going out of the system. Thus, the ratio of energy going into sound waves to energy going into the wake is
\begin{equation}
\label{E:OurRatio}
	1+v^2:-1 \,.
\end{equation}
While sound modes carry energy away from the moving quark, the wake feeds energy in toward the quark.  While this may seem counter-intuitive, in some sense it's obvious: the diffusion wake consists of a flow of the medium forward toward the quark.  The forward-moving momentum in the diffusion wake is the momentum deposited by the quark at earlier times.  Qualitatively, \eqref{E:OurRatio} says that the diffusion wake and the sonic boom have comparable strength.
When comparing \eqref{E:OurRatio} to the scenarios of energy loss in the literature,\cite{CasalderreySolana:2004qm,Chaudhuri:2005vc} one finds that \eqref{E:OurRatio} quantitatively matches a scenario where the relative strength of the wake is so large that it washes out features of jet-splitting associated with the sonic boom.  It may be significant, however, that the medium is infinite and static, both in our work and in the linear hydrodynamic scenario\cite{CasalderreySolana:2004qm} that our results match onto at large length scales.

In light of the difficulties in explaining the data with a ``boom and wake'' model, focused on the hydrodynamic regime, 
it is natural to investigate the effect on hadron production of the region of the medium close to the moving quark where hydrodynamics is inapplicable.  This has been pursued in a series of works,\cite{Noronha:2007xe,Noronha:2008tg,Noronha:2008ef,Noronha:2008un,Gyulassy:2008fa,Betz:2008wy,Betz:2008dm,Torrieri:2009mv} which we briefly summarize in the next few paragraphs.  The first idea is to subtract away the leading order Coulombic contribution to $\langle T_{mn} \rangle$.  Up to an overall multiplicative rescaling, these are the quantities we denoted ${\cal E}_{\rm Coulomb}$ and $\vec{\cal S}_{\rm Coulomb}$ in section~\ref{S:Stress}.  The justification for this is that these fields describe the energy of the energetic parton itself, not the energy lost from it.  The remaining energy density, which we will denote as $\epsilon_{\rm sub}$, can be split up as $\epsilon_{\rm sub} = \epsilon_{\rm bath} + \Delta\epsilon$.  (Note that in contrast to our definitions of ${\cal E}$ and its variants, $\epsilon_{\rm sub}$ explicitly includes the contribution from the bath.)  The Poynting vector $\vec{S}_{\rm sub}$, with the Coulombic contribution subtracted away, is non-zero only because of the presence of the quark.  The basic plan is to use the Cooper-Frye algorithm\cite{Cooper:1974mv} to convert string theory predictions for $(\epsilon_{\rm sub},\vec{S}_{\rm sub})$ into a spectrum of hadrons.

The Cooper-Frye algorithm is based on converting a fluid element at temperature $T$ and with local four-velocity $U^m$ into hadrons according a Maxwell-Boltzmann distribution in the local rest frame:
 \begin{eqnarray}
  f(p_T,\phi) &=& \left. {dN \over p_T dp_T d\phi} \right|_{y=0} = 
    -\int_{\bf R^3} d\Sigma_\mu \, P^\mu 
      e^{U^\mu P_\mu / T}  \nonumber \\
     &=& \int_0^\infty x_\perp dx_\perp \int_0^{2\pi} d\varphi
         \int_{-\infty}^\infty dx_1 \, E
        e^{U^\mu P_\mu / T} \,,
       \label{E:CooperFrye}
 \end{eqnarray}
where $N$ is the number of hadrons, and we have set
 \begin{equation}\label{E:PmuDef}
  P^m = \begin{pmatrix} p_T & p_T \cos(\pi-\phi) &
   p_T \sin(\pi-\phi) & 0 \end{pmatrix}
 \end{equation}
and
 \begin{equation}\label{E:UDef}
  U^m = \begin{pmatrix} U^0 & U^1 & U_\perp \cos\varphi & 
    U_\perp \sin\varphi \end{pmatrix} \,.
 \end{equation}
Note that because of our choice of mostly plus signature, the energy of the hadron in the local rest frame of the fluid is $-U^m P_m$.  Also because of this choice of signature, we are obliged to include an explicit minus sign in the first integral expression of (\ref{E:CooperFrye}).

\begin{figure}
  \centerline{\includegraphics[width=4.5in]{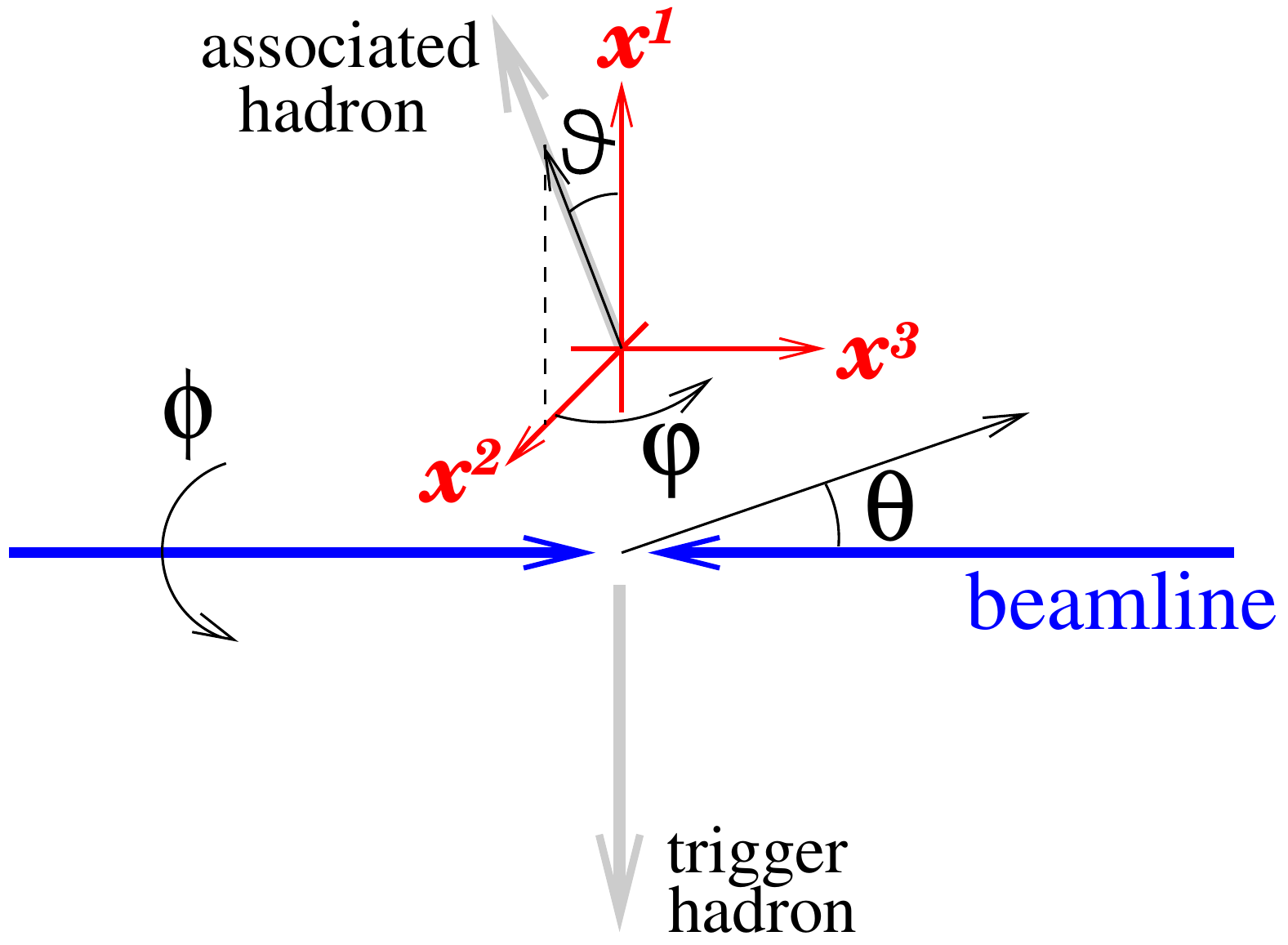}}
  \caption[Short]{(Color online.)  A sketch of the coordinate systems used to describe away-side hadron production.  If the trigger hadron is at $\phi=0$, then a helpful relation at mid-rapidity ($\theta = \pi/2$) is $\vartheta = \pi - \phi$, where $\Delta\phi$ is the azimuthal separation between the trigger and associated hadrons.}\label{F:coords}
 \end{figure}

To understand (\ref{E:CooperFrye})--(\ref{E:UDef}), it helps to refer to figure~\ref{F:coords}.  The momentum of the associated hadron is $P^m$, and (\ref{E:CooperFrye}) is written in the approximation that the associated hadron is massless---an excellent approximation since a typical hadron of interest is a pion with $p_T \sim 3\,{\rm GeV}/c$.  The rapidity $y$ is related to the angle from the beamline $\theta$ by 
 \begin{equation}\label{RapidityDef}
  \tanh y = \cos\theta \,.
 \end{equation}
(Note that rapidity $y$ has nothing to do with the depth coordinate $y=z/z_H$ used in previous sections.)  The freeze-out surface is chosen to be a slice of constant $x^0$ in (\ref{E:CooperFrye}).  This is the best motivated choice for an infinite, asymptotically static medium.  In an expanding medium, a more usual choice is a fixed-temperature surface with the temperature set close to the QCD scale.  For isochronous freeze-out, the measure $d\Sigma^m$ is simply $dx^1 dx^2 dx^3 \, \begin{pmatrix} 1 & 0 & 0 & 0 \end{pmatrix}$, and in passing to the second line of (\ref{E:CooperFrye}) we have simply expressed the metric on ${\bf R}^3$ in radial coordinates.  It is important to realize that the azimuthal angle $\varphi$ around the direction of motion of the parton (assumed to be in the $+x^1$ direction, as usual) is different from the azimuthal angle $\phi$ around the beamline.  We have omitted in (\ref{E:CooperFrye}) a subtraction of the contribution of the bath to hadron production, which depends on $p_T$ but not on $\phi$.  We have also not attempted to normalize $f(p_T,\phi)$: doing so would involve partitioning over the spectrum of hadrons.

The information from string theory enters into (\ref{E:CooperFrye}) in two ways.  First, the local four-velocity $U^m$ is the local rest frame of the medium, in which the Poynting vector vanishes.  Second, the temperature $T$ is the temperature in this local rest frame, deduced by plugging the energy density $T_{mn}^{\rm sub} U^m U^n$ into the equation of state.  Evidently, one needs all components of $T^{mn}_{\rm sub}$ in order to precisely determine $U^m$ and $T$.  This is a problem since only the $m=0$ row of the stress tensor has been computed in full.\cite{Gubser:2007ga,Chesler:2007sv} Another problem is that close to the quark, the medium is presumably far from equilibrium, so using Cooper-Frye seems somewhat perilous.  We will return to a discussion of these two issues below.

To understand how hydrodynamical and non-hydrodynamical effects contribute to the spectrum of produced hadrons, one must have some notion of where the boundary is between hydrodynamical and non-hydrodynamical regimes.  This boundary is presumably not sharp.  Three considerations have gone into identifying an appropriate boundary:
 \begin{enumerate}
  \item The non-hydrodynamical region can be chosen as the region where $\Delta\epsilon / \epsilon_{\rm bath}$ is less than some constant of order unity.  For 
 \begin{equation}
  v=0.9 \,,\quad \lambda = 5.5 \,,\quad
   N=3 \,,\quad T_{\rm SYM} = 200\,{\rm MeV} \,,
   \label{E:QuarkParams}
 \end{equation}
a preferred choice is 
 \begin{equation}
  \Delta\epsilon / \epsilon_{\rm bath} \leq 0.3 \,. \label{E:NeckDef}
 \end{equation}
Here and below, we will describe the region defined by~(\ref{E:NeckDef}) with the parameter choices (\ref{E:QuarkParams}) as the ``Neck.''  It extends roughly over $-1 \leq X_1 \leq 0.5$ and $0 \leq X_\perp \leq 1.7$.  (For $T_{\rm SYM} = 200\,{\rm MeV}$, $X=1$ corresponds to $x = 1 / \pi T_{\rm SYM} = 0.31\,{\rm fm}$.)
  \item The Neck region can be compared with the region where the Knudsen number exceeds some constant of order unity.  An appropriate version of the Knudsen number in the current context is
 \begin{equation}\label{E:KnDef}
  \Kn \equiv \Gamma {|\nabla \cdot \vec{S}_{\rm sub}| \over 
    |\vec{S}_{\rm sub}|} \,.
 \end{equation}
Here the sound attenuation length $\Gamma$ is the same as the one discussed following \eqref{E:constitutive}: $\Gamma = 4\eta / 3 s T = 1/3\pi T$. For the choice of parameters~(\ref{E:QuarkParams}), examination of the near-field expressions~(\ref{E:nearfield}) shows that the region where $\Kn \gsim 1/3$ is somewhat bigger in the $X_1$ direction than the Neck.  However, corrections to the near-field approximation to $\vec{S}_{\rm sub}$ may not be negligible for $X_1$ and/or $X_\perp$ of order unity.
  \item The Neck region can be compared with the region where the constitutive relations of hydrodynamics break down.  Given all components $T^{mn}_{\rm sub}$ of the stress tensor, with the Coulomb field subtracted away, there is a straightforward procedure for testing the constitutive relations.  First determine the local velocity field $U^m$ by passing to the local rest frame of the fluid.  Let $(T^{mn})_L$ be the subtracted stress-energy tensor in the local rest frame.  The energy density is read off immediately as $(T^{00})_L$; the pressure is deduced from the equation of state; and the shear viscosity contribution to the space-space parts of the stress tensor can be determined from $U^m$ and its gradient.  Deviations of the space-space components of $(T^{mn})_L$ from the combined contribution of pressure and shear viscosity are measures of the failure of hydrodynamics.
  
A study\cite{Noronha:2008ef} of the near-field expressions~(\ref{E:nearfield}) for a somewhat different choice of parameters from (\ref{E:QuarkParams}) (namely $v=0.99$, $\lambda=3\pi$, $N=3$) concludes that deviations from hydrodynamics are appreciable out as far as $X \sim 8$.  However, the near-field expressions definitely cannot be trusted at such large distances.\footnote{A computation of the energy radiated from a moving quark\cite{Chesler:2007sv} shows that agreement with linearized hydrodynamics is already fairly good at $X \sim 5$.  Based on results of an earlier study,\cite{Gubser:2007xz} the onset of reasonable agreement with linearized hydro occurs near $X = 3$.}  This analysis could therefore be considerably improved if all components of $T^{mn}_{\rm sub}$ were computed directly from string theory.
 \end{enumerate}

The main conclusion to draw from points 2 and 3 is that in the Neck region, the subtracted stress tensor is essentially unrelated to hydrodynamics.  Instead, the physics may be presumed to be dominated by strong coherent color fields combined with responses of the medium to strong field gradients: hence the term ``chromo-viscous neck.''

\begin{figure}
  \centerline{\includegraphics[width=3in]{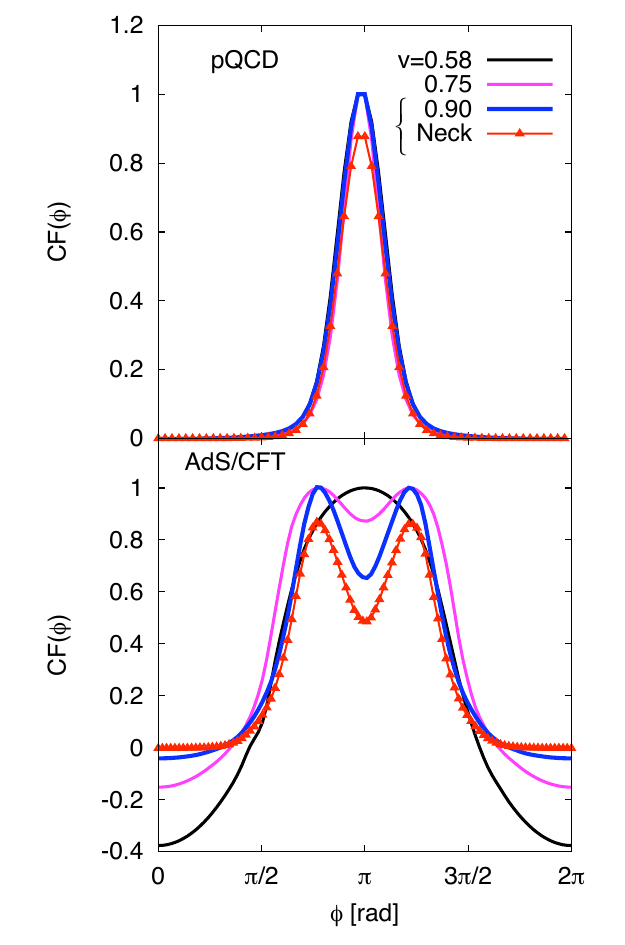}}
  \caption[Short]{(Color online.)  Hadron production based on a Cooper-Frye hadronization of a perturbative QCD calculation and of  trailing string results (AdS/CFT).\cite{Betz:2008wy}  Bath contributions have been subtracted away, and the curves have all been normalized to have the same maximum when the Cooper-Frye integral is carried out over the entire volume accessible to each computation.  The Neck region for the perturbative QCD calculation is again defined as a region close to the quark where deviations from hydrodynamics are significant.
}\label{F:SplitCompare}
 \end{figure}

To return to hadronization: The Cooper-Frye integral over ${\bf R}^3$ can be split into the Neck region and the ``Mach'' region---which is everything else.  In the Mach region, where the energy density comes mainly from the bath, a good approximation to the local rest frame can be found by setting
 \begin{equation}\label{UChoice}
  \vec{U} = {3 \over 4} {\vec{S}_{\rm sub} \over \epsilon_{\rm bath}} \,.
 \end{equation}
In the neck region, this approximation is less reliable, but because space-space components of $T^{mn}$ are not available from a string theory calculation, it is hard to give a better motivated prescription for determining the local rest frame.  With the choice (\ref{UChoice}), the result is that the Neck contribution to the Cooper-Frye integral leads to a distinctive double-peaked structure in $f(p_T,\phi)$ for $p_T$.  This is remarkable when compared to the single-peaked structure emerging from a computation in a perturbative QCD framework based on Joule heating,\cite{Betz:2008wy} which is similarly passed through the Cooper-Frye hadronization algorithm.  See figure~\ref{F:SplitCompare}.

The double-peaked structure from the Neck region of the trailing string stress tensor has nothing to do with the Mach cone.  It doesn't occur at the same angle: for example, at $v=0.58$, the Mach angle $\vartheta = \cos^{-1} c_s / v$ is very nearly zero, but the Neck region still produces a double peak structure (not shown in figure~\ref{F:SplitCompare}) about a radian away from $\phi=\pi$.  When $v$ is very close to $1$, the double peaks get closer to $\phi = \pi$.  This is reminiscent of the structure observed at large $K$ in Fourier space,\cite{Friess:2006aw,Friess:2006fk} but the peaks observed in the predicted hadron spectra are more widely separated than the ones in Fourier space.

While the hadronization studies\cite{Noronha:2007xe,Noronha:2008tg,Noronha:2008ef,Noronha:2008un,Gyulassy:2008fa,Betz:2008wy,Betz:2008dm,Torrieri:2009mv} give valuable insight into the relation of the trailing string to high-angle hadron emission from an energetic parton, it is not claimed that the results are fully realistic, or that a direct comparison to untagged dihadron histograms, like the ones in figure~\ref{F:BroadenOrSplit}, is justified.  Let us review the potential difficulties.  First, the trailing string describes an infinitely massive quark that propagates at a constant velocity through an infinite, static, thermal medium.  Fluctuations leading to stochastic motions of finite mass quarks may significantly affect the results.  Also, there may be an effective ``form factor'' for massive quarks that partially smears out the field close to the quark.  In addition, it is not obvious that Cooper-Frye is justified, because the crucial effect comes from the non-equilibrium part of the medium (the Neck).  Subtracting away the Coulomb field is certainly well-motivated physically, but it is possible to maintain some skepticism about whether it is the correct prescription in combination with Cooper-Frye.  Finally, the approximation (\ref{UChoice}) to the local rest frame is imprecise in the Neck region.

Despite these potentially serious issues, the punchline of the phenomenological studies\cite{Noronha:2007xe,Noronha:2008tg,Noronha:2008ef,Noronha:2008un,Gyulassy:2008fa,Betz:2008wy,Betz:2008dm,Torrieri:2009mv} seems to us likely to be robust: the near-quark region has a substantially greater tendency toward high angle emission in the trailing string treatment than in the perturbative QCD treatment based on Joule heating.  Modulo concerns already expressed, the near-field contribution to high-angle emission is stronger than the contribution of the hydrodynamical regime, and it results in a significant double-hump structure, reminiscent of jet-splitting.

\section{Conclusions}
\label{S:Conclusions}

Let us conclude by addressing the four main questions we raised in the introduction:
 \begin{enumerate}
  \item {\it What is the rate of energy loss from an energetic probe?}

For a heavy quark moving at a velocity $v$, the drag force is $F_{\rm drag} = -{\pi\sqrt\lambda \over 2} T^2 {v \over \sqrt{1-v^2}}$.  This is explained in more detail in section~\ref{S:Magnitude}.\medskip
  \item {\it What is the hydrodynamical response far from the energetic probe?}

There is a sonic boom with Mach angle $\vartheta = \cos^{-1} c_s / v$, where $c_s = 1/\sqrt{3}$ is the speed of sound dictated by conformal invariance.  There is also a diffusion wake of comparable strength.  These points are explained in section~\ref{SSS:hydro} and~\ref{S:Stress}.\medskip
  \item {\it What gauge-invariant information can be extracted using the gauge-string duality about the non-hydrodynamic region 
near the probe?}

The expectation values $\langle T^{m0} \rangle$ of the energy density and the Poynting vector of the gauge theory stress tensor have been computed with uniformly good accuracy across all length scales for several values of the velocity of the heavy quark, as we review in section~\ref{S:Stress}.  Analytic approximations to the space-space components of $\langle T^{mn} \rangle$ are also available at small length scales: see section~\ref{SS:Smallrasymptotics}.\medskip
  \item {\it Do the rate and pattern of energy loss have some meaningful connection to heavy ion phenomenology?}

A suitable translation of parameters from SYM to QCD results in estimates of energy loss for $c$ and $b$ quarks which are not far from realistic, or which may be fully realistic.  We summarize these estimates in section~\ref{S:Magnitude}.

Studies of hadronization starting from the string theory predictions for the energy density and Poynting vector indicate that the trailing string leads to significant high-angle emission from the Neck region, close to the quark, suggestive of jet-splitting.  We describe these studies in section~\ref{S:Hadronization}.  Although it is premature to make detailed comparisons to data, it is clearly worthwhile to extend and refine both the string theory analysis and the phenomenological studies.
 \end{enumerate}

\subsection*{Acknowledgments}

We thank M.~Gyulassy and J.~Noronha for useful correspondence.  This work was supported in part by the Department of Energy under Grant No.\ DE-FG02-91ER40671 and by the NSF under 
award number PHY-0652782.  F.D.R.~was also supported in part by the FCT grant SFRH/BD/30374/2006.

\clearpage
\appendix
\section{{}\ \hskip0.69in Notation}
\label{S:Glossary}

In this appendix we present short explanations of some of the nomenclature and mathematical notations used in the main text.
 \begin{itemize}
  \item[$x^\mu$:] The five spacetime coordinates of AdS${}_5$-Schwarzschild, usually $(t,x^1,x^2,x^3,z)$.
  \item[$x^m$, $p_m$:] The four-vectors for position and momentum in ${\bf R}^{3,1}$.  We use mostly plus signature, so (for example) $\eta^{mn} p_m p_n = -E^2 + \vec{p}^2 = -m^2$.
  \item[$\vec{x}$, $\vec{p}$:] The three-vectors for position and momentum: spatial components of $x^m$ and $p^m$.
  \item[$x_1$:] This could mean either $G_{1\mu} x^\mu = {L^2 \over z^2} x^1$ or $\eta_{1\nu} x^\nu = x^1$.  Our convention is to prefer the latter; likewise $x_2 = x^2$ and $x_3 = x^3$.
  \item[$x_{\bot}$:] The radial distance from the quark in the $x_2$, $x_3$ plane: $x_\perp = \sqrt{x_2^2+x_3^2}$.  Occasionally we consider the two-vector $\vec{x}_\perp = (x_2,x_3)$.
  \item[$z$:] This is the depth coordinate in AdS${}_5$ or AdS${}_5$-Schwarzschild which is $0$ at the boundary and has dimensions of length.  The AdS${}_5$ metric is $ds^2 = {L^2 \over z^2} (-dt^2 + d\vec{x}^2 + dz^2)$.
  \item[$z_H$:] The depth of the horizon in AdS${}_5$-Schwarzschild, related to the temperature by $T = 1/\pi z_H$.
  \item[$y$:] Usually, a rescaled depth coordinate in AdS${}_5$-Schwarzschild, defined by $y = z/z_H$.  But in section~\ref{S:Hadronization} we use $y$ to denote rapidity, i.e.~$\tanh y = p_z / E$ where $p_z$ is the momentum along the beampipe and $E$ is the energy. 
  \item[$r$:] We use $r$ to indicate a radial separation in ${\bf R}^3$.  Some authors use $r$ to denote the depth coordinate $r = L^2/z$ in AdS${}_5$.
  \item[AdS${}_5$:]  Five-dimensional anti-de Sitter space, the maximally symmetric negatively curved spacetime in $4+1$ dimensions. Its metric is given by \eqref{E:LineElement} with $h=1$.
  \item[SYM:] An abbreviation for ``${\cal N}=4$ super-Yang-Mills theory in four dimensions,'' which is the theory controlling the low-energy excitations of D3-branes.
  \item[$N$:] Usually, the number of colors: $N=3$ in QCD.  An exception is that in section~8, we use $N$ to indicate the number of hadrons predicted by the Cooper-Frye algorithm.
  \item[$g_{\rm YM}$:] The gauge coupling of SYM, normalized so that $g_{\rm YM}^2 N = L^4/\alpha'^2$, where $N$ is the number of colors.
  \item[$g_s$:] The gauge coupling of QCD.  We also use $\alpha_s = g_s^2/4\pi$.
  \item[$\lambda$:] The 't~Hooft coupling, $\lambda = g_{\rm YM}^2 N$.
  \item[$L$:] The radius of curvature of AdS${}_5$.
  \item[$\alpha'$:] The Regge slope parameter of fundamental strings, see \eqref{E:NambuGoto}.
  \item[$\kappa$:] The five-dimensional gravitational coupling.
  \item[$G_{\mu\nu}$:] The spacetime metric of AdS${}_5$ or AdS${}_5$-Schwarzschild.
  \item[$R$:] The Ricci scalar in AdS${}_5$ or AdS${}_5$-Schwarzschild.  We also use the Ricci tensor $R_{\mu\nu}$ and the Riemann tensor $R_{\alpha\mu\beta\nu}$.  Our conventions are $R = G^{\mu\nu} R_{\mu\nu}$ and $R_{\mu\nu} = G^{\alpha\beta} R_{\alpha\mu\beta\nu}$, with signs arranged so that $R_{\mu\nu} = -{4 \over L^2} G_{\mu\nu}$ in AdS${}_5$ of radius $L$.
  \item[$h$:] The ``blackening function'' for AdS${}_5$-Schwarzschild, whose metric is given in \eqref{E:LineElement}.  It is given by $h(z) = 1 - z^4/z_H^4$.  We sometimes think of $h$ as a function of the depth $z$, and sometimes as a function of $y = z/z_H$.  $h'$ always means $h'(y) = -4 y^3$.
  \item[$T$:] The temperature in the dual field theory, which is the same as the Hawking temperature of the dual black hole background.  The temperature of the AdS${}_5$-Schwarzschild background \eqref{E:LineElement} is $T = 1/\pi z_H$.
  \item[$g_{\alpha\beta}$:] The worldsheet metric of a string.
  \item[$\sigma^\alpha$:] Coordinates on the string worldsheet.
  \item[$v$:] The speed of a moving quark.  Usually we take this motion to be in the $+x_1$ direction.
  \item[$\xi$:] Gives the shape of the string that describes the quark in the five-dimensional geometry. See \eqref{E:TrailingAnsatz} and \eqref{E:TrailingString}.
  \item[$x_*^\mu(\sigma)$:] The embedding function for a classical string in AdS${}_5$-Schwarzschild.  When no ambiguity is possible, we denote this embedding function more simply as $x^\mu(\sigma)$.
  \item[$R_{AA}$:] The nuclear modification factor, defined as the number of particles produced (usually at a particular value of $p_T$ and in a specified range of rapidity) in a collision of two nuclei with atomic number $A$, divided by the number produced in a proton-proton collision scaled up by the effective number of binary nucleon-nucleon collisions in the heavy-ion collision.
  \item[$\tau_{\mu\nu}$:] The five-dimensional stress-energy tensor, not to be confused with the expectation value of the boundary stress-energy tensor $\langle T_{mn} \rangle$.  For the trailing string, $\tau_{\mu\nu}$ is given in \eqref{E:Gottaumunu}.
  \item[$\tau_{\mu\nu}^K$:] The Fourier components of $\tau_{\mu\nu}$.  See \eqref{E:FourierSpace} and \eqref{E:GottaumunuK}.
  \item[$h_{\mu\nu}$:] Small metric perturbations around AdS${}_5$-Schwarzschild.
  \item[$h_{\mu\nu}^K$:] The Fourier components of $h_{\mu\nu}$ defined by analogy with \eqref{E:FourierSpace}.
  \item[Axial gauge:] A gauge choice for the metric perturbations where $h_{\mu z} = 0$.
  \item[$H_{mn}$:] The Fourier components of $h_{\mu\nu}^K$ in axial gauge, up to a normalization factor.  See \eqref{E:hmunuFourier}.  We think of the $H_{mn}$ as functions of $y = z/z_H$, not of $z$.
  \item[$A$:] The even tensor mode combination of metric perturbations in axial gauge: see (\ref{E:HmnDefA}).  Similar definitions for $B_i$, $C$, $D_i$, and $E_i$ follow, and gauge-invariant combinations $B$, $D$, and $E$ can be found in (\ref{E:BDInvDef})-(\ref{E:EInvDef}).  These are all functions of $y = z/z_H$.
  \item[$\psi_S$:] The scalar master field.  A master field is a gauge-invariant combination of metric fluctuations in AdS${}_5$-Schwarzschild with a simple equation of motion.  We encountered four other master fields: $\psi_V^{\rm even}$, $\psi_V^{\rm odd}$, $\psi_T^{\rm even}$, and $\psi_T^{\rm odd}$: see \eqref{E:masterfields}.
  \item[$\alpha_v$:] A recurring normalization factor given by $\alpha_v=1/\left(2\pi\alpha'\sqrt{1-v^2}\right)$.
 \item[$\vec{K}$:] The momentum conjugate to $\vec{x}/z_H$; used as a dimensionless wave-number to parameterize the three-dimensional Fourier space used to describe the medium's response to the quark.
 \item[$K_\perp$:] The magnitude of the component of $\vec{K}$ perpendicular to the motion of the quark.
 \item[$R_{mn}$:] The asymptotic values of $H_{mn}$ at $y = 0$.  The boundary condition $R_{mn} = 0$ says that the four-dimensional metric which the boundary gauge theory experiences is flat Minkowski space.
 \item[$P_{mn}$:] The coefficients of $y^3$ in a small $y$ expansion of $H_{mn}$.  See \eqref{E:HmnBoundary} and \eqref{E:PXval}.  They are related to the divergent contribution to the boundary stress-energy tensor given in \eqref{E:TmnFromPmn}.
 \item[$Q_{mn}$:] The coefficients of $y^4$ in a small $y$ expansion of $H_{mn}$ when $R_{mn} = 0$.  They are related to the expectation value of the boundary stress-energy tensor by \eqref{E:TmnFromQmn}.
 \item[$g_{mn}$:] The metric of the boundary conformal field theory, usually set equal to the Minkowski metric $\eta_{mn}$ with mostly plus signature.
 \item[$u^m$:] The four-velocity of a heavy quark moving through the thermal medium.
 \item[$\langle T_{mn} \rangle$:] The one-point function of the stress-energy tensor in SYM in the presence of the moving quark.  We find it convenient to decompose it into three pieces given in \eqref{E:TmnKDef}.
 \item[$\Thydro_{mn}$:] A stress tensor which satisfies the hydrodynamic constitutive relations, \eqref{E:constitutive}. $(T^{\rm hydro,SYM})_{mn}$ is the hydrodynamic contribution to the stress tensor of the SYM theory.
 \item[$\langle T_{mn}^K \rangle$:]  The Fourier modes of the contribution of the moving quark to the stress-energy tensor in the dual field theory minus the divergent piece corresponding to the infinite mass of the quark.  See \eqref{E:TmnKDef}.  It can be computed from $Q_{mn}$ through \eqref{E:TmnFromQmnAgain}.
 \item[$f_n$:] The source term for the energy-momentum tensor, $i K_m T^{mn} = f^n$. Various superscripts specify which contribution of the energy-momentum tensor is being sourced. For example, $f_n^{\rm hydro}$ sources $\Thydro_{mn}$.
 \item[$R_X$:]  The coefficient of the leading homogeneous solution for various linear combinations of $H_{mn}$ and their derivatives, such as $A$, $B$, $B_i$, $\psi_T$, etc.  Our boundary condition is $R_X = 0$.
 \item[$P_X$:]  The analog of $P_{mn}$ for various linear combinations of $H_{mn}$ and their derivatives.
 \item[$Q_X$:]  The analog of $Q_{mn}$ for various linear combinations of $H_{mn}$ and their derivatives.  See \eqref{E:GotQmnFromQX} and \eqref{E:GotQXFromPsi}.  Also, we use the notations $Q_D = Q_{D_1}$ and $Q_E = Q_{E_1}$. 
 \item[$p_T$:] The component of momentum perpendicular to the beamline.
 \item[$\vartheta$:] An angular coordinate in momentum space, $\sin\vartheta = K_1/K$, or in real space, $\sin\vartheta = x_1/x$.
 \item[$\varphi$:] The azimuthal angle around the direction of motion of an energetic quark.  Our usual convention is that the energetic quark moves in the $+x^1$ direction, and then $\tan\varphi = K_3/K_2$ or $x_3/x_2$.
 \item[$\theta$:] The angle of a trajectory relative to the beam.  $\theta=\pi/2$ is mid-rapidity.
 \item[$\phi$:] The azimuthal angle around the beam.  The angular variable $\Delta\phi$ in dihadron histograms is the separation in $\phi$ between two hadrons.
 \item[$\mathbf{I},\mathbf{K}$:] Modified Bessel functions of the first and second kind.  $\mathbf{J}$ is a Bessel function of the first kind, and $\mathbf{L}$ is a modified Struve function.
 \item[Neck:] The neck is the region near a moving quark where the response of the medium is non-hydrodynamical.  In practice, for the choice of parameters (\ref{E:QuarkParams}), the Neck can be defined, as in (\ref{E:NeckDef}), as the region where the energy density, excluding the Coulombic contribution, exceeds $1.3$ times the asymptotic energy density of an infinite static bath.
 \item[$U^m$:] The four-velocity of a fluid element, usually defined so that it vanishes in precisely the same Lorentz frame in which the Poynting vector vanishes.
 \item[$\mathcal{E},\vec{\cal S}$:] The energy density and Poynting vector, rescaled to make them dimensionless, with contributions from the thermal bath excluded: see~(\ref{E:calEDef}) and~(\ref{E:calSDef}).  ${\cal E}$ can be decomposed into a sum of contributions from the Coulomb field of the quark, subleading UV effects, IR effects, and a residual quantity ${\cal E}_{\rm res}$, as in~\eqref{E:calESum}.  An analogous decomposition can be performed on $\vec{\cal S}$.
 \end{itemize}

\clearpage
\def\href#1#2{{\tt #2}}
\bibliographystyle{ssg}
\bibliography{qgp4}

\end{document}